\documentclass[11pt]{article}
 \usepackage{amsmath}
\usepackage{amssymb}
\usepackage{bbm}
\usepackage{graphicx}
 \usepackage{fullpage}
\usepackage{multirow}
\usepackage{caption}
\usepackage{array}
\usepackage{wrapfig}
\usepackage{natbib,bm}
\usepackage{pdflscape}
\usepackage{booktabs}
\usepackage{rotating}
\usepackage{tcolorbox}
\usepackage{subfig}
\usepackage{enumitem}
\usepackage{xr}
\usepackage{setspace}
\newcommand{\bi}{\begin{itemize}}
	\newcommand{\ei}{\end{itemize}}
\newcommand{\be}{\begin{enumerate}}
	\newcommand{\ee}{\end{enumerate}}

\begin{document}
\title{Borrowing Strength Across Exposures and Outcomes via Index Models for Multi-pollutant Mixtures}

\author{Glen McGee\\
	  Department of Statistics and Actuarial Science, University of Waterloo, ON, Canada \\ {\tt glen.mcgee@uwaterloo.ca}
	    \and 
	   Joseph Antonelli \\
	   Department of Statistics, University of Florida, FL, USA\\
       {\tt jantonelli@ufl.edu}}

\maketitle

\abstract{An important goal of environmental health research is to assess the health risks posed by mixtures of multiple environmental exposures. In these mixtures analyses, flexible models like Bayesian kernel machine regression and multiple index models are appealing because they allow for arbitrary non-linear exposure-outcome relationships. However, this flexibility comes at the cost of low power, particularly when exposures are highly correlated and the health effects are weak, as is typical in environmental health studies. We propose a multivariate index modelling strategy that borrows strength across exposures and outcomes by exploiting similar mixture component weights and exposure-response relationships. In the special case of distributed lag models, in which exposures are measured repeatedly over time, we jointly encourage co-clustering of lag profiles and exposure-response curves to more efficiently identify critical windows of vulnerability and characterize important exposure effects. We then extend the proposed approach to the multiple index model setting where the true index structure---the number of indices and their composition---is unknown, and introduce variable importance measures to quantify component contributions to mixture effects. Using time series data from the National Morbidity, Mortality and Air Pollution Study, we demonstrate the proposed methods by jointly modelling three mortality outcomes and two cumulative air pollution measurements with a maximum lag of 14 days.}


\section{Introduction}
Quantifying associations between health outcomes and mixtures of environmental pollutants is a public health priority, and addressing the methodological challenges of mixtures analyses has been the focus of much recent research \citep{davalos2017current,stafoggia2017statistical,hamra2018environmental,maitre2022state,yu2022review,tanner2020environmental,joubert2022powering}. 
Common to mixtures studies is a lack of power---owing in part to limited sample sizes, highly correlated exposures, and small effect sizes. This is exacerbated by the fact that environmental exposures may have non-linear relationships with health outcomes, so the functional form of these relationships must also be estimated from limited data.  Moreover, in studies of time-varying exposures, the timing of exposure effects is itself unknown: for example, studies of prenatal exposure to metals attempt to identify  windows of time in which children are most susceptible to exposure effects \citep{barr2000workshop}.

Many methods  address one or more of these issues. To handle high correlation among mixture components in a linear model, Dirichlet Process (DP) priors have been placed on  coefficients \citep{herring2010nonparametric,dunson2008bayesian,lalonde2020discovering}, among other latent variable approaches \citep{zavez2023bayesian}. 
Non-linear relationships have been captured using splines \citep{antonelli2020estimating,bai2022spike} and Gaussian processes  \citep{bobb2015bayesian,ferrari2019identifying}, as well as by single \citep{wang2020singleindex} and multiple index models \citep{mcgee2021bayesian, mcgee2024collapsible}. When exposures are measured repeatedly over time, distributed lag models have smoothed associations over time and characterized lag profiles to identify high risk time windows \citep{gasparrini2010distributed,gasparrini2014modeling,gasparrini2017penalized,wilson2017bayesian,wilson2022kernel,antonelli2024multiple}. While most of these approaches are intended to address multiple exposures, multiple time points, or both, few additionally address the complexity stemming from having multiple outcomes \citep{warren2012bayesian,jain2018multivariate}. 


It is common in environmental research to be interested in multiple health outcomes, but analyzing each separately ignores the fact that associations might be similar across outcomes. For example, in the motivating analysis of cause-specific mortality and air pollution in Chicago (1987--2000), separately modeling lagged effects for each outcome yields wide intervals for associations between pollutants and respiratory/other non-accidental deaths, which could be compatible with either highly positive or negative associations.

In this paper we propose a flexible index modelling framework for analyzing the relationships between environmental mixtures and health outcomes, and develop a co-clustering prior structure that borrows information across multiple outcomes and indices to improve power. The framework contains several important special cases, including a novel multivariate distributed lag non-linear model (MV-DLNM) when exposures are measured over time, in which the proposed co-clustering prior exploits similarity in  exposure-response curves as well as in time-lag profiles. 
In our application to the Chicago air pollution mortality analysis, the proposed approach exploits similarity across several associations, resulting in more efficient inference and evidence of null relationships, while still capturing the strong non-linear effect of ozone on cardiovascular mortality. The framework also extends naturally to other exposure structures, such as exposures measured via multiple biomarkers. When exposures are measured only once, we show how the method can be used to fit multiple index models (MIMs) when the true index structure---the number of indices and their composition---is unknown \textit{a priori}. To aid interpretability, we also propose exposure importance measures that quantify the strength of each component’s contribution to the mixture.
\vspace*{-8pt}

\section{Methods}\label{sec:methods}
\subsection{General Model Formulation and Parameterization}
\label{ss:model}
For the $i^{th}$ study unit ($i=1,\dots,n$), let $\mathbf{y}_i=(y_{i1},\dots,y_{iK})^T$ be a $K$-dimensional outcome vector, and let $\mathbf{x}_i^*$ be a general exposure vector of length $M$. Examples include: (i) $P$ exposures measured at $L$ times ($M=PL)$; (ii) $P$ exposures measured at $B$ different biomarkers ($M=PB$); (iii) $P$ exposures measured cross-sectionally ($M=P$). Finally let $\boldsymbol{z}_i$ be a vector of other covariates. For a given $J$, we write the proposed model:
\begin{align}
y_{ik}&= \sum_{j=1}^J f_{kj}\left([\mathbf{A}_{j}\mathbf{x}^*_{i}]^T{\boldsymbol{\omega}}_{kj}\right) +\boldsymbol{z}_i^T\boldsymbol{\beta}_{Zk}+\xi \sigma_k u_i+\epsilon_{ik} , ~~k=1,\dots,K,\label{eqn:general}
\end{align}
where $\mathbf{A}_j$ is an $r\times M$ design matrix chosen by the analyst. The choice of $\mathbf{A}_j$ determines the model structure; we highlight two special cases in Sections \ref{ss:dlnm} and \ref{ss:amim}, and outline others in Appendix \ref{app:special}. Here $f_{kj}(\cdot)$ is an unknown smooth exposure-response function,  $\boldsymbol{\omega}_{kj}$ is an $r$-vector of unknown weights, and $\boldsymbol{\beta}_{Zk}$ is a vector of coefficients corresponding to covariates $\boldsymbol{z}_i$. We let $\epsilon_{ik}$ represent independent Gaussian errors, $\epsilon_{ik}\overset{iid}{\sim} N(0,\sigma_k^2)$ and allow for dependence across the same  unit's outcomes via shared random effect $u_i\overset{iid}{\sim} N(0,1)$, scaled by the coefficient $\xi$. We adopt a random intercepts structure here for simplicity, but more flexible random effect structures could be included with minor modification to the estimation procedure. To identify both $f_{kj}(\cdot)$ and $\boldsymbol{\omega}_{kj}$, constraints on the weights' magnitude (${\boldsymbol{\omega}}_{kj}^T{\boldsymbol{\omega}}_{kj}=1$) and sign (e.g., ${{\omega}}_{kj1}>0$)  are typically imposed.

 For generality, we approximate ${\boldsymbol{\omega}}_{kj} \approx \mathbf{\Psi} {\boldsymbol{\theta}}_{kj}$, where $\mathbf{\Psi}$ is a known $r\times m$ matrix whose columns define an orthonormal basis as in \cite{wilson2022kernel}, and ${\boldsymbol{\theta}}_{kj}$ is an $m$-dimensional ($m\leq r$) vector of unknown coefficients. Equivalently,  $\tilde{\mathbf{x}}_{ij}^T {\boldsymbol{\theta}}_{kj} \approx [\mathbf{A}_j\mathbf{x}_{i}^*]^T \boldsymbol{\omega}_{kj}$, where $\tilde{\mathbf{x}}_{ij}=\mathbf{\Psi}^T \mathbf{A}_j \mathbf{x}_{i}^* $.  Typically we avoid the approximation and set $\boldsymbol{\omega}_{kj}=\boldsymbol{\theta}_{kj}$, i.e. $\mathbf{\Psi}=\mathbf{I}_{r\times r}$, but we allow for $\mathbf{\Psi}$ with $m<r$ to facilitate dimension reduction and improved computation when $r$ is large.  
We approximate $f_{kj}(\cdot)$ via B-splines: $f_{kj}(\tilde{{\mathbf{x}}}_{ij}^T{\boldsymbol{\theta}}_{kj})\approx \mathbf{b}_{ij,\boldsymbol{\theta}_{kj}}^T \boldsymbol{\beta}_{kj}$ where $\mathbf{b}_{ij,\boldsymbol{\theta}_{kj}}=[b_{1}(\tilde{\mathbf{x}}_{ij}^T \boldsymbol{\theta}_{kj}),\dots, b_{d}(\tilde{\mathbf{x}}_{ij}^T \boldsymbol{\theta}_{kj})]^T$, $b_{1}(\cdot),\dots,b_{d}(\cdot)$ are known  basis functions,
and $\boldsymbol{\beta}_{kj}=(\beta_{kj1},\dots,\beta_{kjd})^T$ is a vector of unknown coefficients. For identifiability, the basis is subject to centering constraints \citep{wood2017generalized}, and we  include outcome-specific intercepts $\beta_{k0}$.

\subsection{Cluster-Inducing Prior Structure}
\label{ss:clustering}

In the proposed model, there are $K\times J$  distinct functions $f_{kj}(\cdot)$ and $K\times J$ distinct weight profiles $\boldsymbol{\omega}_{kj}$, which can be overly flexible. In small samples, or when the correlation across indices is high, as is common in environmental epidemiology, this may lead to instability and low power,  which is already limited in environmental studies prone to weak effect sizes. To rein in this flexibility, we exploit similarity in both  $f_{kj} (\cdot)$ and in ${\boldsymbol{\omega}}_{kj}$ across pairs $(k,j)$. For example when $\mathbf{A}_{j}\mathbf{x}_i^*$ represents the $j^{th}$ exposure  measured at $L$ different times (as in Section \ref{ss:dlnm}), this matches our expectation that exposures have similar exposure-response curves, and the same exposure may have similar associations with different outcomes. 

A natural way to borrow information would be to adopt independent DP priors: $\boldsymbol{\beta}_{kj}\sim G_{\beta}$ where $G_{\beta} \sim DP(\alpha_{\beta}, G_{0, \beta})$,  and $\boldsymbol{\theta}_{kj}\sim G_{\theta}$ where $G_{\theta} \sim DP(\alpha_{\theta}, G_{0, \theta})$. Here $(\alpha_{\beta}, \alpha_{\theta})$ are  concentration parameters, dictating the degree of clustering, and $(G_{0, \beta}, G_{0, \theta})$ are centering distributions. To see how this encourages clustering across exposure-outcome pairs, consider the following latent variable representation \citep{sethuraman1994constructive}:
\begin{align*}
\boldsymbol{\beta}_{kj}&=\sum_{c=1}^{\infty} I(Z^\beta_{kj}=c)\boldsymbol{\beta}^*_{c} &
\boldsymbol{\theta}_{kj}&=\sum_{c=1}^{\infty}  I(Z^\theta_{kj}=c)\boldsymbol{\theta}^*_{c} \\
P(Z^\beta_{kj}=c)&= \pi_c^{\beta}= V^{\beta}_{c} \prod_{j < c} (1 - V^{\beta}_{j})&
P(Z^\theta_{kj}=c)&= \pi_c^{\theta} =V^{\theta}_{c} \prod_{j < c} (1 - V^{\theta}_{j}) ~~ c=1, 2, \dots\\
V^{\beta}_{c} &\sim \text{Beta}(1, \alpha_{\beta}) &
V^{\theta}_{c} &\sim \text{Beta}(1, \alpha_{\theta}), ~~c=1,2,\dots, \\
\boldsymbol{\beta}^*_c &\sim G_{0,\beta} &
\boldsymbol{\theta}^*_c &\sim G_{0,\theta}.
\end{align*}
Here $(Z^\beta_{kj}, Z^\theta_{kj})$ is a discrete latent variable that indicates to which cluster the pair $(k,j)$ belongs for $\boldsymbol{\beta}_{kj}$ and $\boldsymbol{\theta}_{kj}$, respectively. In practice we use a finite mixture with $C$ clusters ($V^{\beta}_{C}=V^{\theta}_{C}=1$), as there are only a finite number of parameters to be estimated, since there are $J \times K$ parameter vectors. Therefore we recommend setting $C = J \times K$ or to some fraction of this in order to speed up computation time without losing model flexibility. We assess sensitivity to this choice in Appendix G.4 and find that results are not sensitive to this choice as long as $C$ is large enough. It is easy to show that the prior probabilities of clustering are given by $P(\boldsymbol{\beta}_{kj}=\boldsymbol{\beta}_{k'j'})=(\alpha_{\beta}+1)^{-1}$ and $P(\boldsymbol{\theta}_{kj}=\boldsymbol{\theta}_{k'j'})=(\alpha_{\theta}+1)^{-1}$, and is increasingly likely with smaller values of $(\alpha_{\beta}, \alpha_{\theta})$. 

Although this encourages clustering separately for $\boldsymbol{\beta}_{kj}$ and $\boldsymbol{\theta}_{kj}$, it does not encourage  borrowing information between the two. In practice, exposures may behave similarly both in terms of the shape of the exposure-response relationship as well as weight profile. For example in DLNMs (where $(k,j)$ corresponds to an outcome-exposure pair, see Section \ref{ss:dlnm}), two pollutants may both have U-shaped effects that decay linearly over the course of a week, or different outcomes may be susceptible to exposures at similar times and in similar manners. With that in mind we extend the DP clustering strategy above to encourage borrowing information across the $\boldsymbol{\beta}_{kj}$ and $\boldsymbol{\theta}_{kj}$ domains. Consider the latent variable representation as above. Instead of independently specifying marginal probabilities of clustering indicators $Z_{kj}^{\beta}$ and $Z_{kj}^{\theta}$, we specify a joint distribution that allows for dependence between the two:
\begin{align*}
P(Z_{kj}^{\beta}=a, Z_{kj}^{\theta}=b)&=\frac{(1+\rho)^{I(a=b)}{\pi}^{\beta}_{a }  {\pi}^{\theta}_{ b}}{\sum_{a'=1}^C \sum_{b'=1}^C (1+\rho)^{I(a'=b')}{\pi}^{\beta}_{a' }  {\pi}^{\theta}_{ b'}}, ~~a,b \in \{1,\dots,C\}.
\end{align*}
Here $\pi_{a}^{\beta}$ and $\pi_b^{\theta}$ are assigned the same stick-breaking priors as above, and  $\alpha_{\beta}$ and $\alpha_{\theta}$ again control the levels of clustering for $\boldsymbol{\beta}_{kj}$ and $\boldsymbol{\theta}_{kj}$. However, $\rho$ now quantifies the degree of dependence in clustering between $\boldsymbol{\beta}_{kj}$ and $\boldsymbol{\theta}_{kj}$. When $\rho=0$, the model reduces to the independent DP case. When $\rho\to \infty$, $P(Z_{\beta, j} = a, Z_{\theta, j} = b) = 1$ when $a=b$ and 0 otherwise; that is, either  both $\boldsymbol{\beta}_{kj}=\boldsymbol{\beta}_{k'j'}$ and $\boldsymbol{\theta}_{kj}=\boldsymbol{\theta}_{k'j'}$, or  $\boldsymbol{\beta}_{kj}\neq\boldsymbol{\beta}_{k'j'}$ and $\boldsymbol{\theta}_{kj}\neq\boldsymbol{\theta}_{k'j'}$.

For $G_{0,\beta}$ we adopt smoothness-inducing priors, as we expect small changes in exposures to lead to small changes in outcomes. We adopt a mean-zero Gaussian with precision matrix $\lambda^{\beta}\boldsymbol{\Sigma}^{-1}_0$, where  $\boldsymbol{\Sigma}_0$ is a matrix of integrated second derivatives of basis functions to penalize overly rough functions \citep{wood2017generalized}, and $\lambda^{\beta}$ is a smoothing parameter. For $G_{0,\theta}$ we adopt ridge penalty priors by default with appropriate constraints (and penalty parameter $\lambda^{\theta}$), but consider an alternative for time-varying exposures in Sec. \ref{ss:dlnm}. See Appendix \ref{app:modelspec} for details. 

\subsection{Posterior sampling}
\label{ss:post}
We sample from the posterior via MCMC. Using a standard Gibbs sampler, sampling most parameters is straightforward (see Appendix \ref{app:sampler}). Sampling from the full conditional for the  index weights $\boldsymbol{\theta}^*_{c}$, however, is complicated by its constrained parameter space. In Appendix \ref{app:post} we present two approaches to sampling $\boldsymbol{\theta}^*_{c}$: the first transforms $\boldsymbol{\theta}^*_{c}$ to an unconstrained space; the second exploits a Taylor approximation to obtain a closed form for the full conditional.

\section{Special Case I: Multivariate Distributed Lag Non-Linear Model} 
\label{ss:dlnm}

Suppose for each unit $i$ we observe each of $P$ exposures at $L$ times; that is, $\mathbf{x}_{ip}=[x_{ip1},\dots,x_{ipL}]^T$ for $p=1,\dots,P$. We write the full vector of exposures  $\mathbf{x}_{i}^*  =\left[\mathbf{x}_{i 1}^{T}, \dots, \mathbf{x}_{i P}^{T}\right]^{T} $ of length $M$=$PL$. The goal is to quantify possibly non-linear associations between the $K$ outcomes and the $P$ time-varying exposures. We set the number of indices to be equal to the number of distinct exposures $J=P$ and choose $A_{kj}=\left[
\mathbf{0}_{L \times (j-1)L} ~|~ \mathbf{I}_{L \times L} ~|~ \mathbf{0}_{L \times (P-j)L}  \right],$ the $L\times PL$ horizontal block matrix with $j^{th}$ block equal to the identity matrix, and all other entries zero.  This yields a multivariate distributed lag non-linear model (MV-DLNM):
\begin{align}
y_{ik}&=f_{k1}({\mathbf{x}}_{i1}^T{\boldsymbol{\omega}}_{k1})+\cdots+f_{kP}({\mathbf{x}}_{iP}^T{\boldsymbol{\omega}}_{kP})+\boldsymbol{z}_i^T\boldsymbol{\beta}_{Zk}+\xi \sigma_k u_i+\epsilon_{ik},\quad k=1,\dots,K.  \label{eqn:dlnmmodel}
\end{align}
Here $f_{kp}(\cdot)$ quantifies the potentially non-linear association between the $k^{th}$ outcome and the $p^{th}$ time-varying exposure $\mathbf{x}_{ip}$, and  ${\boldsymbol{\omega}}_{kp}=({\omega}_{kp1},\dots,{\omega}_{kpL})^T$ quantifies the relative strength of that exposure's association at each of the $L$ times. In particular, adopting the cluster-inducing priors of Section \ref{ss:clustering} here borrows strength across exposure-outcome pairs to exploit similarity in both exposure-response relationships and temporal profiles. 

\textit{Smoothness over time:} Exposure measurements over time may be highly multicollinear, making estimation of $\boldsymbol{\omega}_{kp}$ challenging. One solution is to encourage smoothness over time---i.e., encourage $\boldsymbol{\omega}_{kpl}$ and $\boldsymbol{\omega}_{kpl'}$ to be similar when $|l'-l|$ is small---corresponding to our prior belief that effects should not change drastically over small time periods. A Gaussian prior with precision matrix $\lambda^{\omega}\mathbf{D}_s^T\mathbf{D}_s$ (where $\mathbf{D}_s$ is a known  $[L-s] \times L$ difference matrix of order $s$) shrinks differences between coefficients at similar lags, with the degree of shrinkage controlled  by $\lambda^{\omega}$, a parameter to be estimated \citep{gasparrini2017penalized}. However this leaves the magnitude of ${\boldsymbol{\omega}}_{kp}$ unconstrained. To address this, we propose an analogous Fisher-Bingham \citep{kent1982fisher} prior $\boldsymbol{\omega}_{kj} \sim FB(\mathbf{0},\lambda^{\omega} \mathbf{D}_s^T\mathbf{D}_s)$, which corresponds to the same Gaussian constrained to the unit $L$-sphere. To further satisfy the sign constraint, we can restrict this prior to the half hypersphere where $\omega_{kpL}\geq 0$. 

To facilitate sampling, we can optionally incorporate the dimension reduction approach approximating
${\boldsymbol{\omega}}_{kp} \approx \mathbf{\Psi} {\boldsymbol{\theta}}_{kp}$, 
 yielding the proposed prior $\boldsymbol{\theta}_{kp} \sim FB(\mathbf{0},\lambda^{\omega} \mathbf{\Psi}^T \mathbf{D}_s^T\mathbf{D}_s \mathbf{\Psi})$, 
and we can again impose a sign constraint ${{\theta}}_{kpm}>0$, restricting to the half-sphere. We adopt this as the centering distribution $G_{0, \theta}$ for the cluster-inducing priors.

 \section{Special Case II: Adaptive Multiple Index Model} 
\label{ss:amim}
Suppose we observe each of $P$ exposures at one time only, and let $\mathbf{x}_{i}^*=[x_{i1},\dots,x_{iP}]^T$ be the $P$-vector of exposures measured once each ($M=P$). Setting $J=1$ and $\mathbf{A}_1=\mathbf{I}_{P\times P}$ yields  a (multivariate) single index model (SIM):
\begin{align*}
y_{ik}&= f_{k}({\mathbf{x}}_{i}^T{\boldsymbol{\omega}}_{k})+\boldsymbol{z}_i^T\boldsymbol{\beta}_{Zk} +\xi \sigma_k u_i+\epsilon_{ik},   
\end{align*}
for $k=1,\dots,K$. This SIM can be restrictive, requiring that the associations between each exposure and the $k^{th}$ outcome have the same shape but differ in magnitude. Further adopting the cluster-inducing priors then encourages borrowing strength across the $K$ outcomes.

As a flexible alternative to SIMs,  multiple index models (MIMs) for environmental mixtures partition the exposure vector ${\mathbf{x}}_{i}$ \citep{mcgee2021bayesian}, but this requires knowing the true index structure \textit{a priori}.  Instead, our proposed framework naturally allows for unknown index structures. First set $J>1$ to be the maximum order of the index, then let $\mathbf{A}_j=\mathbf{I}_{P\times P}$ for $j=1,\dots,J$. The (multivariate) adaptive multiple index model of maximum order $J$ is: \begin{align}
y_{ik}&= \sum_{j=1}^J f_{kj}({\mathbf{x}}_{i}^T{\boldsymbol{\omega}}_{kj}) +\boldsymbol{z}_i^T\boldsymbol{\beta}_{Zk}+\xi \sigma_k u_i+\epsilon_{ik} , ~~k=1,\dots,K.\label{eqn:mim}
\end{align}
In contrast to model (\ref{eqn:dlnmmodel}), the summation is over multipollutant \textit{indices} that share common exposures  $\mathbf{x}_i$. Applying the priors
proposed in Section \ref{ss:clustering} now encourages clustering across outcomes and multipollutant indices rather than different 
exposures. This has two key consequences. First, it again improves efficiency by borrowing strength across outcomes. Second, clustering across indices facilitates collapsing to a lower order ($<J$) index model. For example if $J$=2, then when $Z^{\theta}_{k1}=Z^{\theta}_{k2}$ and $Z^{\beta}_{k1}=Z^{\beta}_{k2}$, model (\ref{eqn:mim}) reduces to a single index model---without assuming this a priori. This marks a major advantage over standard MIMs (like in \citealp{mcgee2021bayesian}), as we need only specify the \textit{maximum} order $J$, rather than the true index structure (i.e., the number of indices and their composition).

\subsection{Exposure Importance Measures}
\label{ssec:ExposureVIMs}

For some choices of $\mathbf{A}_j$, weight vectors $\boldsymbol{\omega}_{kj}$ have intuitive interpretations: in the MV-DLNM they quantify the strength of association at different lags; in a SIM, they quantify the strength of association between mixture components and outcome. In a MIM with unknown index structure as in Section \ref{ss:amim}, however, they do not have a natural interpretation, and it is difficult to infer which exposures are driving the mixture association. 
To facilitate interpretation, we propose alternative exposure importance measures that can be used for complex mixture models like the proposed MIM. 

Rewriting $f^*_k(\mathbf{x})=\sum_{j=1}^J f_{kj}({\mathbf{x}}^T{\boldsymbol{\omega}}_{kj})$ in (\ref{eqn:mim})  and omitting the subscript $i$, we define
\begin{align}
    \Phi_{kp} = 1 - \frac{Var_{\mathbf{x}_{-p}} \bigg\{ E_{x_p \mid \mathbf{x}_{-p}} \Big[ f^*_k(x_p,\mathbf{x}_{-p}) \Big] \bigg\}}{Var[f^*_k(\mathbf{x})]}, \label{eqn:psi_jk}
\end{align}
where $\mathbf{x}_{-p}$ represents all but the $p^{th}$ exposure.  The denominator represents the variability explained by the mixture, and  $\Phi_{kp}$ represents the proportion of this variability explained \textit{only} by the $p^{th}$ exposure. Naturally, $\Phi_{kp}$ will be small if $x_p$ is not predictive of the outcome. It will also be small if $x_p$ is highly collinear with $\mathbf{x}_{-p}$, as its association can be easily captured by other components.  $\Phi_{kp}$ thus quantifies the importance of the $p^{th}$ exposure to the mixture effect on the $k^{th}$ outcome. In Appendix \ref{app:VIM} we generalize this to arbitrary {groups} of exposures.

Computing these quantities is straightforward given posterior samples of $f_k^*(\mathbf{x})$. Marginal variances are calculated using the empirical distribution of the exposures. Estimating $\Phi_{kp}$ requires integrating over the conditional distribution of $x_p$ given the other exposures. If the dimension of $\mathbf{x}$ is small, we can estimate the conditional mean in (\ref{eqn:psi_jk}) non-parametrically: 
$$\widehat{E}_{x_p \mid \mathbf{x}_{-p}}\{f^*_k(x_p,\mathbf{x}_{-p})\} = \frac{\sum_{i=1}^n f^*_k(x_{ip}, \mathbf{x}_{(-p)}) K(\mathbf{x}_{i(-p)} - \mathbf{x}_{-p})}{\sum_{i=1}^n K(\mathbf{x}_{i(-p)} - \mathbf{x}_{-p})}$$
where $K(\cdot)$ is an appropriate kernel. Alternatively, when $\mathbf{x}$ is high-dimensional, one could make parametric assumptions, such as assuming that $x_p$ given $\mathbf{x}_{-p}$ is normally distributed, and the conditional mean can be estimated using any regression approach.

\section{Simulation studies}\label{sec:sims}

We demonstrate the effectiveness of the proposed methods in two different simulation studies.
In Simulation A, we consider time-varying exposures and demonstrate the benefit of borrowing strength across exposures and outcomes in DLNMs. In B, we consider cross-sectional exposure measurements  and try to capture complex, non-linear relationships between a mixture of exposures using MIMs. In both, we average results across 250 simulated datasets.

\subsection{Simulation A: Distributed lag models with multiple exposures}

We first focus on settings where exposures are measured repeatedly over time as in Section \ref{ss:dlnm}. Throughout, we set the number of outcomes to be $K = 4$, the number of exposures to be $P = 5$,  and the number of time points to be $L = 52$. We varied the sample size $n = \{500,1500\}$. We generated  exposures from a vector autoregressive model in order to induce autocorrelation within each exposure as well as  dependence among different exposures. We assume exposures have correlation 0.85 across successive time points and correlation $0.6^{|j - j'|}$ between exposures $j$ and $j'$ at each time point (see Appendix \ref{appSec:simAdetails} for full details). 

Responses were generated from a multivariate normal distribution with mean:
\begin{align*}
E[\boldsymbol{Y} \mid \boldsymbol{X}]&=\left[\begin{array}{l}
f_1\left(\mathbf{x}_1^{T} \boldsymbol{\omega}_1\right)+f_2\left(\mathbf{x}_2^{T} \boldsymbol{\omega}_2\right)+f_2\left(\mathbf{x}_3^{T} \boldsymbol{\omega}_3\right)+f_2\left(\mathbf{x}_4^{T} \boldsymbol{\omega}_1\right)+f_1\left(\mathbf{x}_5^{T} \boldsymbol{\omega}_1\right) \\
f_2\left(\mathbf{x}_1^{T} \boldsymbol{\omega}_1\right)+f_1\left(\mathbf{x}_2^{T} \boldsymbol{\omega}_2\right)+f_1\left(\mathbf{x}_3^{T} \boldsymbol{\omega}_3\right)+f_1\left(\mathbf{x}_4^{T} \boldsymbol{\omega}_1\right)+f_2\left(\mathbf{x}_5^{T} \boldsymbol{\omega}_1\right) \\
f_3\left(\mathbf{x}_1^{T} \boldsymbol{\omega}_1\right)+f_3\left(\mathbf{x}_2^{T} \boldsymbol{\omega}_2\right)+f_4\left(\mathbf{x}_3^{T} \boldsymbol{\omega}_3\right)+f_4\left(\mathbf{x}_4^{T} \boldsymbol{\omega}_1\right)+f_4\left(\mathbf{x}_5^{T} \boldsymbol{\omega}_1\right) \\
f_4\left(\mathbf{x}_1^{T} \boldsymbol{\omega}_1\right)+f_4\left(\mathbf{x}_2^{T} \boldsymbol{\omega}_2\right)+f_3\left(\mathbf{x}_3^{T} \boldsymbol{\omega}_3\right)+f_3\left(\mathbf{x}_4^{T} \boldsymbol{\omega}_1\right)+f_3\left(\mathbf{x}_5^{T} \boldsymbol{\omega}_1\right)
\end{array}\right], 
\end{align*}
where $f_1(a)=0.04a$, $f_2(a)=0.24(0.3a)^2$, $f_3(a)=0.09 a$, $f_4(a)=2\sin(0.2a)$; $\boldsymbol{\omega}_1 $ is flat, $\boldsymbol{\omega}_2$ is decreasing, and 
$\boldsymbol{\omega}_3$  is increasing.
The lag profiles are the same across outcomes, and the exposure response functions are shared across some, but not all of the exposures.

To each simulated dataset we then fit four variations of the proposed approach: (1) the MV-DLNM with dimension reduction (DLNM dimension reduction); (2) the MV-DLNM with no dimension reduction (DLNM); (3) the MV-DLNM without cluster-inducing priors (DLNM no clustering); (4) DLNM applied to each outcome separately (DLNM separate), which allows for clustering  across exposures but not outcomes.
We report mean squared error (MSE) for the posterior mean of $f_{kp}(\cdot)$ and $\boldsymbol{\omega}_{kp}$, as well as  95\% credible interval coverage,  averaged over all  datasets. For the proposed MV-DLNM, we also report pairwise clustering probabilities---$P(\boldsymbol{\beta}_{kp}=\boldsymbol{\beta}_{k'p'}|\boldsymbol{y})$ and $P(\boldsymbol{\omega}_{kp}=\boldsymbol{\omega}_{k'p'}|\boldsymbol{y})$---averaged across datasets.

\begin{figure}[hbtp!]
    \centering
    \includegraphics[width=0.99\linewidth]{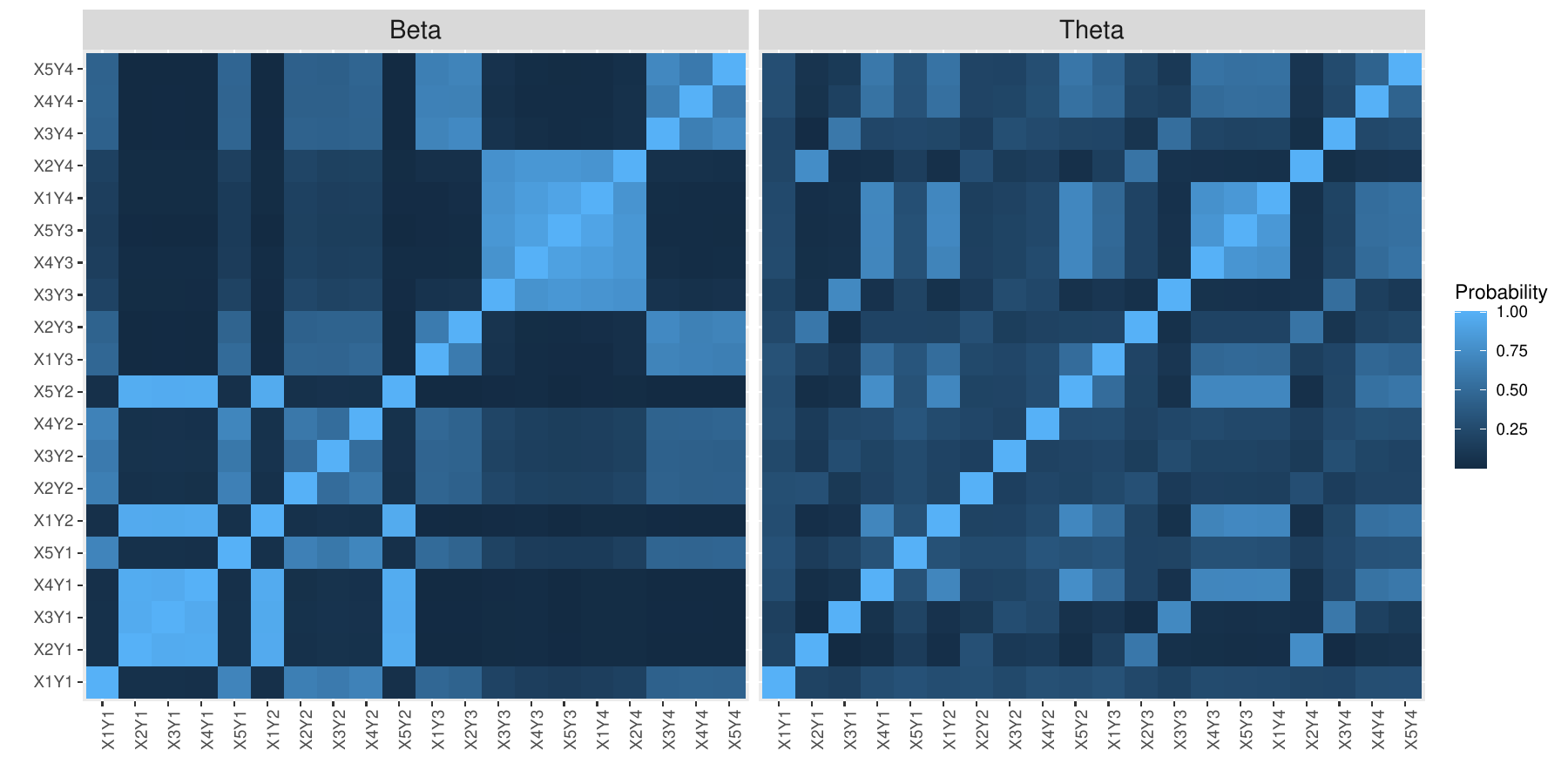}
    \includegraphics[width=0.99\linewidth]{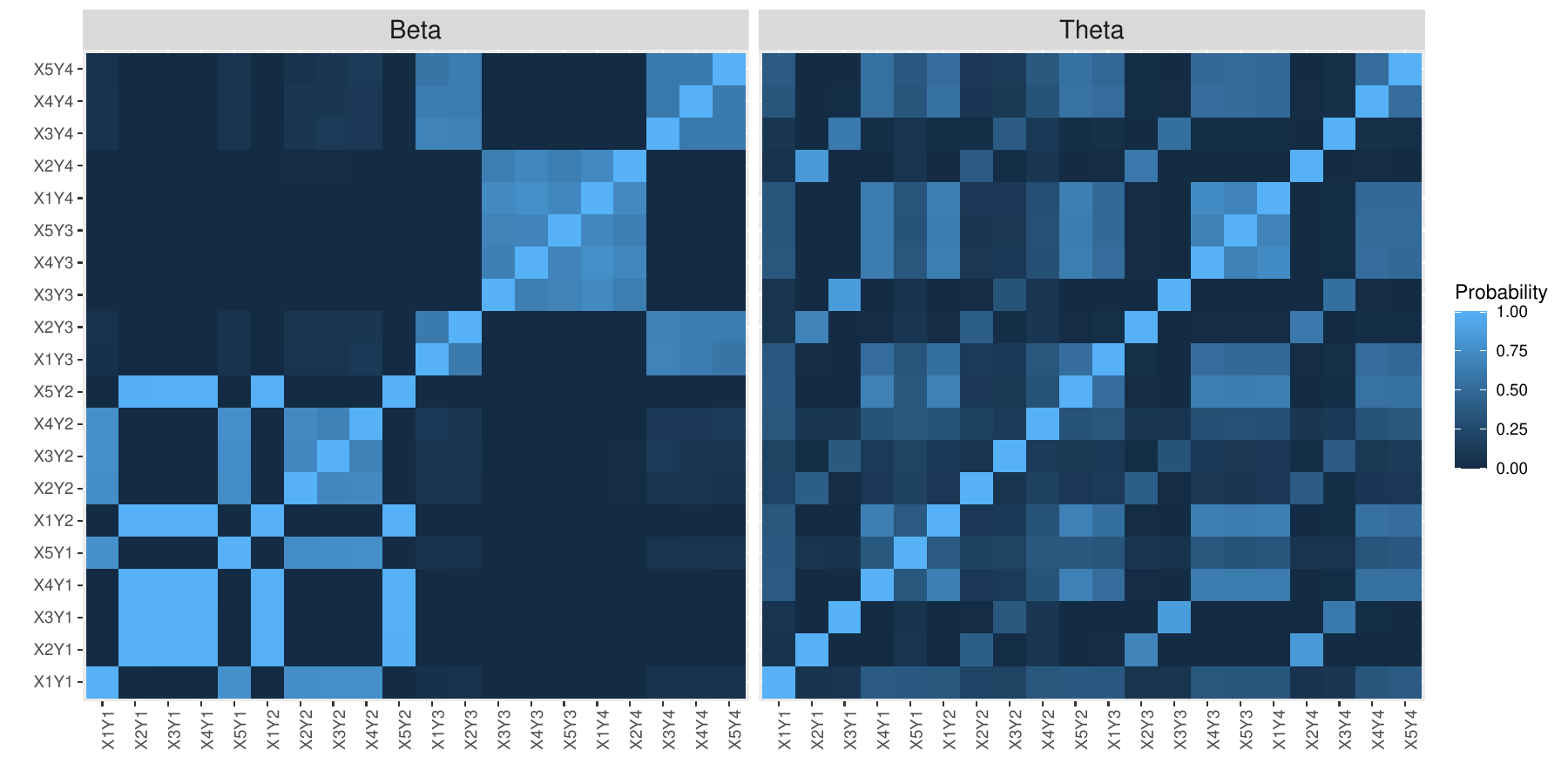}
    \caption{Posterior probabilities of pairwise clustering for both sets of model parameters, averaged across simulated datasets when $n=500$ (top) and $n=1500$ (bottom).}
    \label{fig:DLAG_clustering}
\end{figure}

Fig. \ref{fig:DLAG_clustering} displays average pairwise clustering probabilities. 
Overall the proposed framework was able to capture similarity in both $f_{kp}(\cdot)$ and $\boldsymbol{\omega}_{kp}$. There was a high degree of clustering for the $\boldsymbol{\beta}_{kp}$ parameters across exposure-outcome pairs with the same exposure-response function, and a low degree of clustering elsewhere. One exception was that there was a moderate degree of clustering between exposure-outcome pairs with $f_1(\cdot)$ as the true exposure response function and those with $f_3(\cdot)$. This is due to the fact that both had the same \textit{shape} of exposure response relationship (both linear), despite being of different strengths. This improves substantially with higher sample size where erroneous clustering occurs much less frequently (bottom left; Fig. \ref{fig:DLAG_clustering}). Pairwise clustering probabilities were somewhat lower for $\boldsymbol{\omega}_{kp}$ (top right and bottom right panel; Fig. \ref{fig:DLAG_clustering}), as the lag weights were only weakly identified due to high autocorrelation (and exacerbated by correlation among different exposures).

\begin{figure}[hbtp!]
    \centering
    \includegraphics[width=0.9\linewidth]{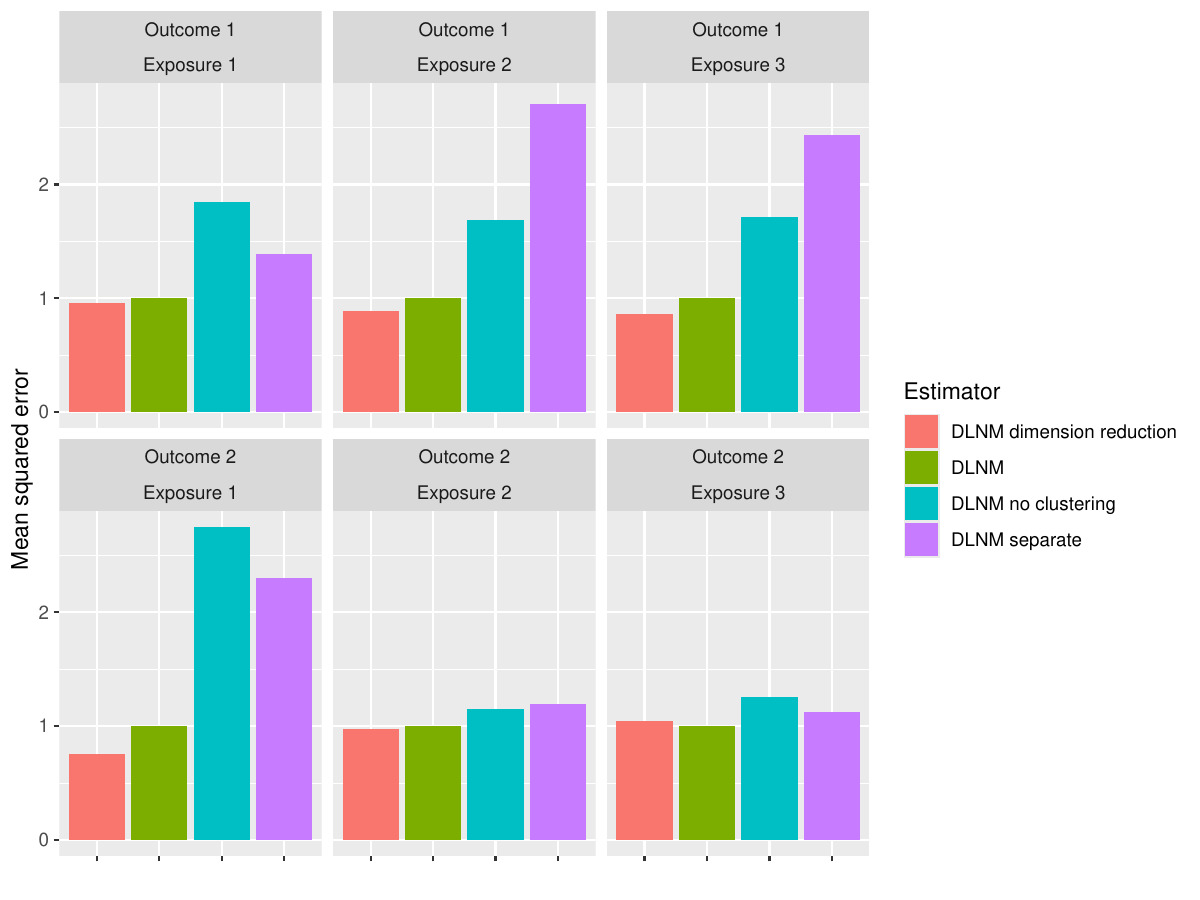}
    \caption{Mean squared error for estimating $\boldsymbol{\omega}_{kp}$ across all estimators. MSE values are scaled so that DLNM has an MSE of 1 in all scenarios. }
    \label{fig:DLAG_MSE}
\end{figure}

In Fig. \ref{fig:DLAG_MSE} we report MSE for $\boldsymbol{\omega}_{kp}$ when $n=500$. We highlight results for the first two outcomes and first three exposures (with others being similar).  
Clustering generally improved estimation of $\boldsymbol{\omega}_{kp}$: models with more clustering (DLNM, DLNM dimension reduction) had substantially lower MSE than others. We also observe that the DLNM with dimension reduction not only improves computation time, but can at times improve estimation error as well, likely due to the reduced number of parameters being estimated. We found similar results for estimation of $f_{kp}(\cdot)$, where clustering generally provides much lower mean squared error (see Appendix \ref{appSec:Simulations}). Appendix Fig. \ref{fig:DLAG_COV} shows that all four models yielded near nominal coverage for most of the exposure-outcome pairs, though coverage was lower for $(k,p)=$ (2,2) and (3,2). Lag weights $\boldsymbol{\omega}_{kp}$ for these pairs were commonly estimated to be flat as a result of being incorrectly clustered with other exposure-outcome pairs. This shows the bias-variance trade-off inherent to clustering: MSE was generally lower for the clustering models, but this came from a significant reduction in variance and a slight increase in bias. In larger samples, clustering improved and led to nominal coverage rates. 

\subsection{Simulation B: Single time point mixture modeling}

We now consider a setting where exposures are observed at a single time point as in Section \ref{ss:amim}. We generated datasets with $K$=4 outcomes and $P$=10 exposures and varied the sample size $n \in \{200,600,1000\}$.
The exposures $\mathbf{x}$ were generated from a multivariate normal distribution with a moderate autoregressive correlation structure  ($\rho$=0.5). Exposure-outcome associations were weak to moderate, with signal to noise ratios between 0.07 and 0.37, and we  considered three scenarios that vary in their favorability to MIMs: \\

    \begin{tabular}{lrl}
\hspace{-0.5cm}\textbf{Scenario 1:} \multirow{4}{*}{$ E[\boldsymbol{Y} \mid \boldsymbol{X}]=\left[\begin{array}{c}0.5f_1(\mathbf{x})\\0.5 f_1(\mathbf{x})\\0.5f_2(\mathbf{x})\\0.3f_2(\mathbf{x})\end{array}\right]$} 
& & $f_1(\mathbf{x}) = \mathbf{x}^T \boldsymbol{\alpha}_1,$ \\
&& $f_2(\mathbf{x}) = \mathbf{x}^T \boldsymbol{\alpha}_1 + (\mathbf{x}^T \boldsymbol{\alpha}_2)^2,$ \\
& & $\boldsymbol{\alpha}_1 = (0.1, 0.1, 0.2, 0.2, 0.25, 0.1, 0.05, 0.08, 0.3, 0.1)^T$ \\
&& $\boldsymbol{\alpha}_2 = (0.3, 0.05, 0.1, 0.2, 0.1, 0.2, \text{-}0.2, \text{-}0.2, 0.1, 0.2)^T.$ \\  
    \end{tabular}

   \begin{tabular}{lrl}
\hspace{-0.5cm}\textbf{Scenario 2:} \multirow{4}{*}{$E[\boldsymbol{Y} \mid \boldsymbol{X}]=\left[\begin{array}{c}0.25f_1(\mathbf{x})\\0.25 f_1(\mathbf{x})\\0.3f_2(\mathbf{x})\\0.3f_2(\mathbf{x})\end{array}\right]$} 
& \multirow{2}{*}{} & \multirow{2}{*}{$f_1(\mathbf{x}) = e^{0.5x_1} + \dfrac{1.5 e^{1.2 x_2}}{1 + e^{1.2 x_2}} - 0.5 x_3^2,$} \\
&& \multirow{2}{*}{$f_2(\mathbf{x}) = x_9 - 0.75x_{10}^2.$} \\  && \\ &&\\
    \end{tabular}

    \begin{tabular}{lrl}
\hspace{-0.5cm}\textbf{Scenario 3:} \multirow{4}{*}{$E[\boldsymbol{Y} \mid \boldsymbol{X}]=\left[\begin{array}{c}f_1(\mathbf{x})\\f_1(\mathbf{x})\\ \sqrt{0.5}f_1(\mathbf{x})\\ \sqrt{0.5}f_1(\mathbf{x})\end{array}\right]$} 
& & {$f_1(\mathbf{x}) = (\mathbf{x}^T \boldsymbol{\alpha}_1) \times (\mathbf{x}^T \boldsymbol{\alpha}_2),$} \\
& & {$\boldsymbol{\alpha}_1,$ $\boldsymbol{\alpha}_2$ are as above.} \\  && \\&& \\ && \\
    \end{tabular}

To each dataset we fit: (1) the proposed MIM separately for each outcome  (MIM separate); (2) the proposed multivariate MIM, clustering across outcomes (MIM clustering); (3) Bayesian kernel machine regression (BKMR, \citealp{bobb2015bayesian}), a widely used Gaussian process approach for modelling non-linear and non-additive relationships with variable selection on the exposures; and (4)  weighted quantile sum regression (WQS, \citealp{carrico2015characterization}), a popular mixtures approach that fits a linear index model on exposure quantiles.

We also computed exposure importance metrics $\Phi_{kj}$ proposed in Section \ref{ssec:ExposureVIMs} for the MIM. To estimate the distribution of $x_p$ given $\mathbf{x}_{-p}$ we assumed normality and estimated the mean of $x_p$ given $\mathbf{x}_{-p}$ via random forests to allow for flexible associations between exposures.
We report MSE for the posterior mean of the exposure-response surface averaged over 200 randomly chosen locations, and we also report the average exposure importance metrics. 

\begin{figure}[hbtp!]
    \centering
    \includegraphics[width=0.95\linewidth]{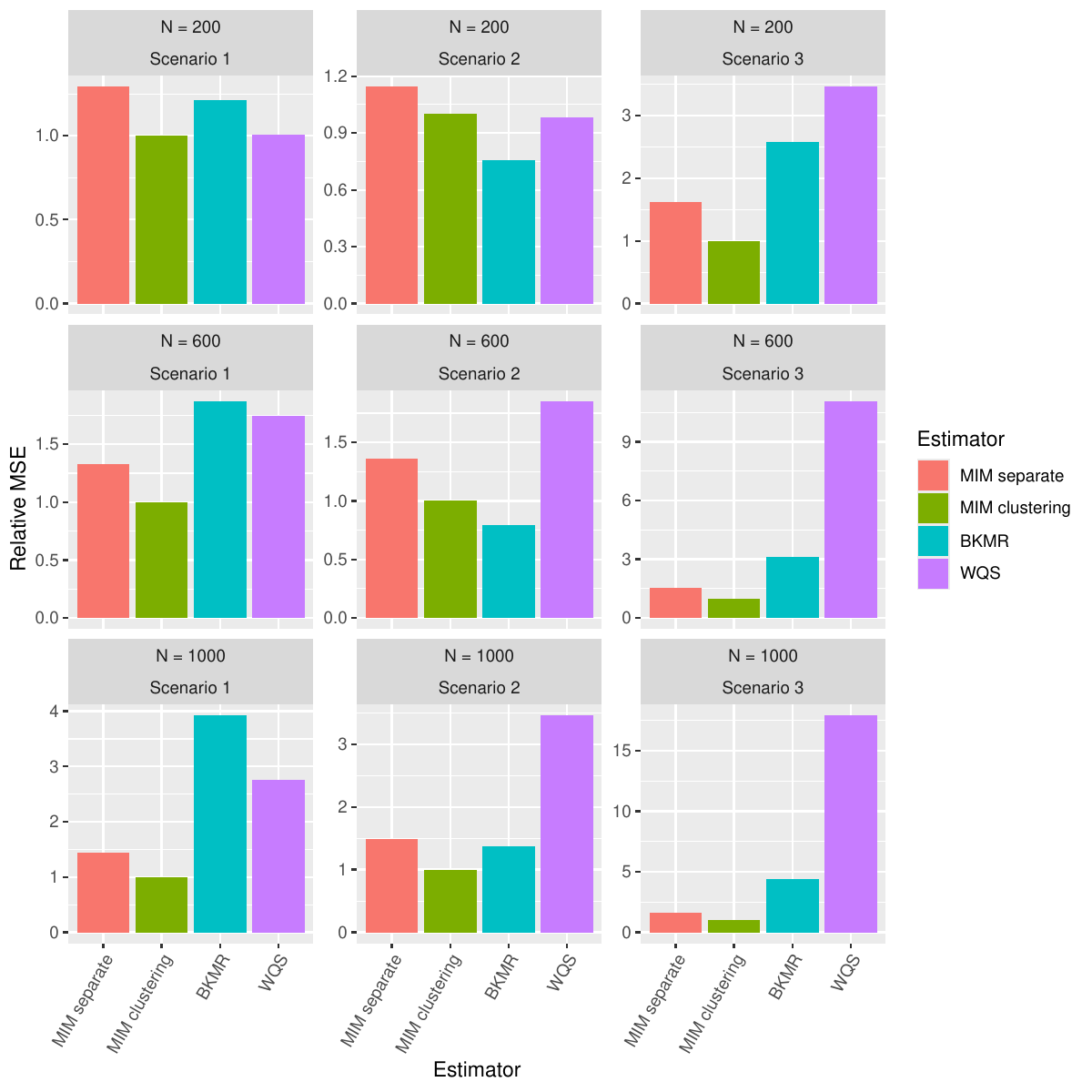}
    \caption{Mean squared error for estimating $f^*(\mathbf{x})$ at 200 randomly chosen locations for $\mathbf{x}$ across all estimators, sample sizes, and scenarios. MSE values are scaled so that MIM clustering has an MSE of 1 in all scenarios. }
    \label{fig:MIM_MSE}
\end{figure}

We report results for MSE in Fig. \ref{fig:MIM_MSE}. In Scenario 1, the true exposure-response relationships fit into the index framework, and the MIM approaches naturally performed best. By contrast, Scenario 2 is favorable to BKMR, as the regression surfaces are sparse (70-80\% of  effects are null), and each non-null effect has a different functional form. As such, BKMR performed best here in small sample sizes. Surprisingly, however, the proposed MIM approaches began to outperform BKMR in larger samples, as BMKR tended to incorrectly exclude important exposures in some datasets. Scenario 3, characterized by many pairwise interactions,  is not particularly favorable to either approach. Here the MIMs---despite having an additive structure---greatly outperformed the other approaches. Across all scenarios and sample sizes, MIM clustering successfully exploited shared structure across outcomes, and outperformed MIM separate (which did not borrow information across outcomes). Overall, the proposed approach fared well in a range of settings: under sparsity, it performed reasonably well, and when many exposures had small effects, it greatly outperformed competitors.

\begin{figure}[hbtp!]
    \centering
    \includegraphics[width=0.95\linewidth]{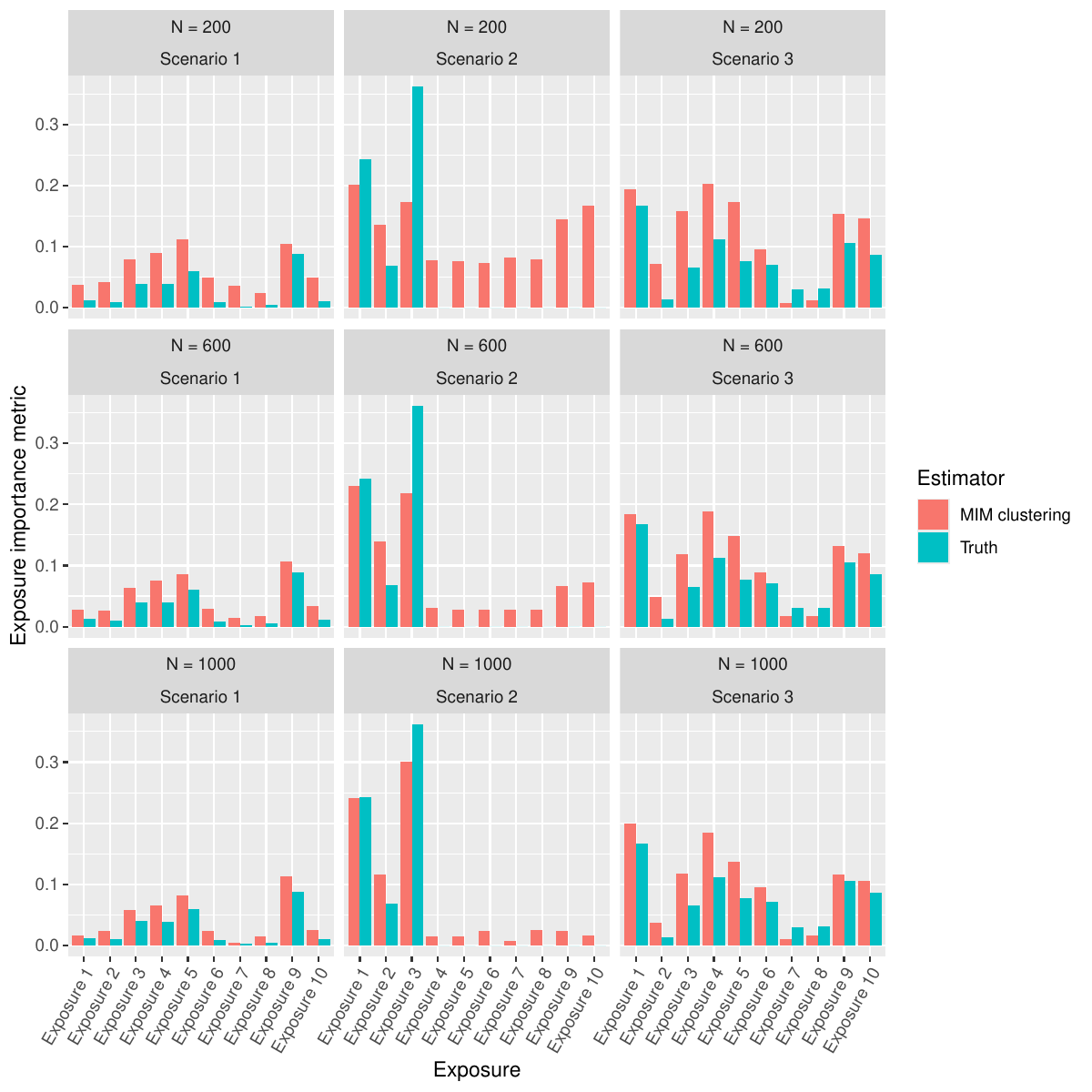}
    \caption{Exposure importance metrics for the proposed approach across all scenarios and sample sizes. }
    \label{fig:MIM_VIM}
\end{figure}

In Fig. \ref{fig:MIM_VIM},  we see that estimation of exposure importance metrics is difficult in small sample sizes. In particular when $n=200$ we observe substantial overestimation in the highly sparse Scenario 2, in part because when $\Phi_{kp}=0$ it can never be underestimated (as $\Phi_{kp}\geq 0$ by definition). Ultimately, as sample size increased, estimation improved significantly in all three scenarios. This shows that the proposed exposure metrics can be reasonably estimated in order to provide additional interpretability to the proposed MIM framework.

\section{NMMAPS Data Analysis}\label{sec:NMMAPS}

The National Morbidity, Mortality and Air Pollution Study (NMMAPS) collected time series data on daily mortality, air pollution and meteorological variables in Chicago between 1987 and 2000.  We observe daily  cardiovascular deaths, respiratory deaths and other non-accidental deaths, as well as median PM$_{10}$ and O$_3$ measurements. Full study details can be found in \cite{samet2000national,samet2000nationalb}. Mortality outcomes were log transformed and standardized. Exposures were standardized, and we excluded one two week period in 1988 with a PM$_{10}$ measurement 15 standard deviations above the mean, leaving a dataset of $n=5,100$ days, and we considered a maximum lag of $L=14$ days. We note that for this time series analysis, the unit of study is the day; for the $i^{th}$ day ($i=1,\dots,5100$), we observe  outcomes $\mathbf{y}_{i}=(y_{i1},y_{i2},y_{i3})^T$, exposures $\mathbf{x}_{i1}=(x_{i1,1},\dots,x_{i1,14})^T$ and $\mathbf{x}_{i2}=(x_{i2,1},\dots,x_{i2,14})^T$, as well as covariates $\mathbf{z}_i$, including date and meteorological covariates. A preliminary additive model found no associations with same-day exposures, but this ignored delayed exposure effects; a lagged WQS \citep{gennings2020lagged} analysis suggested negative associations with the exposure mixture (Appendix figures \ref{fig:gam} \& \ref{fig:lwqs}), though this ignored uncertainty in time-varying mixture weights. 

To quantify non-linear relationships between exposures and outcomes, we fit the proposed MV-DLNM with $K=3$ and $P=2$. Following \cite{chen2019distributed}, we adjusted for day of week, seasonality, year, and mean daily temperature and dew point temperature per day and averaged over the prior three days, and we allowed different covariate associations for each outcome (see Appendix \ref{sec:appendix_chicago}).
We compared the proposed MV-DLNM with cluster-inducing priors to separate DLNMs for each outcome, reflecting a standard approach that did not borrow strength across outcomes. We also considered the MV-DLNM without clustering as a middle ground between the two (with results in  Appendix \ref{sec:appendix_chicago}).

\begin{sidewaysfigure}[htbp!]
    \centering
    \subfloat[PM$_{10}$. Left: separate models; right: proposed MV-DLNM.]{\includegraphics[width=0.45\linewidth]{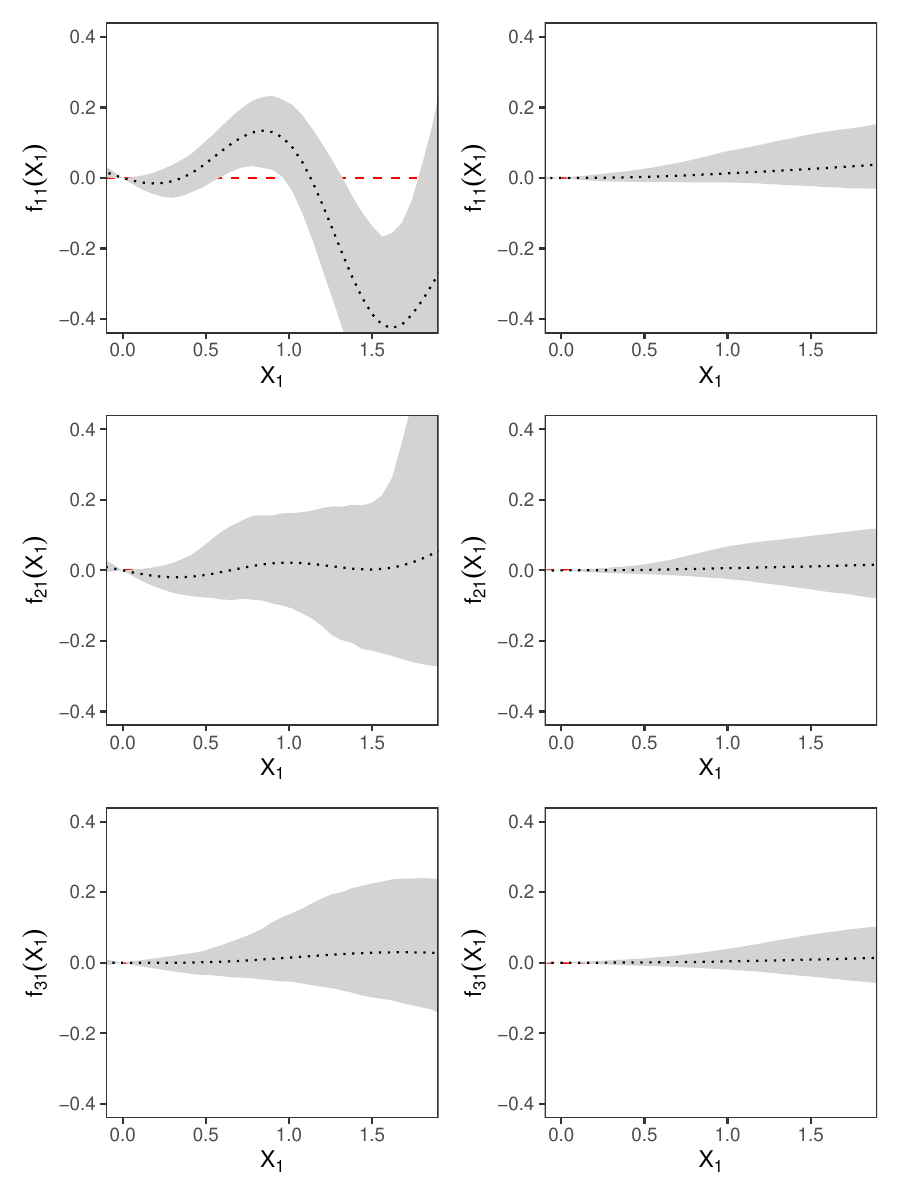}}
    \hspace{2cm}
    \subfloat[O$_{3}$. Left: separate models; right: proposed MV-DLNM.]{\includegraphics[width=0.45\linewidth]{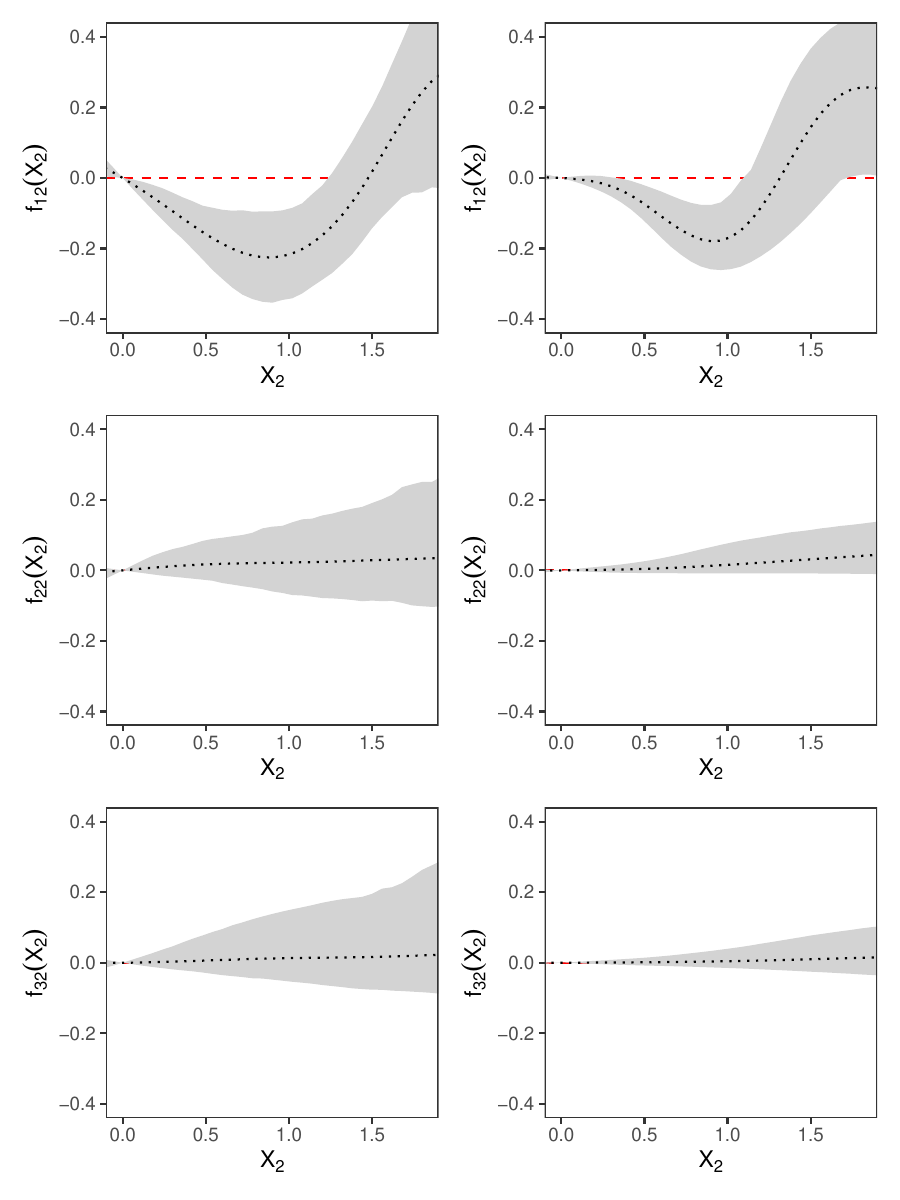}}
    \caption{Mean difference in outcomes increasing standardized PM$_{10}$ (panel a) and O$_{3}$ (panel b)  from their means. Three rows correspond to (log) cardiovascular, respiratory and other (non-accidental) deaths. }
    \label{fig:erfs}
\end{sidewaysfigure}

The separate modelling approach indicated a strong non-linear association between O$_3$ and cardiovascular deaths (Fig. \ref{fig:erfs}b, left column). This relationship was mostly negative and linear for exposures less than 1SD above the mean but turned sharply positive beyond that. The relationship between PM$_{10}$ was also non-linear, with associations first increasing, then decreasing, then increasing again for exposures above the mean. Associations with respiratory and other deaths were more uncertain, and intervals were wide enough to include strong positive or negative associations (Fig. \ref{fig:erfs}, rows 2--3). 

The proposed MV-DLNM also captured the non-linear O$_3$--cardiovascular relationship (Fig. \ref{fig:erfs}b, right), this time with statistically significantly more deaths at the 95th percentile vs. mean exposure. In contrast with the  separate modelling strategy, other associations were close to null with much tighter intervals. These differences are explained by the posterior clustering probabilities (Appendix Fig. \ref{fig:chicagoHeat2}). Pairwise clustering probabilities for $\boldsymbol{\beta}_{kp}$ were all above 0.90. Those for $\boldsymbol{\theta}_{kp}$ were more modest, but five of six exposure-outcome pairs exhibited substantial clustering, with the exception of the O$_3$ -- cardiovascular relationship. The upshot is that the proposed approach exploited similarity in the weaker associations to gain efficiency while allowing the distinct non-linear relationship between O$_3$ and cardiovascular deaths to shine through. We note that the PM$_{10}$--cardiovascular relationship was flattened in this approach, possibly due to instability in the separate modelling estimates.

\begin{figure}[htbp!]
    \centering
    \includegraphics[width=0.9\linewidth]{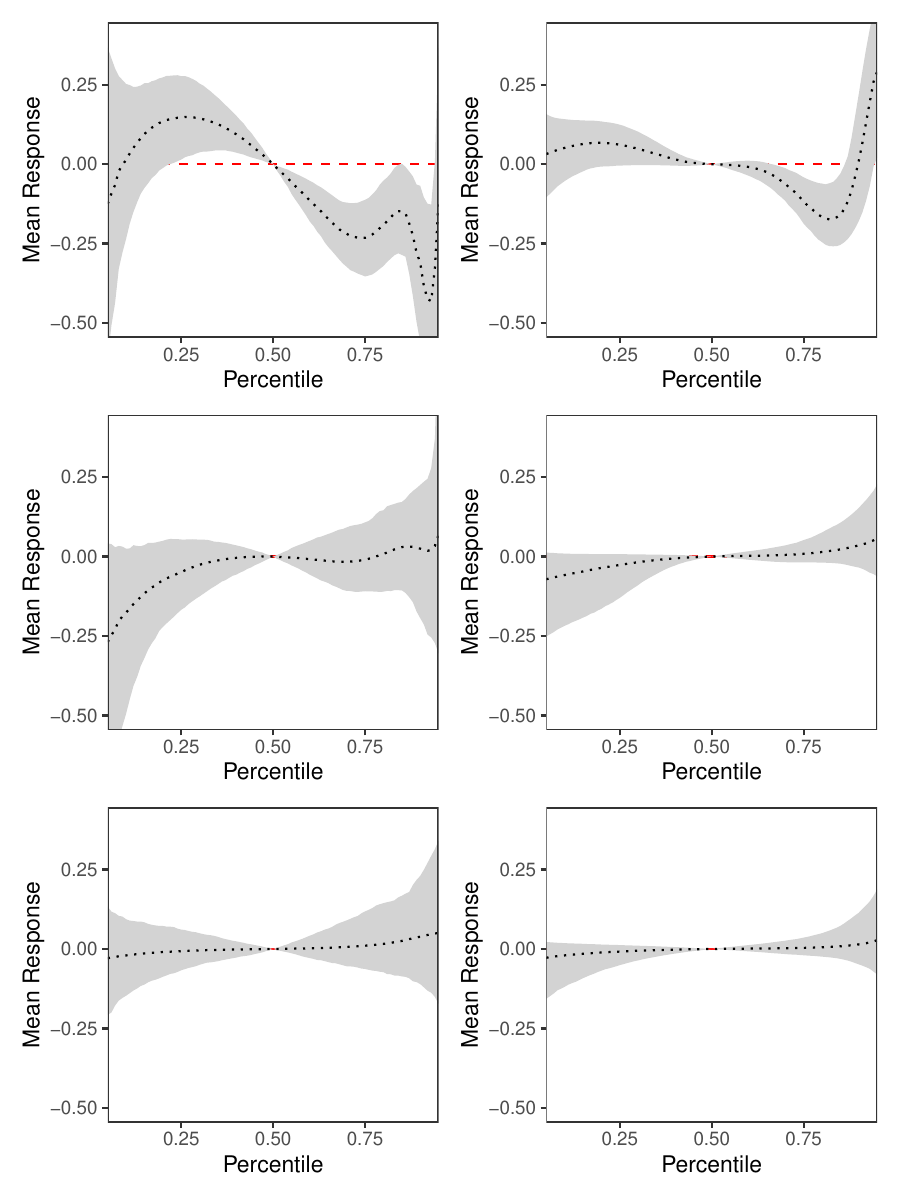}
    \caption{Overall exposure effects: estimated mean outcomes setting all exposures to different quantiles (vs median). Three rows correspond to (log) cardiovascular, respiratory and other (non-accidental) deaths. Left column is separate models; right column is proposed MV-DLNM.}
    \label{fig:overall}
\end{figure}

Overall mixture effects are often of interest, and in Fig. \ref{fig:overall} we plot estimated mean differences when exposures are simultaneously set to different quantiles. Counterintuitively, the separate model for cardiovascular deaths showed a strong negative relationship with the overall mixture. The proposed MV-DLNM reflected a much more modest negative relationship, and exhibited a sharp  upturn at  the highest exposure levels. Separate model fits for the other outcomes exhibited wide intervals that could be consistent with null, positive, or negative relationships. The proposed MV-DLNM showed much tighter intervals, indicating more evidence of no relationship with the overall mixture.

To quantify temporal profiles we estimate mean outcome differences corresponding to an increase in a pollutant from 1SD below its mean to 1SD above at each lag $l$, holding exposures at all other lags constant (as in \citealp{chen2019distributed}). Estimates were similar across models here (Appendix figures \ref{fig:contrast1}--\ref{fig:contrast2}), with weak associations across all three outcomes, in part because they represent single day changes in exposure while holding constant the other 13 days' exposures (which are highly correlated). Nevertheless, separate models yielded intervals on average more than twice the width of those of the proposed MV-DLNM.

Ultimately, the proposed MV-DLNM fit better than modelling outcomes separately (WAIC  40159.8 vs 40282.6). We also fit the MV-DLNM without clustering, which had slightly worse fit (40161.4) and gave estimates more similar to separate model fits (Appendix \ref{sec:appendix_chicago}). The cluster-inducing priors enabled the MV-DLNM to capture the non-linear O$_3$-cardiovascular relationship while exploiting similarity in the other weak associations to improve efficiency.

\section{Discussion} \label{sec:discussion}
Model (\ref{eqn:dlnmmodel}) is a special case of our framework, but more flexible parameterizations could be considered, including non-parametric functional additive models \citep{james2005functional,mclean2014functional}, though the low signal of mixtures analyses motivates more structured approaches. A limitation of our MV-DLNM is that it assumes additivity. In  Appendix \ref{app:nonadditivity} we discuss how our methods could be extended to non-additive settings; in our experience with environmental mixtures in settings with repeated exposures over time, however, we have found there is seldom enough signal to support identifying interactions, particularly in non-linear models. Nevertheless, alternative DLNMs could be considered, like the non-separable parameterizations of \cite{gasparrini2017penalized}. In Appendix \ref{app:nonseparableDLNM} we extend our approach to models of that type and demonstrate similar performance in simulations. Recently, \cite{li2025dynamic} proposed a model that combines exposures into a single index and allows for a time-varying function of this index---though it restricts contributions of each exposure to be constant over time, and for each to share the same  exposure-response function. 
Ultimately, our goal is not to argue the supremacy of any one specification but to present a flexible framework for borrowing strength across outcomes and exposures facilitated by our MV-DLNM parameterization. Ongoing work aims to develop methods for comparing and combining inferences from these different types of DLNMs.

In this paper we proposed a multivariate index model framework for modeling environmental mixtures with a novel co-clustering strategy to exploit similarity across exposures and outcomes, and we developed new exposure-importance metrics to aid interpretation. 
Though we highlighted the MV-DLNM for time-varying exposures, the proposed framework extends to other exposure structures. Rather than measuring $P$ pollutants at multiple times, for example, \cite{bauer2020associations} analyzed five biomarkers (in hair, blood, urine, nails, and saliva) of four  metals, for each of three outcomes via separate models. Our approach can naturally be applied to this setting to borrow strength in an analogous way to the MV-DLNM (omitting smoothness inducing priors over time). 
With univariate exposure measurements, our  framework extends (multivariate) additive models by encouraging similar exposure-response curves across different exposure-outcome pairs (see Appendix \ref{app:special}). 
Even if the outcome is univariate ($K=1$),  the proposed approach may still offer a benefit by borrowing strength across a large number of  exposures.

 
\section*{Data Availability Statement}
The data analyzed in Section \ref{sec:NMMAPS} are available in the {\tt dlnm} package \citep{dlnmpackage}.\vspace*{-8pt}

\section*{Supplementary Materials}
R code for running the proposed methods are  available at github.com/glenmcgee/MVmixture.\vspace*{-8pt}

\section*{Acknowledgements}
We acknowledge the support of the Natural Sciences and Engineering Research Council of Canada (NSERC) [RGPIN-2022-03068, DGECR-2022-00433]. Research described in this article was conducted under contract to the Health Effects Institute (HEI), an organization jointly funded by the United States Environmental Protection Agency (EPA) (Assistance Award No. CR-83590201) and certain motor vehicle and engine manufacturers. The contents of this article do not necessarily reflect the views of HEI, or its sponsors, nor do they necessarily reflect the views and policies of the EPA or motor vehicle and engine manufacturers.\vspace*{-8pt}

\bibliographystyle{apalike}
\bibliography{references}

\clearpage 
\section*{APPENDIX}
\appendix
\renewcommand*\thefigure{\thesection\arabic{figure}}
\renewcommand*{\thetable}{\thesection\arabic{table}}

\section{Special Cases}
\label{app:special}

Choice of $\mathbf{A}_j$ determines the model structure, and we briefly outline here several examples.
\begin{enumerate}
    \item Suppose for each unit $i$ we observe each of $P$ exposures at $L$ times; that is, $\mathbf{x}_{ip}=[x_{ip1},\dots,x_{ipL}]^T$ for $p=1,\dots,P$. We write the full $PL$-vector of exposures  $\mathbf{x}_{i}^*  =\left[\mathbf{x}_{i 1}^{T}, \dots, \mathbf{x}_{i P}^{T}\right]^{T} $. Setting $J$ to be equal to the number of distinct exposures ($J=P$) and choosing $A_{kj}=\left[\mathbf{0}_{L \times (j-1)L} ~|~ \mathbf{I}_{L \times L} ~|~ \mathbf{0}_{L \times (P-j)L}  \right]$ (the $L\times PL$ horizontal block matrix with $j^{th}$ block equal to the identity matrix, and all other zero), one obtains a (multivariate) DLNM. We elaborate on this further in the main text, including specification of smoothness inducing centering distribution $G_{0,\theta}$. 

\item Suppose for each unit $i$ we observe each of $P$ exposures at $B$ biomarkers. For example, \cite{bauer2020associations} analyzed five different biomarkers (in hair, blood, urine, nails, and saliva) of four different metals ($P=4$, $B=5$). Let $\mathbf{x}_{ip}=[x_{ip1},\dots,x_{ipB}]^T$ for $p=1,\dots,P$. We write the full $PB$-vector of exposures  $\mathbf{x}_{i}^*  =\left[\mathbf{x}_{i 1}^{T}, \dots, \mathbf{x}_{i P}^{T}\right]^{T} $. Setting $J$ to be equal to the number of distinct exposures ($J=P$) and choosing $A_{kj}=\left[\mathbf{0}_{B \times (j-1)B} ~|~ \mathbf{I}_{B \times B} ~|~ \mathbf{0}_{B \times (P-j)B}  \right]$ (the $B\times PB$ horizontal block matrix with $j^{th}$ block equal to the identity matrix, and all other zero). This is analogous to the DLNM above, however we would replace the smoothness inducing centering distribution $G_{0,\theta}$ with a default ridge penalty prior.

\item Suppose now we observe $P$ exposures at a single time point, and let $\mathbf{x}_{i}^*  =\left[{x}_{i 1}, \dots, {x}_{i P}\right]^{T}$. Setting $J=P$ and $\mathbf{A}_j=\mathbf{e}_j$ (i.e. the $P$-vector of zeroes except for the $j^{th}$ element, equal to to one) yields a (multivariate)  additive model:
\begin{align*}
y_{ik}&=f_{k1}({{x}}_{i1})+\cdots+f_{kP}({{x}}_{iP})+\boldsymbol{z}_i^T\boldsymbol{\beta}_{Zk}+\xi \sigma_k u_i+\epsilon_{ik},~~k=1,\dots,K.
\end{align*} Here the weights are fixed ($\omega_{kp}=1$), but we can still improve performance by  exploiting similarity in the functions $f_{kp}(\cdot)$ across similar exposure-outcome pairs. Even with univariate outcomes ($K=1$) this encourages borrowing strength across multiple exposure-response curves.

\item Suppose again we observe $P$ exposures at a single time point, and let $\mathbf{x}_{i}^*  =\left[{x}_{i 1}, \dots, {x}_{i P}\right]^{T}$. Setting $J=1$ and $\mathbf{A}_1=\mathbf{I}_{P\times P}$ yields  a multivariate single index model (SIM).
\begin{align*}
y_{ik}&= f_{k}({\mathbf{x}}_{i}^T{\boldsymbol{\omega}}_{k})+\boldsymbol{z}_i^T\boldsymbol{\beta}_{Zk} +\xi \sigma_k u_i+\epsilon_{ik},   
\end{align*}
for $k=1,\dots,K$. Cluster-inducing priors allow borrowing information across different outcomes here, but not exposures. 

\end{enumerate}

\clearpage
\section{Model Specification}\label{app:modelspec}
\begin{align*}
y_{i1}&\approx \beta_{10} + \mathbf{b}_{i1,\boldsymbol{\theta}_{11}}^T\boldsymbol{\beta}_{11}+\cdots+\mathbf{b}_{iP,\boldsymbol{\theta}_{1P}}^T \boldsymbol{\beta}_{1P}+\boldsymbol{z}_i^T\boldsymbol{\beta}_{Z1}+\xi \sigma_1 u_i+\epsilon_{i1}  \\
&\vdots \\
y_{iK}&\approx \beta_{K0} + \mathbf{b}_{i1,\boldsymbol{\theta}_{K1}}^T\boldsymbol{\beta}_{K1}+\cdots+\mathbf{b}_{iP,\boldsymbol{\theta}_{KP}}^T \boldsymbol{\beta}_{KP}+\boldsymbol{z}_i^T\boldsymbol{\beta}_{ZK}+\xi \sigma_K u_i+\epsilon_{iK}  \\ ~\\
\epsilon_{ik} &\overset{iid}{\sim} N(0,\sigma^2_k) \\
u_{i} &\overset{iid}{\sim} N(0,1) 
\end{align*}

	\subsubsection*{Likelihood and Parameterization}
    {\small
	\begin{align*}
	\mathcal{L}&= \prod_{k=1}^K \prod_{i=1}^n f(y_{ik};\boldsymbol{\beta}_{k1},\dots,\boldsymbol{\beta}_{kP},\boldsymbol{\theta}_{k1},\dots,\boldsymbol{\theta}_{kjP},u_i) \\
\text{ where }~~ \boldsymbol{\beta}_{kj}&=\sum_{c=1}^C I(Z^\beta_{kj}=c)\boldsymbol{\beta}^*_{c} \text{ and } 
\boldsymbol{\theta}_{kj}=\sum_{c=1}^C I(Z^\theta_{kj}=c)\boldsymbol{\theta}^*_{c} \\
\implies 	\mathcal{L}&=\prod_{k=1}^K  \left( \prod_{a_1=1}^C\cdots \prod_{a_p=1}^C  \prod_{b_1=1}^C\cdots \prod_{b_p=1}^C\left[\prod_{i=1}^n f(y_{ik};\boldsymbol{\beta}^*_{a_1},\dots,\boldsymbol{\beta}^*_{a_p},\boldsymbol{\theta}^*_{b_1},\dots,\boldsymbol{\theta}^*_{b_p},u_i)\right]^{\prod_{j=1}^pI(Z^{\beta}_{kj}=a_j)\prod_{j'=1}^p I(Z^{\theta}_{kj'}=b_j)} \right) \\
\text{or }&= \prod_{k=1}^K \left[\prod_{i=1}^n f(y_{ik};\boldsymbol{\beta}^*_{Z^{\beta}_{k1}},\dots,\boldsymbol{\beta}^*_{Z^{\beta}_{kP}},\boldsymbol{\theta}^*_{Z^{\theta}_{k1}},\dots,\boldsymbol{\theta}^*_{Z^{\theta}_{kP}},u_i)\right]
	\end{align*}
	}
	\subsubsection*{Priors}
\begin{align*}
P(Z^{\beta}_{kj}=a,Z^{\theta}_{kj}=b)&=\pi^*_{a,b}=\frac{\pi_{a,b}}{\sum_{a'=1}^C \sum_{b'=1}^C \pi_{a',b'}}, ~~a,b \in \{1,\dots,C\} \\
\pi_{a,b} &= (1+\rho)^{I(a=b)}{\pi}^{\beta}_{a \cdot}  {\pi}^{\theta}_{\cdot b} \\
\pi^{\beta}_{1 } &=V^{\beta}_{1} \qquad \qquad \qquad ~~ \pi^{\theta}_{  1}=V^{\theta}_{1} \\
\pi^{\beta}_{c } &= V^{\beta}_{c} \prod_{j < c} (1 - V^{\beta}_{j}) \qquad  
\pi^{\theta}_{  c} = V^{\theta}_{c} \prod_{j < c} (1 - V^{\theta}_{j})~~c=2,\dots,C  \\
V^{\beta}_{c} &\sim Beta(1, \alpha_{\beta}) \qquad \qquad V^{\theta}_{c} \sim Beta(1, \alpha_{\theta}), ~~c=1,\dots,C-1 \\
V^{\beta}_{C}&=1 \qquad\qquad\qquad \qquad  V^{\theta}_{C}=1 \\
\alpha_{\beta} &\sim Gamma(a_{\beta},b_{\beta})  \qquad 
\alpha_{\theta} \sim Gamma(a_{\theta},b_{\theta}) \\
\boldsymbol{\beta}^*_c &\sim N(\mu_0,[\lambda^{\beta}]^{-1}\Sigma_0) \qquad
\boldsymbol{\theta}^*_c \sim  FB(\tau_{\theta}\mathbf{1},~-\frac{1}{2}\lambda^{\theta} \Sigma_{\theta}^{-1}) \\
\lambda^{\beta}&\sim Gamma(a_{\lambda}^{\beta},b_{\lambda}^{\beta})\qquad
\lambda^{\theta}\sim Gamma(a_{\lambda}^{\theta},b_{\lambda}^{\theta})\\
\alpha_{\beta} &\sim Gamma(a_{\beta},b_{\beta})  \qquad 
\alpha_{\theta} \sim Gamma(a_{\theta},b_{\theta}) \\
\rho &\sim Gamma(a_{\rho},b_{\rho}) \\
u_i &\sim N(0,1) \\
\xi  &\sim InvGamma(a_{\xi},b_{\xi}) \\
\sigma^2_k &\sim InvGamma( a_{\sigma},b_{\sigma})  \\
P(\beta_{Zk})&\propto 1 \\
P(\beta_{0k})&\propto 1
\end{align*}

Note we set $\tau_{\theta}=0$ if using the polar transformation method; otherwise computing the appropriate normalizing constant necessary for updating $\lambda^{\theta}$ is challenging.

\section{Posterior sampling of index parameters}\label{app:post}

Sampling from the full conditional for the  index weights $\boldsymbol{\theta}^*_{c}$  is complicated by its constrained parameter space. We present two approaches to sampling $\boldsymbol{\theta}^*_{c}$: the first transforms $\boldsymbol{\theta}^*_{c}$ to an unconstrained space; the second exploits a Taylor approximation to obtain a closed form for the full conditional distribution and ignores the sign constraint. 

\subsection{Polar Coordinate Reparameterization}
\label{ss:polar}

The constraints on $\boldsymbol{\theta}^*_c$  makes choosing reasonable proposal distributions challenging in practice. One path forward is to absorb the constraints into the parameterization by transforming to polar coordinates. Following the approach of \cite{park2005bayesian}, we let 
\begin{align*}
\theta^*_{c1}&=\sin(\phi^*_{c1}),
~~\theta^*_{c{\ell}}=\sin(\phi^*_{c{\ell}})\prod_{j=1}^{\ell-1} \cos(\phi^*_{cj}),~\ell=2,\dots,L-1,~~
\text{and} ~\theta^*_{cL}=\prod_{j=1}^{L-1} \cos(\phi^*_{cj}), 
\end{align*}
where $\phi^*_{cj}\in[-\frac{\pi}{2},\frac{\pi}{2}]$ for $j=1,\dots,{L-1}$, which satisfies both constraints by default.  
 Whereas the entire vector $\boldsymbol{\theta}^*_c$ is inextricably linked by its magnitude constraint,  the domains of $\phi^*_{cj}$  do not depend on one another, so we can  draw  each separately from its full conditional via grid sampling or Metropolis Hastings with scaled and shifted beta proposals.

\subsection{Approximate Sampling via Taylor Expansion}
The approach of Section \ref{ss:polar} can be slow to mix in high-dimensional settings. Alternatively, we propose the following approximate method, ignoring the sign constraint and using the Fisher-Bingham distribution on the full hypersphere for $G_{0,\theta}$. A first order Taylor expansion of $f_{kj}(\tilde{\mathbf{x}}_{ij}^T\boldsymbol{\theta}^*_{c})$ around the previous value $\tilde{\boldsymbol{\theta}}^{*}_{c}$ 
implies that the sum of squares is approximately
\begin{align*}
\sum_{i=1}^n \left[y_{ik}-\beta_{0k}-\sum_{j=1}^P f_{kj}(\tilde{\mathbf{x}}_{ij}^T\boldsymbol{\theta}^{*}_{Z_{kj}^{\theta}})-\boldsymbol{z}_i^T\boldsymbol{\beta}_{Zk}-\xi \sigma_k u_i\right]^2 
&\approx [\hat{\mathbf{y}}_k-\hat{\mathbf{X}}^T\boldsymbol{\theta}^*_{c}]^T [\hat{\mathbf{y}}_k-\hat{\mathbf{X}}^T\boldsymbol{\theta}^*_{c}], 
\end{align*} 
where $\hat{\mathbf{y}}_k$ is a vector with $i^{th}$ element equal to
\begin{align*}
y_{ik}-\beta_{0k}-\sum_{j:z^{\theta}_{kj} \neq c} f_{kj}(\tilde{\mathbf{x}}_{ij}^T\boldsymbol{\theta}^*_{Z^{\theta}_{kj}})-\sum_{j:z^{\theta}_{kj} = c}f_{kj}(\tilde{\mathbf{x}}_{ij}^T\tilde{\boldsymbol{\theta}}^*_{c})-\boldsymbol{z}_i^T\boldsymbol{\beta}_{Zk}-\xi \sigma_k u_i+\sum_{j:z^{\theta}_{kj} = c}f'_{kj}(\tilde{\mathbf{x}}_{ij}^T\tilde{\boldsymbol{\theta}}^*_{c})\tilde{\mathbf{x}}_{ij}^T \tilde{\boldsymbol{\theta}}^*_{c},
\end{align*}
and $\hat{\mathbf{X}}_k$ is a new design matrix with $i^{th}$ row equal to:
$   \sum_{j:z^{\theta}_{kj} = c}f'_{kj}(\tilde{\mathbf{x}}_{ij}^T\tilde{\boldsymbol{\theta}}^*_{c})\tilde{\mathbf{x}}_{ij}^T.
$
Both quantities require derivatives of $f_{kj}(\cdot)$, but this is straightforward to compute, as derivatives of B-splines are B-splines. The full conditional for ${\boldsymbol{\theta}}^*_{c}$ is then approximately proportional to
\begin{align}
 &\text{I}({\boldsymbol{\theta}^*_{c}}^T\boldsymbol{\theta}^*_{c}=1)\exp (\boldsymbol{\eta}^T\boldsymbol{\theta}^*_{c}  
	-\frac{1}{2}{\boldsymbol{\theta}^*_{c}}^T   \boldsymbol{\Lambda}^{-1} \boldsymbol{\theta}^*_{c}  ) \label{eq:fb} \\
\text{where} ~{\boldsymbol{\eta}}&=\tau_{\theta}\mathbf{1}+\sum_{k,j:z^{\theta}_{kj} = c}{\sigma^{-2}_k} \hat{\mathbf{X}}^T_k \hat{\mathbf{y}}_k ~~~\text{and}~~~~ 
\boldsymbol{\Lambda}^{-1}= \lambda^{\theta} \boldsymbol{\Sigma}_{\theta}  + \sum_{k,j:z^{\theta}_{kj} = c}{\sigma^{-2}_k}\hat{\mathbf{X}}_k^T \hat{\mathbf{X}}_k    \nonumber.
\end{align}
This corresponds to a Fisher Bingham distribution, $\boldsymbol{\theta}^*_c \sim FB({\boldsymbol{\eta}}^T,-\frac{1}{2}\boldsymbol{\Lambda}^{-1})$, which can be conveniently sampled from via the {\tt simdd} package in {\tt R} \citep{kent2018new}. Despite its closed form, directly sampling from a Fisher-Bingham distribution can be computationally challenging in high dimensions. Absent the indicator function, the density in (\ref{eq:fb}) is proportional to a multivariate normal distribution for $\boldsymbol{\theta}^*_c$. As such, in high-dimensional settings we propose to use an approximate update, where we first draw a random variable from a
$N\left(\boldsymbol{\Lambda}{\boldsymbol{\eta}},~\boldsymbol{\Lambda}\right)$ distribution and then project this value to the unit sphere. Empirically we have found that this approximation leads to very similar samples from the posterior distribution of $\boldsymbol{\theta}^*_c$ and is significantly faster to compute.

\section{MCMC Sampler}\label{app:sampler}
	
\begin{enumerate}

		\item For $k=1,\dots,K$, $j=1,\dots,p$, sample cluster IDs:
 
	\begin{align*}
P(Z^{\beta}_{kj}=a|...)&=\frac{\pi_{a,Z^{\theta}_{kj}} \tilde{f}^{\beta}(a)}{\sum_{c=1}^C \pi_{c,Z^{\theta}_{kj}}\tilde{f}^{\beta}(c) }, ~~a \in \{1,\dots,C\}, \\
 \text{where } \tilde{f}^{\beta}(a)&=\left[\prod_{i=1}^n f(y_{ik};\boldsymbol{\beta}^*_{Z^{\beta}_{k1}},\dots,\boldsymbol{\beta}^*_{Z^{\beta}_{k(j-1)}},\boldsymbol{\beta}^*_{a},\boldsymbol{\beta}^*_{Z^{\beta}_{k(j+1)}},\dots,\boldsymbol{\beta}^*_{Z^{\beta}_{kP}},\boldsymbol{\theta}^*_{Z^{\theta}_{k1}},\dots,\boldsymbol{\theta}^*_{Z^{\theta}_{kP}},u_i)\right] \\
P(Z^{\theta}_{kj}=b|...)&=\frac{\pi_{Z^{\beta}_{kj},b} \tilde{f}^{\theta}(b)}{\sum_{c=1}^C \pi_{Z^{\beta}_{kj},b}\tilde{f}^{\theta}(c) }, ~~b \in \{1,\dots,C\} \\
 \text{where } \tilde{f}^{\theta}(b)&=\left[\prod_{i=1}^n f(y_{ik};\boldsymbol{\beta}^*_{Z^{\beta}_{k1}},\dots,\boldsymbol{\beta}^*_{Z^{\beta}_{kP}},\boldsymbol{\theta}^*_{Z^{\theta}_{k1}},\dots,\boldsymbol{\theta}^*_{Z^{\theta}_{k(j-1)}},\boldsymbol{\theta}^*_{b},\boldsymbol{\theta}^*_{Z^{\theta}_{k(j+1)}},\dots,\boldsymbol{\theta}^*_{Z^{\theta}_{kP}},u_i)\right] 
	\end{align*}

	\item For $c=1,\dots, C-1$, sample weights $V^{\beta}_c$ and $V^{\theta}_c$ via MH:
	\begin{align*}
	(V^{\beta}_c|...)&\propto (1-V^{\beta}_c)^{\alpha_{\beta}-1} \prod_{k=1}^K \prod_{j=1}^p \pi^*_{z^{\beta}_{kj},z^{\theta}_{kj}}\\
	(V^{\theta}_c|...)&\propto (1-V^{\theta}_c)^{\alpha_{\theta}-1} \prod_{k=1}^K \prod_{j=1}^p \pi^*_{z^{\beta}_{kj},z^{\theta}_{kj}}.
	\end{align*}
We can either do grid sampling or MH. As proposals we could use the posterior under independent and non-informative clustering ($\rho=0$):
\begin{align*}
V^{\beta}_c|\rho=\gamma^x=\gamma^y=0... &\sim Beta(1+n^{\beta}_c,\alpha_{\beta}+\sum_{c'=c+1}^C n^{\beta}_{c'}) \\
V^{\theta}_c|\rho=\gamma^x=\gamma^y=0... &\sim Beta(1+n^{\theta}_c,\alpha_{\theta}+\sum_{c'=c+1}^C n^{\theta}_{c'})
\end{align*} where $n^{\beta}_c$ is the size of the $c^{th}$ cluster for $\beta$, and analogously for ${\theta}$.

\item For $c=1,\dots,C$, if $\sum_{kj} Z^{\beta}_{kj}=0$, then draw $\boldsymbol{\beta}^*_c$ from the prior. Otherwise sample $\boldsymbol{\beta}^*_c$ from full conditional: 
\begin{align*}
\boldsymbol{\beta}^*_c|... &\sim N(\boldsymbol{\mu}^*,\boldsymbol{\Sigma}^*) \\
\boldsymbol{\mu}^*&= 
\boldsymbol{\Sigma}^* \times 
\left( \boldsymbol{\mu}_0^T \lambda^{\beta}\boldsymbol{\Sigma}_0^{-1} +\sum_{k}^K \sigma^{-2}_k\left[ \sum_{j:Z^\beta_{kj}=c}\mathbf{B}_{j,\boldsymbol{\theta}_{kj}}^T\right]\left[\mathbf{y}_{ k}-\beta_{0k}\mathbf{1}-\xi \sigma_k \mathbf{u}-\sum_{j:Z^\beta_{kj}\neq c}\mathbf{B}_{j,\boldsymbol{\theta}_{kj}} \boldsymbol{\beta}_{kj} \right]\right) \\
\boldsymbol{\Sigma}^*&=
\left(\lambda^{\beta}\boldsymbol{\Sigma}_0^{-1}+\sum_{k}^K \sigma^{-2}_k\left[ \sum_{j:Z^\beta_{kj}=c}\mathbf{B}_{j,\boldsymbol{\theta}_{kj}}^T\right]\left[\sum_{j:Z^\beta_{kj}=c}\mathbf{B}_{j,\boldsymbol{\theta}_{kj}}\right]\right)^{-1}
\end{align*}

\item For $c=1,\dots,C$ if $\sum_{kj} Z^{\theta}_{kj}=0$, then draw $\boldsymbol{\theta}^*_c$ from the prior. Otherwise sample $\boldsymbol{\theta}^*_c$ via 
one of the two following methods:
\begin{enumerate}
    \item MH on polar coordinates:
    \begin{align*}
\theta^*_{c1}&=\sin(\phi^*_{c1}) \\
\theta^*_{c2}&=\sin(\phi^*_{c2})\cos(\phi^*_{c1}) \\
&\vdots \\
\theta^*_{c{L-1}}&=\sin(\phi^*_{c{L-1}})\prod_{j=1}^{L-2} \cos(\phi^*_{cj}) \\
\theta^*_{cL}&=\prod_{j=1}^{L-1} \cos(\phi^*_{cj})
\end{align*}
where $\phi^*_{cj}\in[-\frac{\pi}{2},\frac{\pi}{2}]$ $j=1,\dots,{L-1}$. 

\begin{align*}
(\phi^*_{cj}|...) &  \propto  \exp\left(\tau_{\theta}\mathbf{1}^T\boldsymbol{\theta}^*_c-\frac{1}{2}
\lambda^{\theta}{\boldsymbol{\theta}_c^{*}}^T\Sigma_{\theta}^{-1} \boldsymbol{\theta}_c^{*} \right) \\
&\qquad\qquad\times \prod_{k=1}^K  \left[\prod_{i=1}^n f(y_{ik};\boldsymbol{\beta}^*_{Z^{\beta}_{k1}},\dots,\boldsymbol{\beta}^*_{Z^{\beta}_{kP}},\boldsymbol{\theta}^*_{Z^{\theta}_{k1}},\dots,\boldsymbol{\theta}^*_{Z^{\theta}_{kP}},u_i)\right]^{ I(\sum_j z^{\theta}_{kj}>0)} \\
& \qquad\qquad \times |\cos(\phi^*_{cj})^{L-j}|
\end{align*}
(where the last component comes from change of variable). We use scaled/shifted Beta proposals: $ Beta(a_{\phi},b_{\phi})$ with the mode set to be the previous value: i.e. $b_{\omega}=((1-\left[\frac{\phi^*_{cj}+\frac{\pi}{2}}{\pi}\right])a_{\phi}+2 \left[\frac{\phi^*_{cj}+\frac{\pi}{2}}{\pi}\right]-1)/\left[\frac{\phi^*_{cj}+\frac{\pi}{2}}{\pi}\right]$.

\item Taylor Approximation + Fisher Bingham

Approximating $f_{kj}(\tilde{\mathbf{x}}_{ij}^T\boldsymbol{\theta}^*_{c})$  around the current value $\tilde{\boldsymbol{\theta}}^{*}_{c}$ yields:
\begin{align*}
f_{kj}(\tilde{\mathbf{x}}_{ij}^T\boldsymbol{\theta}^*_{c}) &\approx f_{kj}(\tilde{\mathbf{x}}_{ij}^T\tilde{\boldsymbol{\theta}}^{*}_{c})+f'_{kj}(\tilde{\mathbf{x}}_{ij}^T\tilde{\boldsymbol{\theta}}^{*}_{c})\tilde{\mathbf{x}}_{ij}^T (\boldsymbol{\theta}^{*}_{c}-\tilde{\boldsymbol{\theta}}^{*}_{c})  \implies 
\end{align*}
{\small
\begin{align*}
\sum_{i=1}^n \left[\left(y_{ik}-\beta_{0k}-\sum_{j:z^{\theta}_{kj}\neq c}f_{kj}(\tilde{\mathbf{x}}_{ij}^T\boldsymbol{\theta}_{kj})-\xi \sigma_k u_i\right)-\sum_{j:z^{\theta}_{kj}= c}f_{kj}(\tilde{\mathbf{x}}_{ij}^T\boldsymbol{\theta}^*_{c})\right]^2 
&\approx [\hat{\mathbf{y}}_k-\hat{\mathbf{X}}^T\boldsymbol{\theta}^*_{c}]^T [\hat{\mathbf{y}}_k-\hat{\mathbf{X}}^T\boldsymbol{\theta}^*_{c}] 
\end{align*}}
where $\hat{\mathbf{y}}_k$ is a vector with $i^{th}$ element 
\begin{align*}
 \left(y_{ik}-\beta_{0k}-\sum_{j:z^{\theta}_{kj}\neq c}f_{kj}(\tilde{\mathbf{x}}_{ij}^T\boldsymbol{\theta}_{kj})-\xi \sigma_k u_i\right)-\sum_{j:z^{\theta}_{kj} = c} f_{kj}(\tilde{\mathbf{x}}_{ij}^T\tilde{\boldsymbol{\theta}}_{kj})+\sum_{j:z^{\theta}_{kj} = c}f'_{kj}(\tilde{\mathbf{x}}_{ij}^T\tilde{\boldsymbol{\theta}}_{kj})\tilde{\mathbf{x}}_{ij}^T \tilde{\boldsymbol{\theta}}_{kj}
\end{align*}

 $\hat{\mathbf{X}}_k$ is a new design matrix with $i^{th}$ row equal to:
 \begin{align*}
    \sum_{j:z^{\theta}_{kj} = c}f'_{kj}(\tilde{\mathbf{x}}_{ij}^T\tilde{\boldsymbol{\theta}}_{kj})\tilde{\mathbf{x}}_{ij}^T
 \end{align*}

Then we can approximate the full conditional by
{\small
	\begin{align*}
	(\boldsymbol{\theta}^*_c|...) & \propto I(||\boldsymbol{\theta}^*_{c}||=1)\exp \left( \left[\tau_{\theta}\mathbf{1}+\sum_{k,j:z^{\theta}_{kj} = c} \sigma^{-2}_k\hat{\mathbf{X}}^T_k \hat{\mathbf{y}}_k \right]^T\boldsymbol{\theta}^*_{c}  
	  \right. \\
	&\qquad\qquad\qquad\qquad\qquad -\frac{1}{2}{\boldsymbol{\theta}^*_{c}}^T   \left[ \sum_{k,j:z^{\theta}_{kj} = c}\sigma^{-2}_k\hat{\mathbf{X}}_k^T \hat{\mathbf{X}}_k \right] \boldsymbol{\theta}^*_{c}    -\frac{1}{2}\lambda^{\theta} {\boldsymbol{\theta}_c^{*}}^T\Sigma_{\theta}^{-1} \boldsymbol{\theta}_c^{*}\Biggr) \\
	\boldsymbol{\theta}^*_c|... &\sim FB\left(\bar{\boldsymbol{\mu}}=\left[\tau_{\theta}\mathbf{1}+\sum_{k,j:z^{\theta}_{kj} = c} \sigma^{-2}_k\hat{\mathbf{X}}^T_k \hat{\mathbf{y}}_k \right]^T ,  ~~\bar{\mathbf{A}}=-\frac{1}{2}  \left[ \sum_{k,j:z^{\theta}_{kj} = c}\sigma^{-2}_k\hat{\mathbf{X}}_k^T \hat{\mathbf{X}}_k \right]    -\frac{1}{2}\lambda^{\theta}\Sigma_{\theta} ^{-1}\right)
	\end{align*}
}
In practice sampling from this distribution can be challenging, so we approximate it by sampling from $N\left(\hat{\boldsymbol{\mu}},~~\hat{\boldsymbol{\Sigma}}\right)$ and then standardizing appropriately:
\begin{align*}
 \hat{\boldsymbol{\mu}}&= \left[   \left[ \sum_{k,j:z^{\theta}_{kj} = c}\sigma^{-2}_k\hat{\mathbf{X}}_k^T \hat{\mathbf{X}}_k \right]    +\lambda^{\theta}\Sigma_{\theta}^{-1} \right]^{-1} \left[\tau_{\theta}\mathbf{1}+\sum_{k,j:z^{\theta}_{kj} = c} \sigma^{-2}_k\hat{\mathbf{X}}^T_k \hat{\mathbf{y}}_k \right]^T\\
\hat{\boldsymbol{\Sigma}}&= \left[  \left[ \sum_{k,j:z^{\theta}_{kj} = c}\sigma^{-2}_k\hat{\mathbf{X}}_k^T \hat{\mathbf{X}}_k \right]    +\lambda^{\theta}\Sigma_{\theta}^{-1} \right]^{-1}
\end{align*}

\end{enumerate}

 \item Sample $\alpha_{\beta}$ and $\alpha_{\theta}$ from full conditionals:
\begin{align*}
\alpha_{\beta}|... &\sim Gamma\left(a_{\beta}+C-1,b_{\beta}-\sum_{c=1}^{C-1} \log(1-V^{\beta}_c)\right)
\end{align*}
\begin{align*}
\alpha_{\theta}|... &\sim Gamma\left(a_{\theta}+C-1,b_{\theta}-\sum_{c=1}^{C-1} \log(1-V^{\theta}_c)\right)
\end{align*}

 \item Sample $\log(\rho)$ via random walk metropolis:
\begin{align*}
(\log(\rho)|...)&\propto  Gamma(a_{\rho},b_{\rho}) \times \prod_{k=1}^K \prod_{j=1}^p \pi^*_{z^{\beta}_{kj},z^{\theta}_{kj}} \times \rho 
\end{align*} 
(where the last component comes from the change of variable).

 \item Sample penalty term $\lambda^{\beta}$ from full conditional
\begin{align*}
\lambda^{\beta}|\dots &\sim Gamma\left(a_{\lambda}^{\beta}+\frac{Cd}{2}, ~~~ b_{\lambda}^{\beta}+\frac{1}{2}\sum_{c=1}^C {\boldsymbol{\beta}_c^{*}}^T \boldsymbol{\Sigma}_0^{-1}{\boldsymbol{\beta}_c^{*}} \right)
\end{align*}
where $d$ is the dimension of $\boldsymbol{\beta}^*_c$.

 \item Sample penalty term $\log \lambda^{\theta}$ via  random walk metropolis (if using DLM smoothness penalty):
\begin{align*}
(\log (\lambda^{\theta})|\dots) &\propto Gamma(a_{\lambda}^{\theta},b_{\lambda}^{\theta})\times[C_0(\tau_{\theta}\mathbf{1} ,-\frac{1}{2}\lambda^{\theta}\Sigma_{\theta}^{-1} )]^{-C}\exp\left(-\frac{1}{2}\lambda^{\theta}\sum_{c=1}^C{\boldsymbol{\theta}_c^{*}}^T\Sigma_{\theta}^{-1} \boldsymbol{\theta}_c^{*} \right)  \times \lambda^{\theta} \\ 
\end{align*}
where $C_0(\cdot)$ is the normalizing constant for the FB distribution (which can be obtained via {\tt fb.saddle()} in the {\tt Directional} package).  

\item For $i=1,\dots,n$ sample random effects $u_i$ from full conditional:
\begin{align*}
u_i|... &\sim N\left(\left[1+\sum_{k=1}^K \frac{\xi^2 \sigma_k^2}{\sigma^2_k} \right]^{-1}  \sum_{k=1}^K \frac{\xi \sigma_k}{\sigma^2_k}\left[ y_{ik} - \beta_{0k}-\sum_{j=1}^p\mathbf{b}_{i,\boldsymbol{\theta}_{kj}}^T \boldsymbol{\beta}_{kj}\right]  ,~
\left[1+\sum_{k=1}^K \frac{\xi^2 \sigma_k^2}{\sigma^2_k} \right]^{-1} \right)
\end{align*}

\item For $k=1,\dots,K$, sample random effects coefficient/covariance component $\xi $ via random walk MH on the log scale:
{\small
 \begin{align*}
( \text{log}(\xi) |...) &\propto InvGamma(a_{\xi},b_{\xi})
 \times \exp\left( -\frac{1}{2}\sum_{k=1}^K \frac{1}{\sigma^2_k}\sum_{i=1}^n \left[y_{ik} - \beta_{0k}-\sum_{j=1}^p\mathbf{b}_{i,\boldsymbol{\theta}_{kj}}^T \boldsymbol{\beta}_{kj}-\xi \sigma_k u_i  \right]^2   \right)
 \times \xi 
\end{align*}}where the last component comes from the change of variable.

\item For $k=1,\dots,K$, sample error variances  $\sigma^2_k$ via random walk MH on the log scale
{\small
\begin{align*}
 (\text{log}(\sigma^2_k)|...) & \propto InvGamma\left(a_{\sigma}, b_{\sigma}\right) 
 \times (\sigma^2_k)^{-\frac{n}{2}}\exp\left( -\frac{1}{2\sigma^2_k}\sum_{i=1}^n  \left[y_{ik}-\beta_{0k}-\sum_{j=1}^p\mathbf{b}_{i,\boldsymbol{\theta}_{kj}}^T \boldsymbol{\beta}_{kj}-\xi \sigma_k u_i\right]^2\right)
 \times \sigma^2_k
\end{align*}}where the last component comes from the change of variable.

\item For $k=1,\dots,K$, sample intercept $\beta_{0k}$ from full conditional:

\begin{align*}
    \beta_{0k}|... &\sim N\left(\frac{1}{n}\sum_{i=1}^n \left[y_{ik}-\sum_{j=1}^p\mathbf{b}_{i,\boldsymbol{\theta}_{kj}}^T \boldsymbol{\beta}_{kj}-\xi \sigma_k u_i\right] ,~~ \sigma^2_k/n\right)
\end{align*} 

\item For $k=1,\dots,K$ sample confounder coefficients $\boldsymbol{\beta}_{Zk}$ from full conditional:

\begin{align*}
\boldsymbol{\beta}_{Zk}| \cdots \sim N\left(\left[\mathbf{Z}^{T} \mathbf{Z} \right]^{-1}\mathbf{Z}^{T}\left[\textbf{y}_{\cdot k}-\beta_{0k}-\sum_{j=1}^p\mathbf{b}_{i,\boldsymbol{\theta}_{kj}}^T \boldsymbol{\beta}_{kj}-\xi \sigma_k \mathbf{u},\right] , ~~\sigma_k^2\left[\mathbf{Z}^{T}, \mathbf{Z}\right]^{-1}\right)
\end{align*}

\end{enumerate}

\section{Default Hyperparameter Settings}
\label{app:default_hyperparams}
Users may provide specify their own hyperparameters using the provided software. Below we detail the defaults used in our simulations and analyses unless otherwise stated. 

\begin{itemize}
    \item Priors hyperparameters:
    \begin{itemize}
        \item $a_{\beta}=b_{\beta}=1$
        \item $a_{\theta}=b_{\theta}=1$
        \item $a_{\rho}=b_{\rho}=1$
        \item  $a_{\lambda}^\beta=b_{\lambda}^\beta=1$
        \item $a_{\lambda}^\theta=1$, $b_{\lambda}^\theta=0.001$
        \item $a_{\xi}=b_{\xi}=0.01$
        \item $a_{\sigma}=b_{\sigma}=0.01$
    \end{itemize}
    \item MCMC settings:
    \begin{itemize}
        \item Random walk SD=1 for $\log(\rho)$,  $\log(\lambda^{\theta})$, $\xi$, $\log(\sigma^2_k)$
        \item Gridsize=10 for grid sampling $V_c^{\beta}$ and $V_c^{\theta}$
    \end{itemize}

\item Simulation A: Distributed lag models with multiple exposures
\begin{itemize}
    \item In models (1) \& (2) we set $C=10$ and in model (4) $C=5$. In model (3) there is no clustering.
    \item Order of difference matrix ($D_s$) $s=2$
    \item Dimension of basis $\boldsymbol{\Psi}$ is 6 (when using dimension reduction technique in S2.1) 
\end{itemize}

\item Simulation B: Single time point mixture modeling
\begin{itemize}
    \item Order of index model $P=3$
    \item Maximum number of clusters $C=KP$
\end{itemize}

\item NMMAPS Data Analysis
\begin{itemize}
    \item Maximum number of clusters $C=6$
    \item  $a_{\lambda}^\theta=1$, $b_{\lambda}^\theta=0.001$
\end{itemize}
\end{itemize}

\section{Exposure Importance Metrics: Details and Extension to Exposure Groups}
\label{app:VIM}
When interest lies not in individual exposures but in a \textit{group} of exposures $\boldsymbol{x}^{\dagger}$---for example a class of highly correlated exposures---our proposed importance measure $\Psi_{kj}$ generalizes to any partition $\boldsymbol{x}^T=[{\boldsymbol{x}^{\dagger}}^T,{\boldsymbol{x}^{o}}^T]$: 
$$\Psi_{k}(\boldsymbol{x}^{\dagger}) = 1 - \frac{Var_{\boldsymbol{x}^{o}} \bigg\{ E_{\boldsymbol{x}^{\dagger} \mid \boldsymbol{x}^{o}} \Big[ f^*_k(\boldsymbol{x}^{\dagger},\boldsymbol{x}^{o}) \Big] \bigg\}}{Var[f^*_k(\boldsymbol{x})]}.$$
This now represents the importance of the group of exposures found in $\boldsymbol{x}^{\dagger}$. Computation follows similarly to the univariate case.

The proposed $\Psi_{kj}$ has connections to a number of related variable importance metrics in the literature. It is closely related to those proposed in \cite{shin2024treatment}, which instead target effect modification and identify variables that modify associations the most. It is also related to metrics used in nonparametric regression such as Shapley values \citep{shapley1953value} or leave-one-covariate-out (LOCO; \citealp{lei2018distribution}) that target the extent to which mean squared error is reduced when a covariate is omitted. One important distinction is that our proposed metric does not require fitting multiple models (e.g., omitting exposure $j$). Rather, it uses a single fit along with an estimate of the distribution of the exposures. While the distribution of the exposures can be misspecified, we view $\Psi_{kj}$ as a complementary piece of information to aid interpretation, even if the exposure distribution is imperfectly estimated.

\section{Additional simulation details}
\label{appSec:Simulations}

\subsection{Details on Autocorrelated Exposures for Simulation A}
\label{appSec:simAdetails}

We drew exposures at the first time point $\mathbf{x}_{\cdot 1}=(x_{11},\dots,x_{P1})^T$ from a $\boldsymbol{\epsilon}_1 \sim \mathcal{MVN}(\boldsymbol{0}, \boldsymbol{\Sigma})$ and  generated subsequent exposures $\mathbf{x}_{\cdot l}=(x_{1l},\dots,x_{Pl})^T$ for $l=2, \dots, L$, according to $\mathbf{x}_{\cdot l} = \boldsymbol{M} \mathbf{x}_{\cdot  l-1} + \boldsymbol{\epsilon}_l$, where $\boldsymbol{\epsilon}_l \sim \mathcal{MVN}(\boldsymbol{0}, \boldsymbol{\Sigma})$.  
We set $\boldsymbol{M} = 0.85 \boldsymbol{I}_P$ to induce a high degree of correlation over time within exposures, and we set $\boldsymbol{\Sigma}_{pp'} = 0.6^{|p - p'|}$ to induce dependence across exposures, as is typical in  mixtures of environmental exposures. 

\begin{figure}[htbp!]
    \centering
    \includegraphics[width=0.9\linewidth]{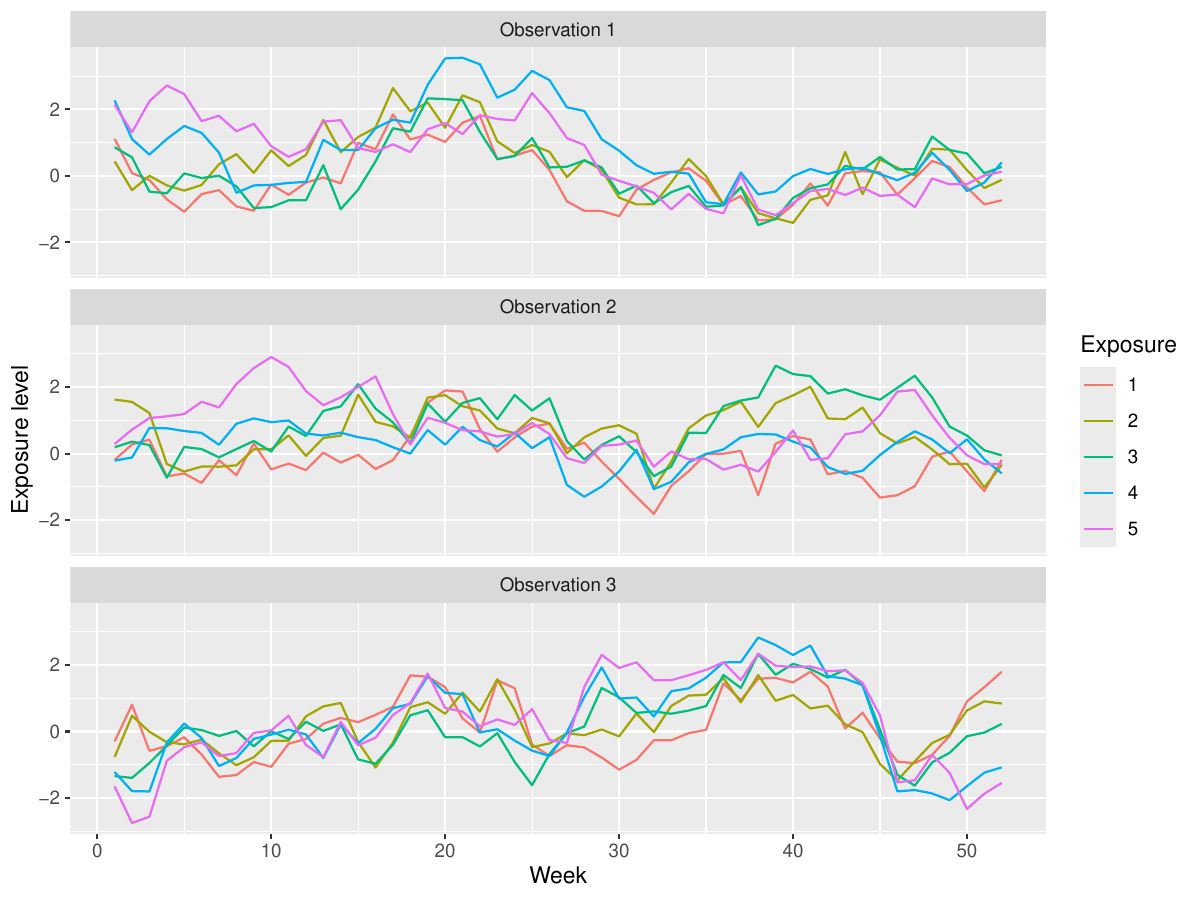}
    \caption{Exposure trajectories over time for three randomly chosen observations from one of the simulated data sets in Section 5.1.}
    \label{fig:ExpTrajectories}
\end{figure}

Here we provide additional information on the data generating process for the simulations involving distributed lag exposures in Section 5.1. Figure \ref{fig:ExpTrajectories} shows three realizations (three observations) of the exposure vectors measured across time. We see that there is a moderate degree of temporal dependence due to the autoregressive structure with a large correlation coefficient. Additionally, the five exposures tend to increase (decrease) together over time due to the dependence induced by the vector autoregressive process generating the data, which imposes correlation across exposures at each point in time. 

\clearpage

\subsection{Additional results}

\begin{figure}[htbp!]
    \centering
    \includegraphics[width=0.9\linewidth]{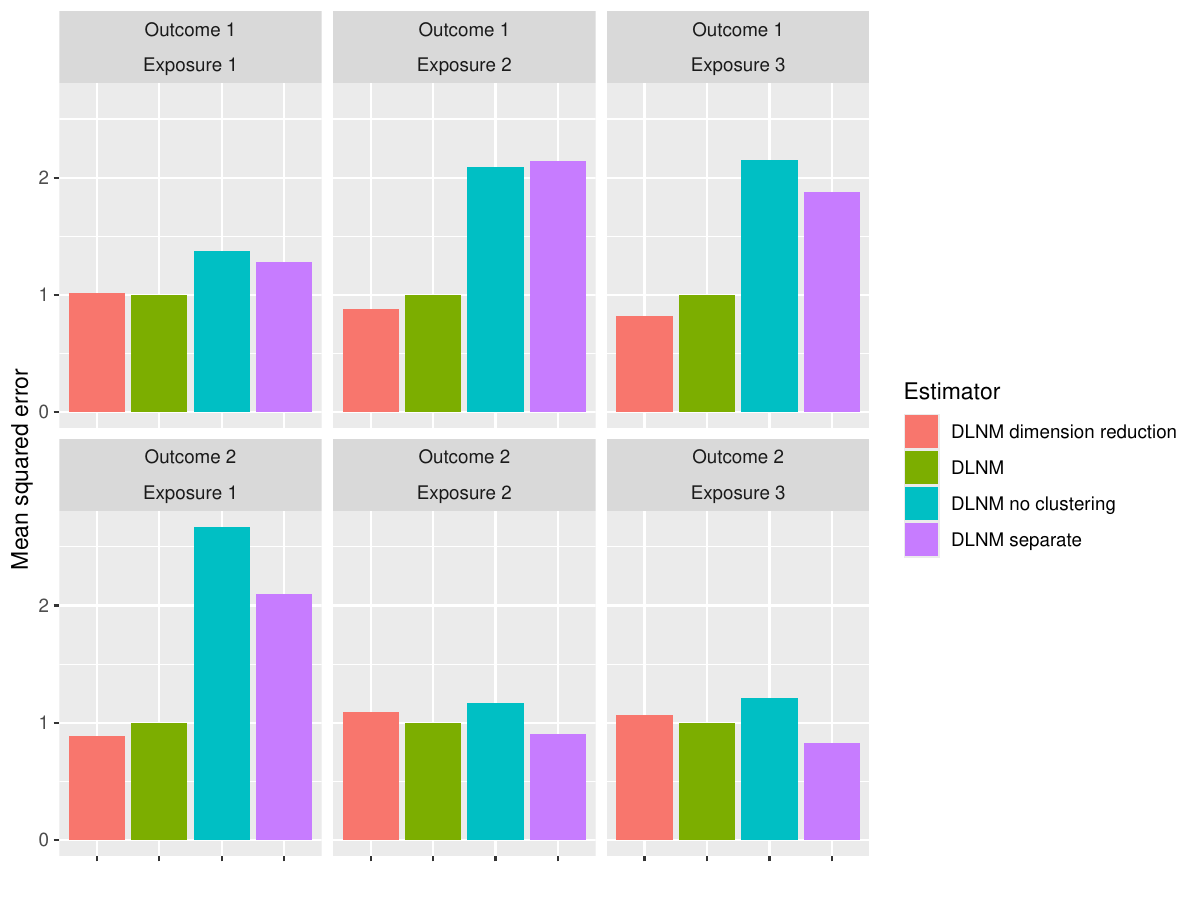}
    \caption{Mean squared error for estimating $f_{kj}(\cdot)$ across all estimators. MSE is scaled so that the DLNM estimator has an MSE of 1 in all scenarios. }
    \label{fig:DLAG_MSEfx}
\end{figure}

\begin{figure}[hbtp!]
    \centering
    \includegraphics[width=0.9\linewidth]{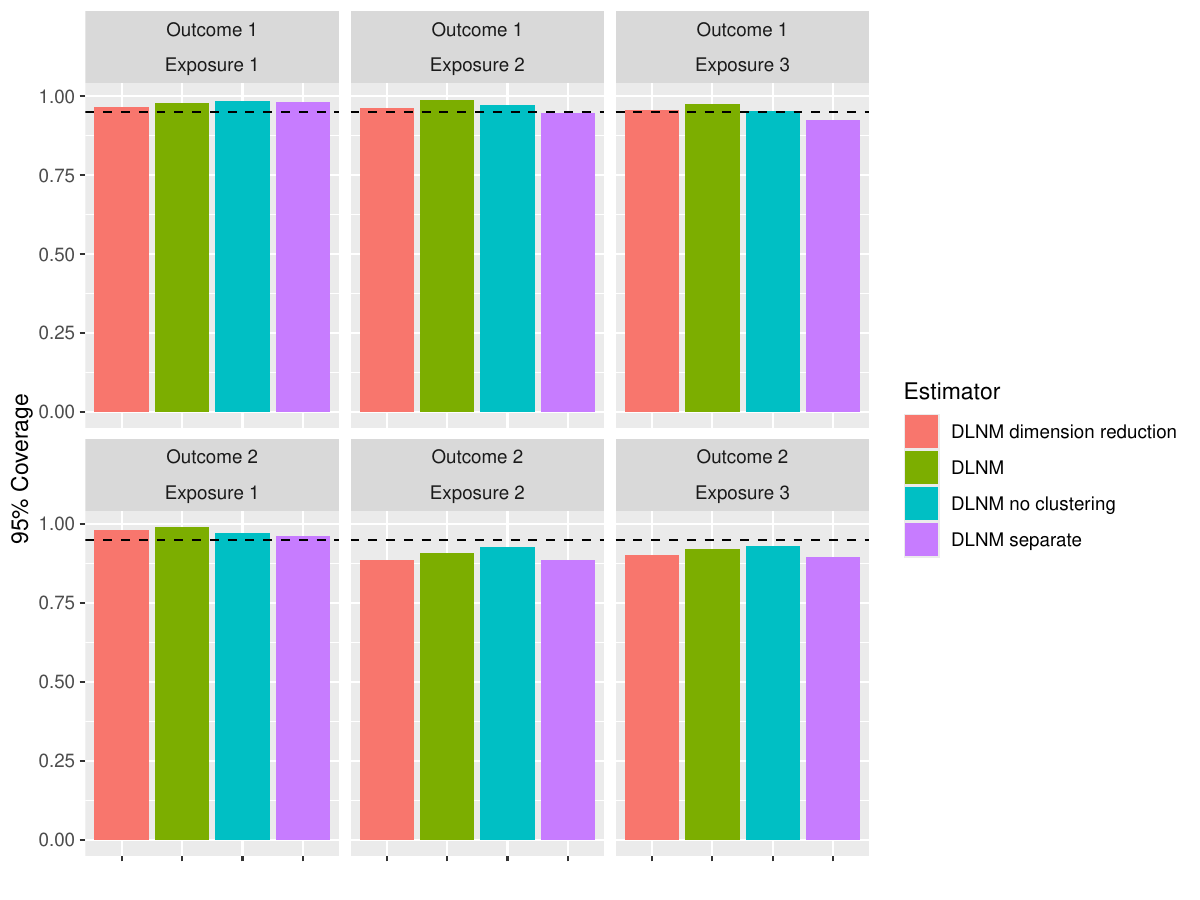}
    \caption{95\% credible interval coverage for estimating $\boldsymbol{\omega}_{kj}$ across all estimators.}
    \label{fig:DLAG_COV}
\end{figure}

\begin{figure}[hbtp!]
    \centering
    \includegraphics[width=0.9\linewidth]{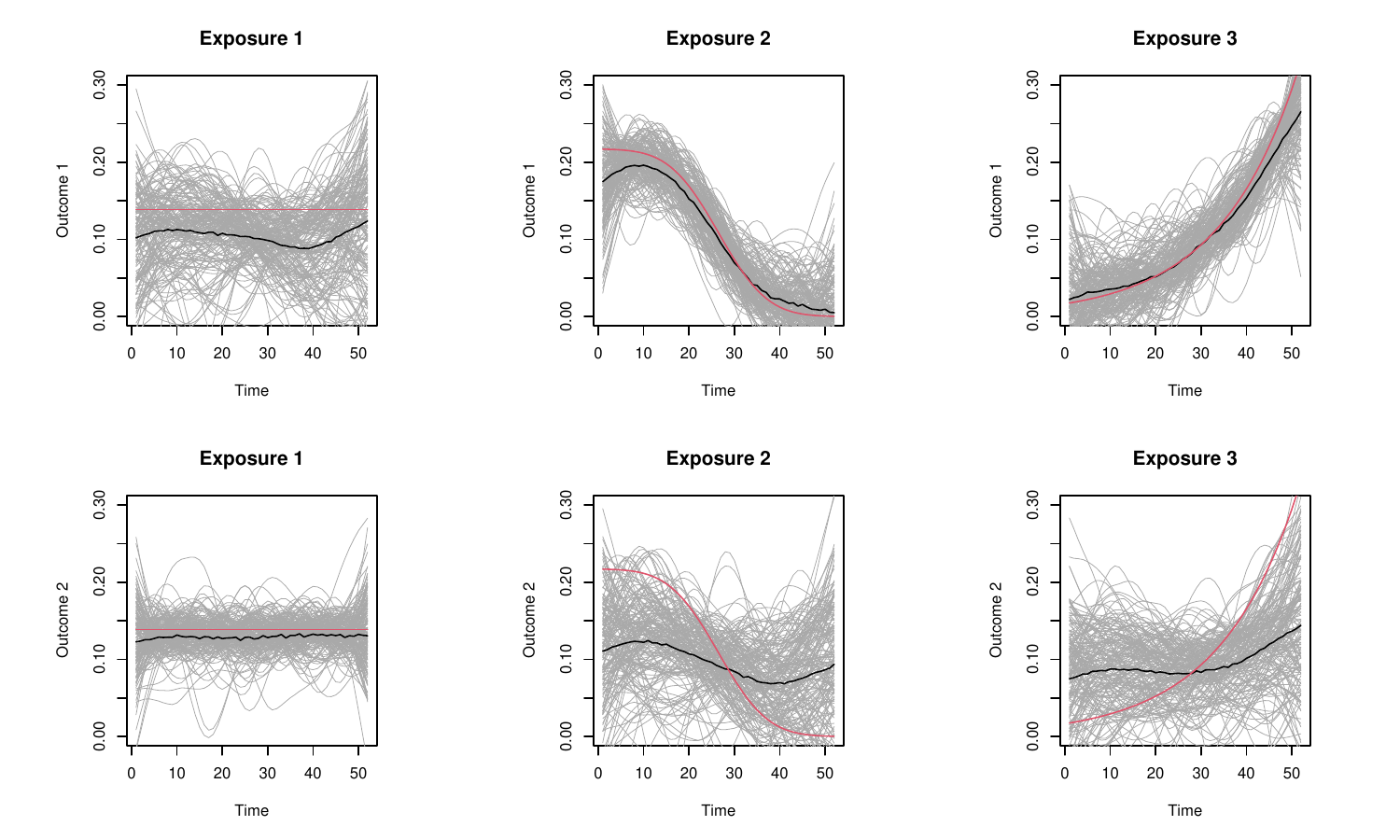}
    \caption{Estimates of $\boldsymbol{\omega}_{kj}$ for the DLNM estimator. Individual grey lines correspond to individual data sets, while the red line represents the truth, and the black line represents the mean across data sets. }
    \label{fig:DLAG_estimates}
\end{figure}

Here we provide additional results for the simulation studies found in Section 5.1. Specifically, we now provide 95\% credible interval coverage for the $\boldsymbol{\omega}_{kj}$ parameters (Figure \ref{fig:DLAG_COV}), estimates of the individual $\boldsymbol{\omega}_{kj}$ trajectories across data sets and their corresponding means (Figure \ref{fig:DLAG_estimates}), and mean squared error of estimation of $f_{kj}(\cdot)$ (Figure \ref{fig:DLAG_MSEfx}). Overall, the results echo those seen in Section 5.1, but provide some additional insight into clustering, and when it is beneficial. We see that the interval coverages are generally at the desired 95\% level, however, for outcome 2 under both exposure 2 and 3, we see some slight under-coverage below the nominal level. We see that this issue is reduced when we do not perform clustering as the DLNM no clustering estimator has credible interval at the nominal level. Figure \ref{fig:DLAG_estimates} provides intuition for this, as we can see that in many data sets, exposures 2 and 3 are placed into the wrong cluster for $\boldsymbol{\omega}_{kj}$. This generally highlights the bias-variance trade-off that comes from our clustering algorithm. Clustering generally improves overall MSE of these parameters, which are difficult to estimate in small sample sizes, but the shrinkage across exposures and outcomes induces some bias that can lead to reduced credible interval coverages. Lastly, when estimating $f_{kj}(\cdot)$, we see similar results as for $\boldsymbol{\omega}_{kj}$, though there are certain instances where it is slightly better to not perform clustering across outcomes, as seen in the bottom panel of Figure \ref{fig:DLAG_MSEfx}. This issue, along with the reduced credible interval coverages for certain combinations of exposures and outcomes, disappears for larger sample sizes, which we show in the following section.

\clearpage

\subsection{Larger sample sizes}
\label{ss:largersamplesizes}

\begin{figure}[hbtp!]
    \centering
    \includegraphics[width=0.9\linewidth]{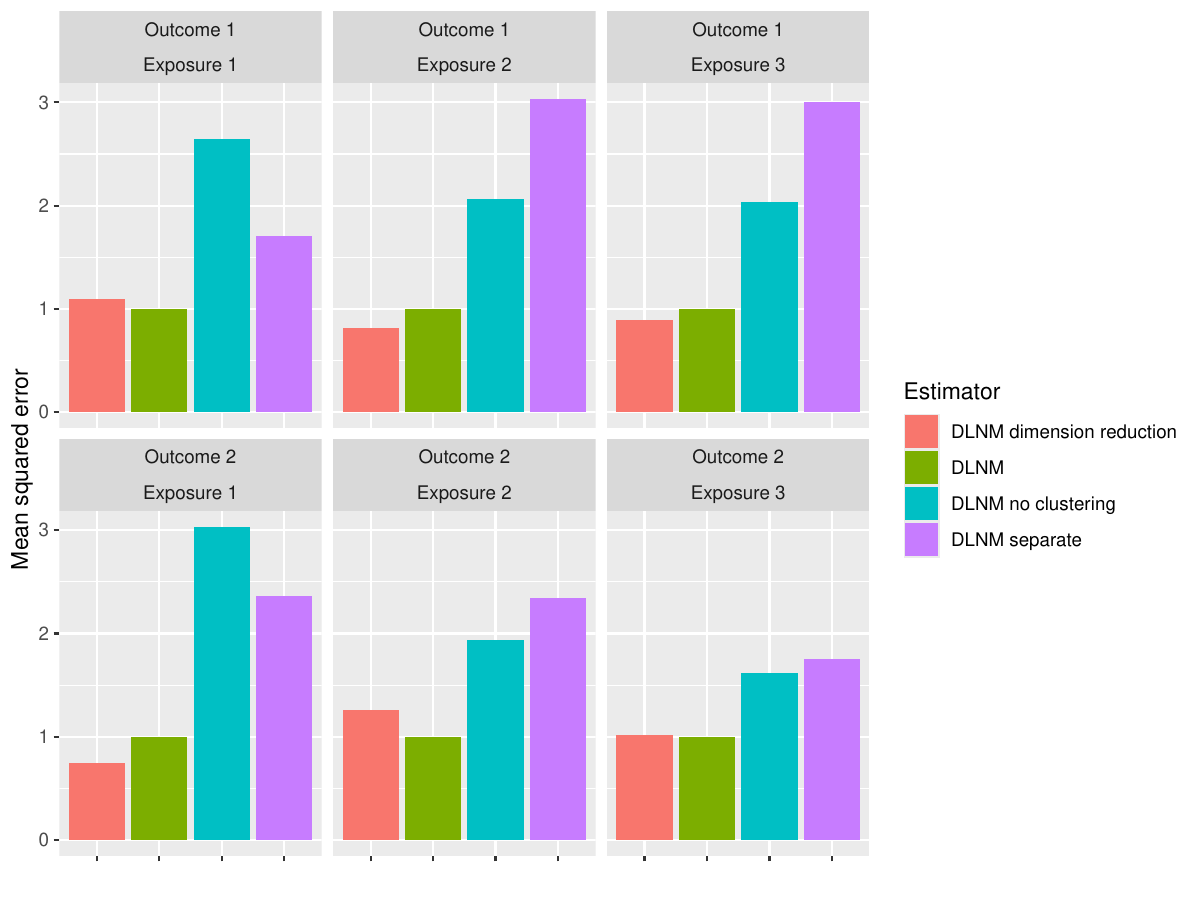}
    \caption{Mean squared error for estimating $\boldsymbol{\omega}_{kj}$ across all estimators when the sample size is $n=1500$. MSE is scaled so that the DLNM estimator has an MSE of 1 in all scenarios. }
    \label{fig:DLAG_MSElarge}
\end{figure}

\begin{figure}[hbtp!]
    \centering
    \includegraphics[width=0.99\linewidth]{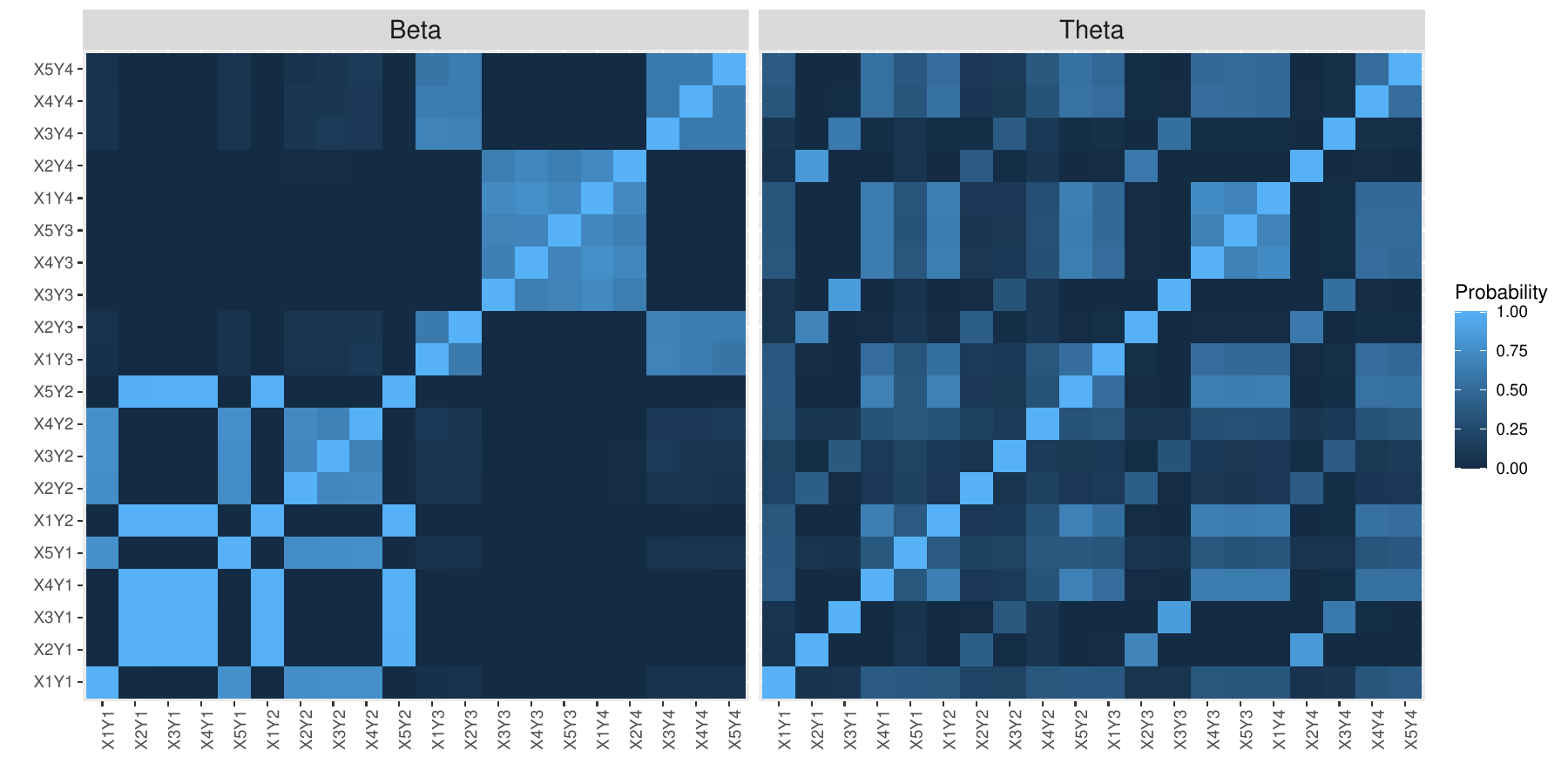}
    \caption{Posterior probabilities of pairwise clustering for both sets of model parameters, averaged across simulated datasets when the sample size is $n=1500$.}
    \label{fig:DLAG_clusteringlarge}
\end{figure}

\begin{figure}[hbtp!]
    \centering
    \includegraphics[width=0.9\linewidth]{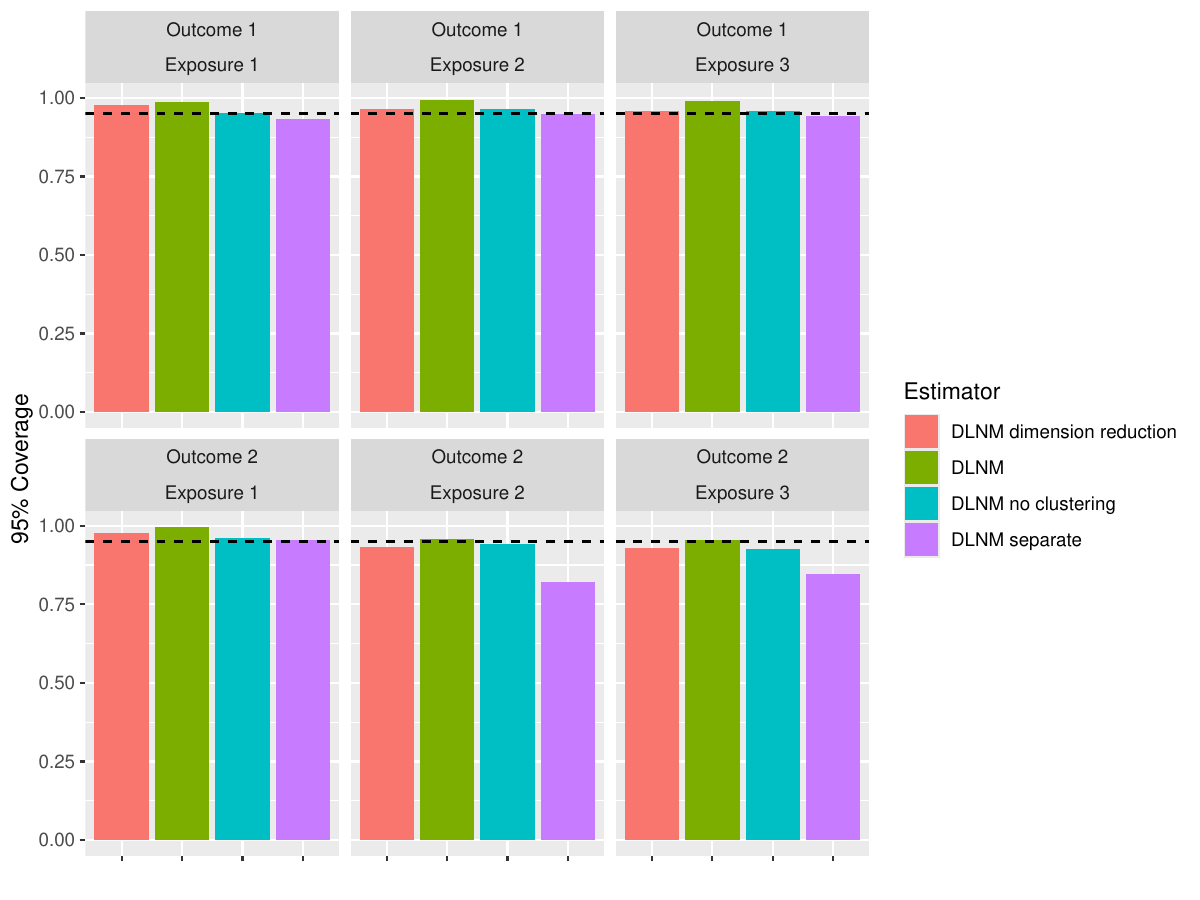}
    \caption{95\% credible interval coverage for estimating $\boldsymbol{\omega}_{kj}$ across all estimators when the sample size is $n=1500$.}
    \label{fig:DLAG_COVlarge}
\end{figure}

\begin{figure}[hbtp!]
    \centering
    \includegraphics[width=0.9\linewidth]{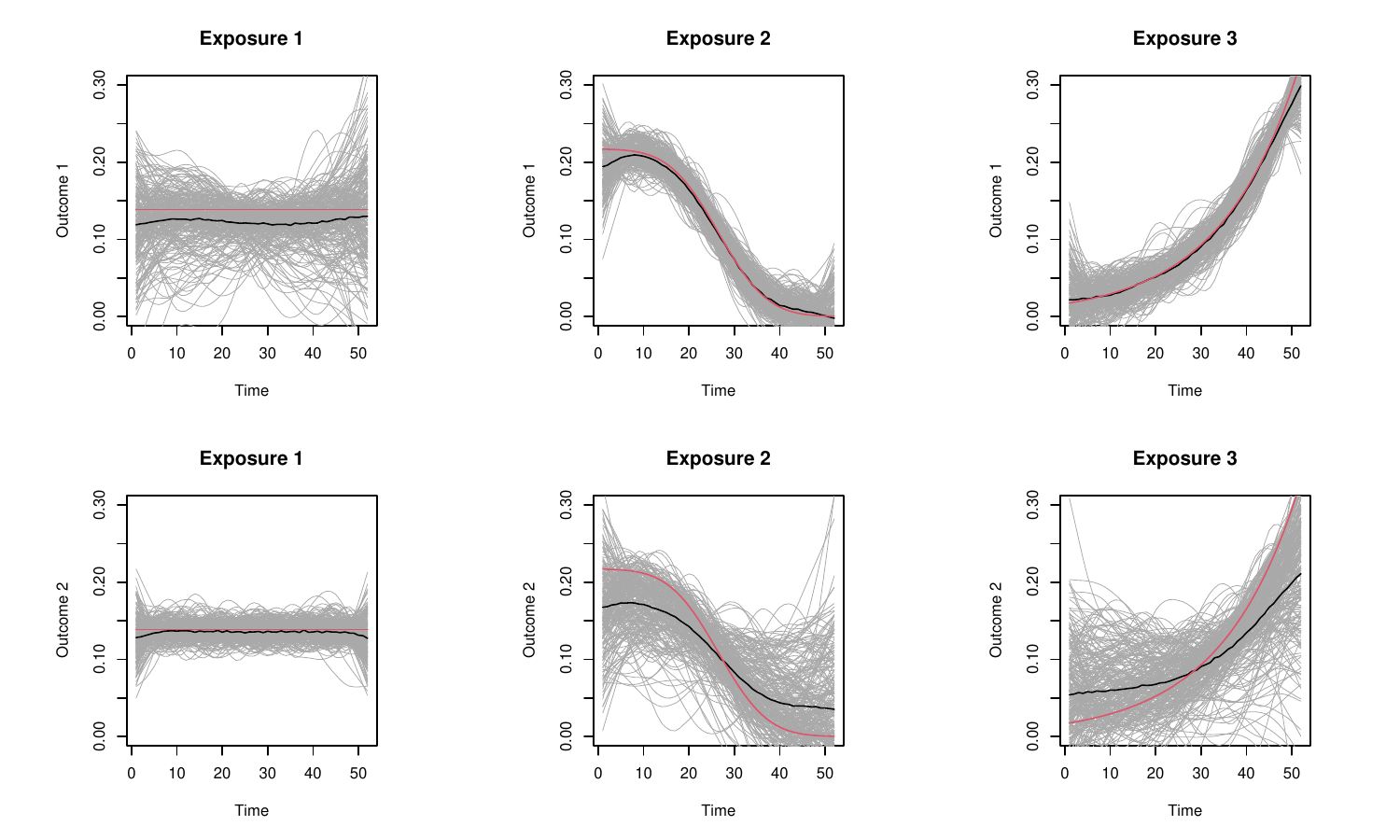}
    \caption{Estimates of $\boldsymbol{\omega}_{kj}$ for the DLNM estimator when the sample size is $n=1500$. Individual grey lines correspond to individual data sets, while the red line represents the truth, and the black line represents the mean across data sets. }
    \label{fig:DLAG_estimateslarge}
\end{figure}

\begin{figure}[htbp!]
    \centering
    \includegraphics[width=0.9\linewidth]{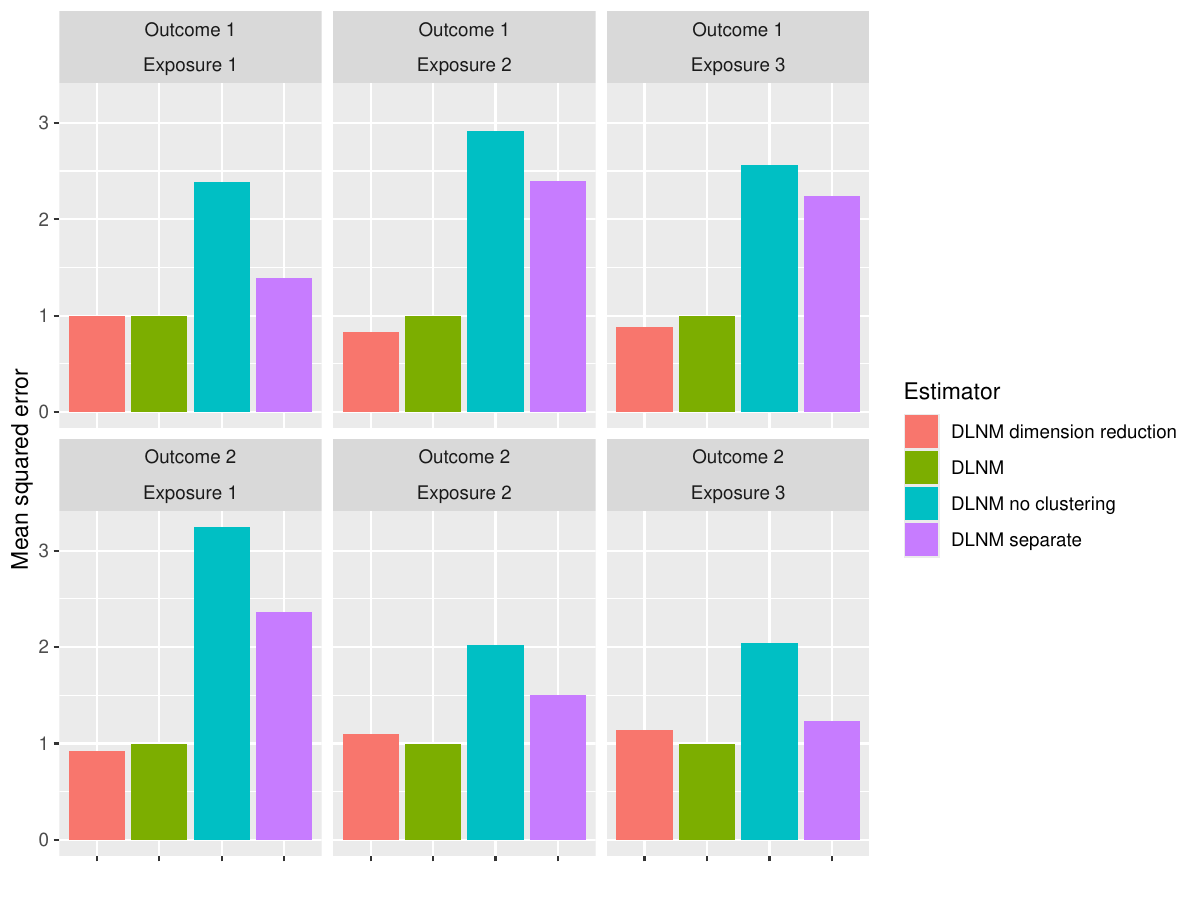}
    \caption{Mean squared error for estimating $f_{kj}(\cdot)$ across all estimators when the sample size is $n=1500$. MSE is scaled so that the DLNM estimator has an MSE of 1 in all scenarios. }
    \label{fig:DLAG_MSEfxlarge}
\end{figure}

Here we present results from the same data generating process as the distributed lag simulation study in Section 5.1, however, we increase the sample size to $n=1500$. The results can be found across Figures \ref{fig:DLAG_MSElarge} - \ref{fig:DLAG_MSEfxlarge}. The results are largely similar to what was seen for smaller sample sizes, with a few expected, but notable, changes. Overall, the results point to the same finding that our proposed DLNM estimator (or DLNM with dimension reduction) generally outperforms estimators that do not perform clustering. One key difference in these results is that the clustering for the $\boldsymbol{\omega}_{kj}$ terms is improved relative to the smaller sample size setting. This can be seen in Figure \ref{fig:DLAG_clusteringlarge} where there is clearer clustering structure, or in Figure \ref{fig:DLAG_estimateslarge} where there are fewer data sets for which the estimated distributed lag curves are placed in the wrong cluster. This leads to improved credible interval coverages, as seen in Figure \ref{fig:DLAG_COVlarge}. We also see that it is preferable (in terms of MSE) to use our DLNM model with clustering across both exposures and outcomes for estimating $f_{kj}(\cdot)$, as seen in Figure \ref{fig:DLAG_MSEfxlarge}. As a general rule, the $\boldsymbol{\omega}_{kj}$ parameters are only weakly identified, and are very difficult to estimate from small sample sizes, because of the high degree of correlation in the exposures across time. These issues improve as the sample size increases, which leads to the improved clustering performance, and the improved results when performing inference on the $\boldsymbol{\omega}_{kj}$ parameters. 

\subsection{Sensitivity to choice of $C$}

\begin{figure}[hbtp!]
    \centering
    \includegraphics[width=0.9\linewidth]{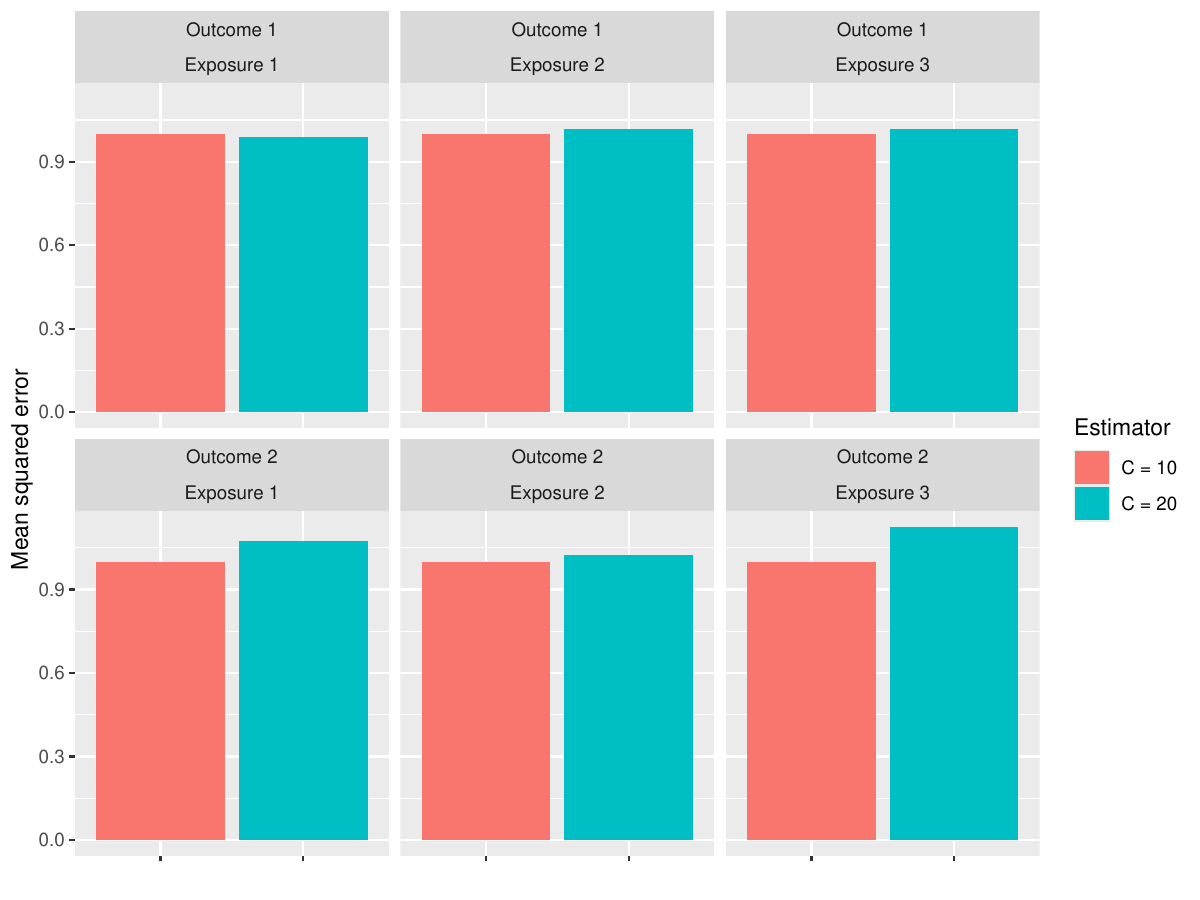}
    \caption{Mean squared error for estimating $\boldsymbol{\omega}_{kj}$ across all estimators using different values of $C$. }
    \label{fig:DLAG_MSE_Cchoice}
\end{figure}

\begin{figure}[hbtp!]
    \centering
    \includegraphics[width=0.9\linewidth]{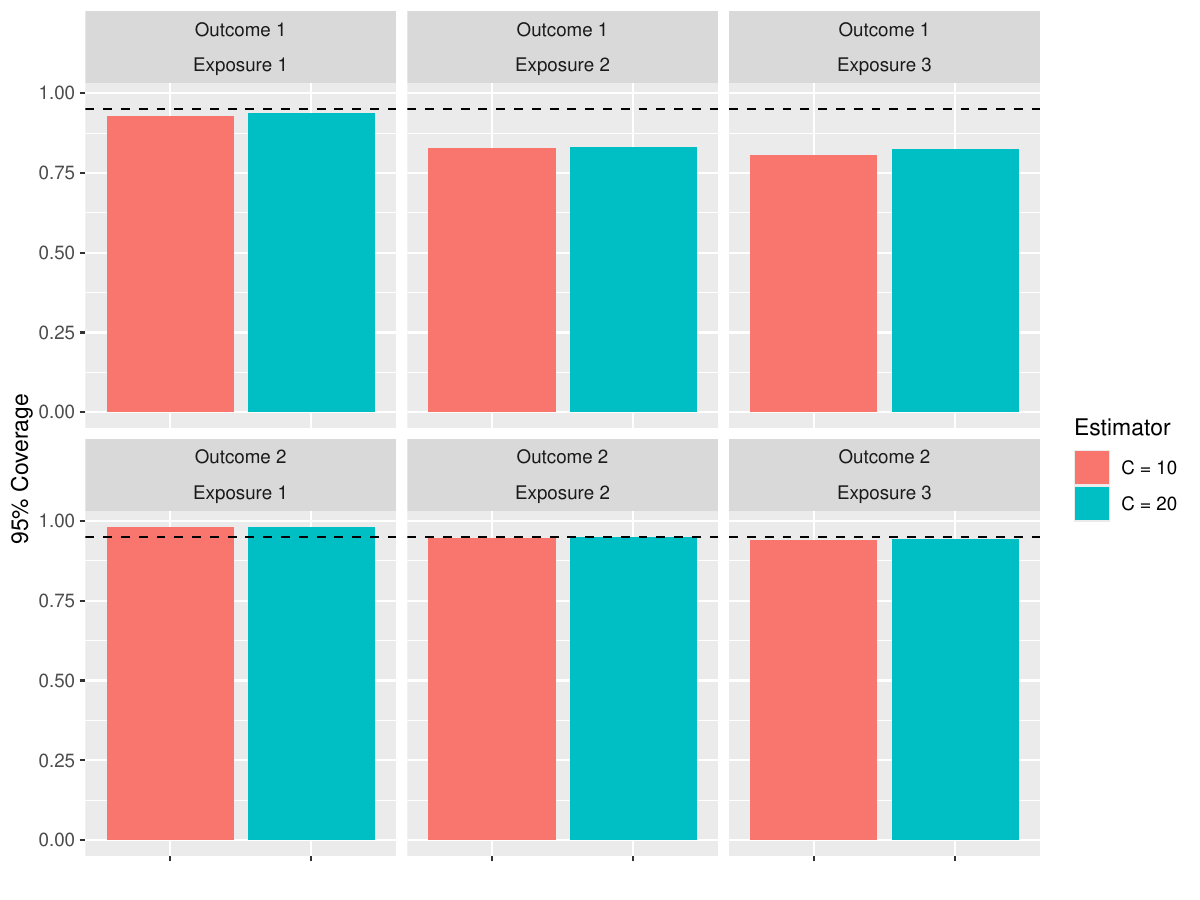}
    \caption{95\% credible interval coverage for estimating $\boldsymbol{\omega}_{kj}$ across all estimators using different values of $C$. }
    \label{fig:DLAG_COV_Cchoice}
\end{figure}

\begin{figure}[htbp!]
    \centering
    \includegraphics[width=0.9\linewidth]{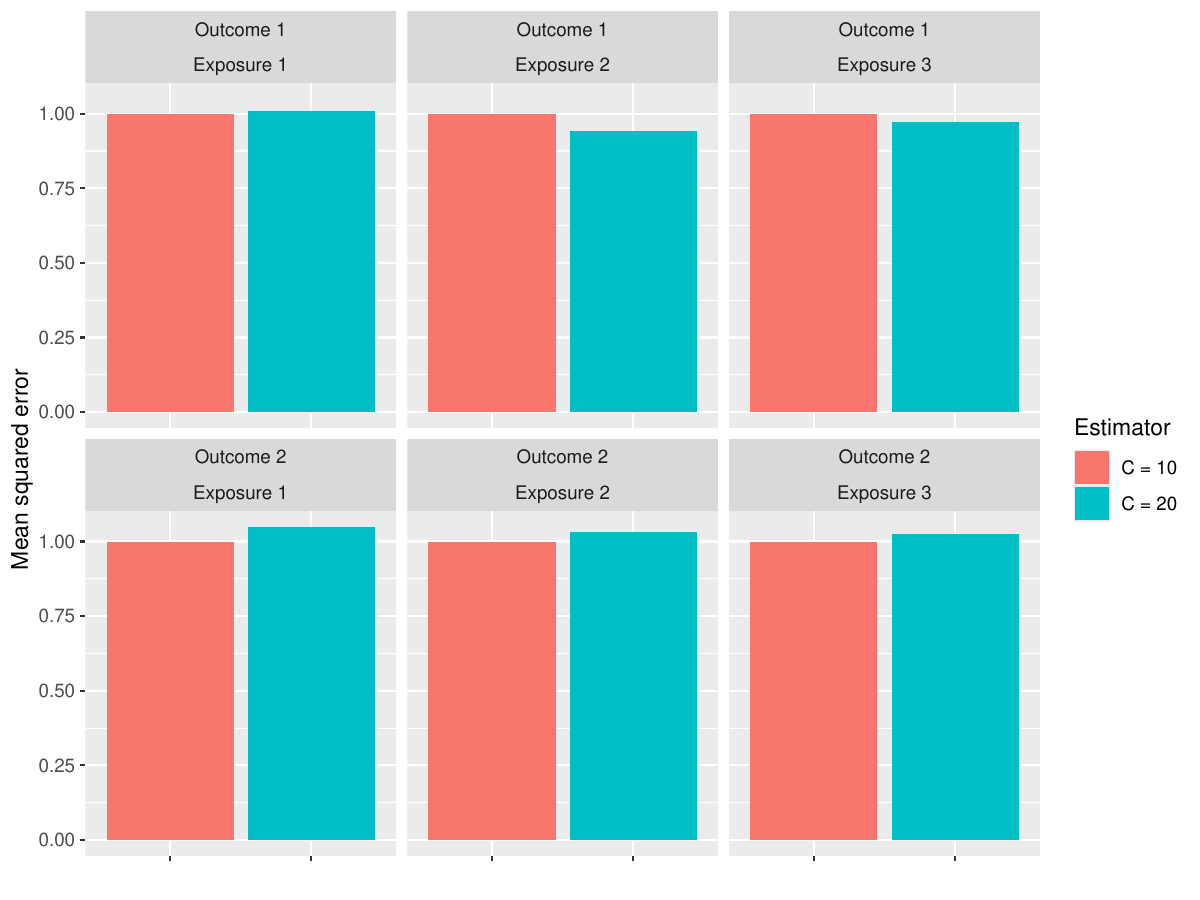}
    \caption{Mean squared error for estimating $f_{kj}(\cdot)$ across all estimators when using different values of $C$. }
    \label{fig:DLAG_MSEfx_Cchoice}
\end{figure}

Here we assess the robustness of our approach to the choice of $C$, the number of mixtures within our finite mixture model. If one is considered about ensuring a large enough value of $C$ is chosen, then we recommend setting $C = K \times P$, which is the maximum number of distinct exposure-response functions that are needed, even when there is no clustering across exposures or outcomes. Ideally, if $C$ is chosen to be sufficiently large to ensure that enough distinct groups are available, then increasing $C$ should have little bearing on the final results as the model's clustering properties should remove the effects of any unnecessary groups in the finite mixture model. To assess this empirically, we run the simulations from Section 5.1 in the manuscript, though we vary the choice of $C \in \{ 10, 20\}$. In this setting, we have $K=4$ and $P=5$, so 20 clusters is the maximum possible number of distinct clusters. In the simulation study in the manuscript, we fixed $C=10$ throughout, and here we evaluate the sensitivity to this choice. The results can be found in Figures \ref{fig:DLAG_MSE_Cchoice}, \ref{fig:DLAG_COV_Cchoice}, and \ref{fig:DLAG_MSEfx_Cchoice}. We see that the choice of $C$ does not make a large impact on the overall results. Both models have nearly identical 95\% coverage rates across all exposures and outcomes examined, and the mean squared error for either estimating $\boldsymbol{\omega}_{kj}$ or $f_{kj}(\cdot)$ are similar between the two results. The MSE is slightly higher generally for $C=20$, though there are cases where it is higher when $C=10$ and the differences between the two are rather small. This is expected due to the clustering properties of the model that ensure unnecessary clusters do not have a large impact on the resulting estimates. More problematic would be to consider values of $C$ that are too small, which would force parameters to be clustered together that otherwise would not be, though this is easy to avoid by setting $C$ to be a relatively large value, where relative is based on the value of $K \times P$.

\clearpage
\section{Additional NMMAPS Analysis Results}
\label{sec:appendix_chicago}

Following \cite{chen2019distributed}, we adjusted for: day of week (categorical), seasonality (cyclic basis splines for month, dim=4), year (thin plate splines, dim=6), and mean daily temperature and dew point temperature per day (thin plate splines, dim=6) and averaged over the prior three days (thin plate splines, dim=3), and we allowed different covariate associations for each outcome.

\begin{figure}[htbp!]
    \centering
    \includegraphics[width=0.9\linewidth]{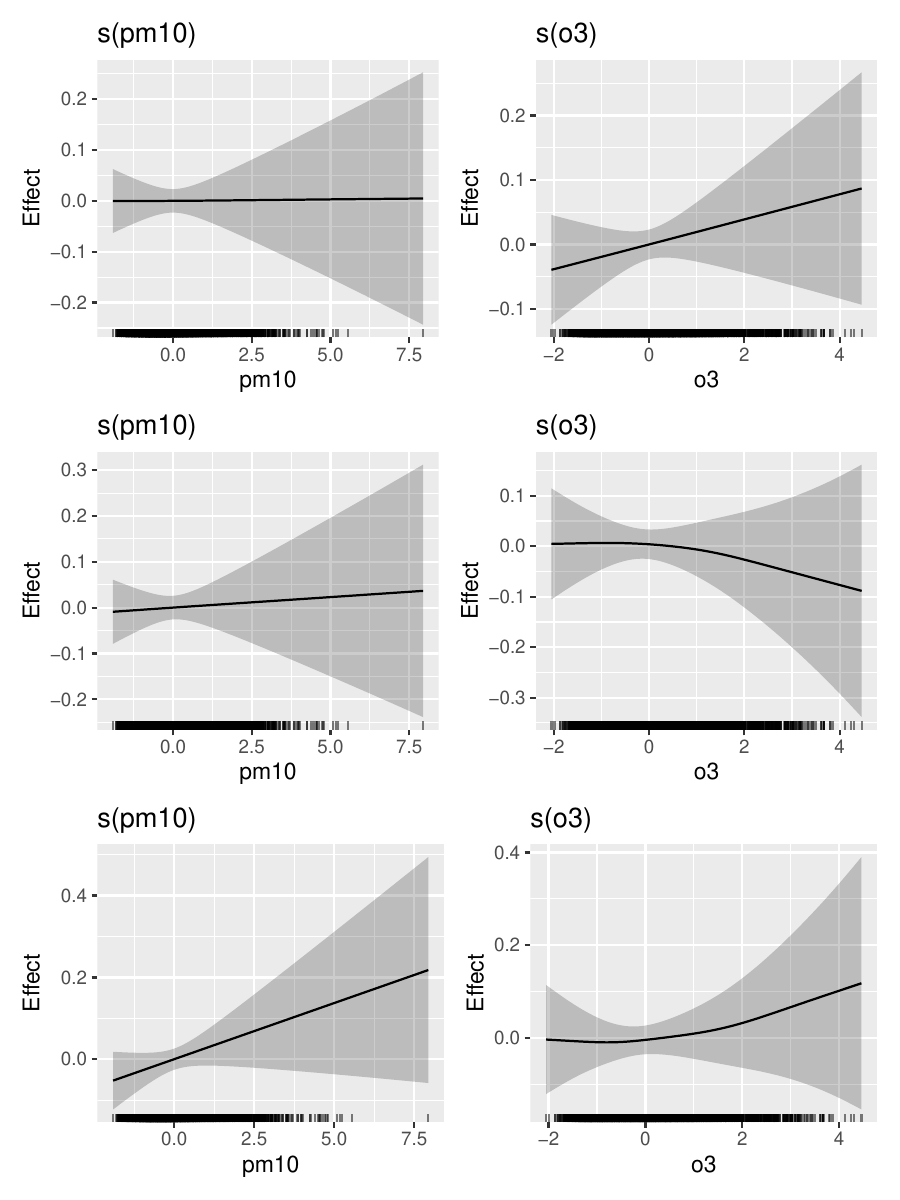}
    \caption{Results of preliminary fits of separate models based on single-day exposures. Three rows correspond to (log) cardiovascular, respiratory and other (non-accidental) deaths. Left column corresponds to associations with same-day PM$_{10}$; right is same-day O$_3$.  }
    \label{fig:gam}
\end{figure}

In Figure \ref{fig:gam} we report results of a preliminary analysis using separate additive models based on single-day exposures. We found no statistically significant associations with same day exposure, but this ignores the fact that exposures may have delayed effects, hence motivating lagged exposure models like DLNMs. 

\begin{figure}[htbp!]
    \centering
    \includegraphics[width=0.9\linewidth]{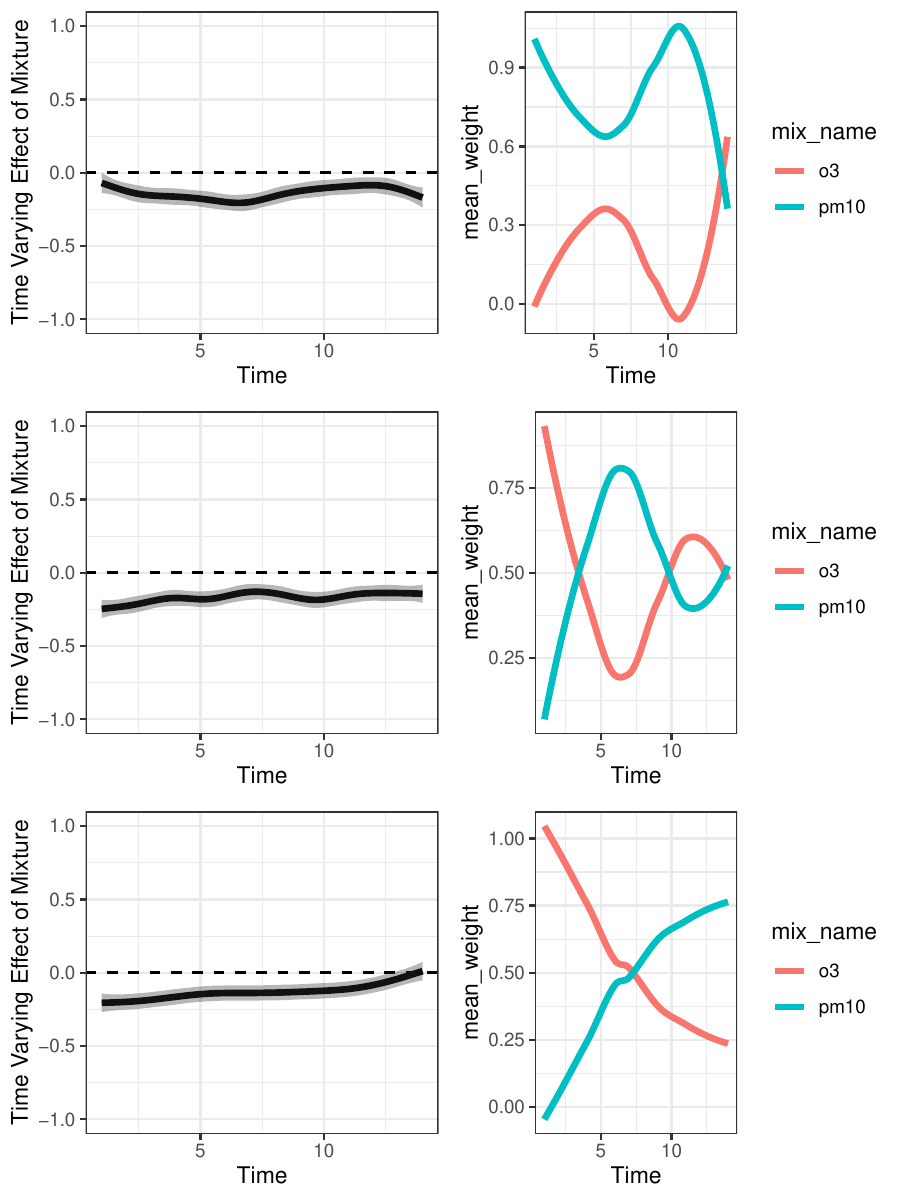}
    \caption{Results of preliminary fits of separate models via lagged WQS. Three rows correspond to (log) cardiovascular, respiratory and other (non-accidental) deaths. }
    \label{fig:lwqs}
\end{figure}

In Figure \ref{fig:lwqs} we report results of a preliminary analysis using separate lagged WQS models for each outcome (using quintiles, and $b=100$ bootstrap resamples). Results suggest weak negative associations with mixture, but uncertainty in time-varying exposure weights are ignored.

\begin{figure}[htb!]
    \centering
    \includegraphics[width=1\linewidth]{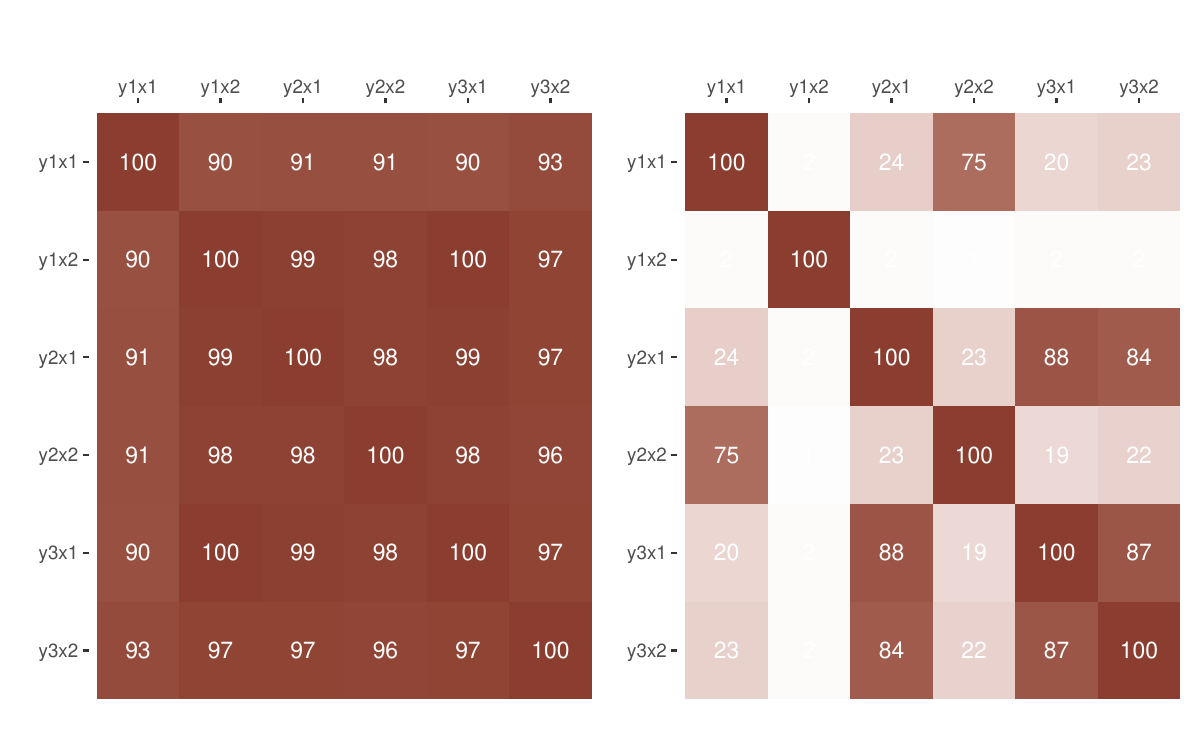}
    \caption{Pairwise posterior clustering probabilities for Chicago analysis. Left plot is for $\boldsymbol{\beta}$; right plot for $\boldsymbol{\theta}$}.
    \label{fig:chicagoHeat2}
\end{figure}

In Figure \ref{fig:chicagoHeat2} we report pairwise posterior clustering probabilities for the proposed MV-DLNM.

\begin{figure}[htbp!]
    \centering
    \includegraphics[width=0.9\linewidth]{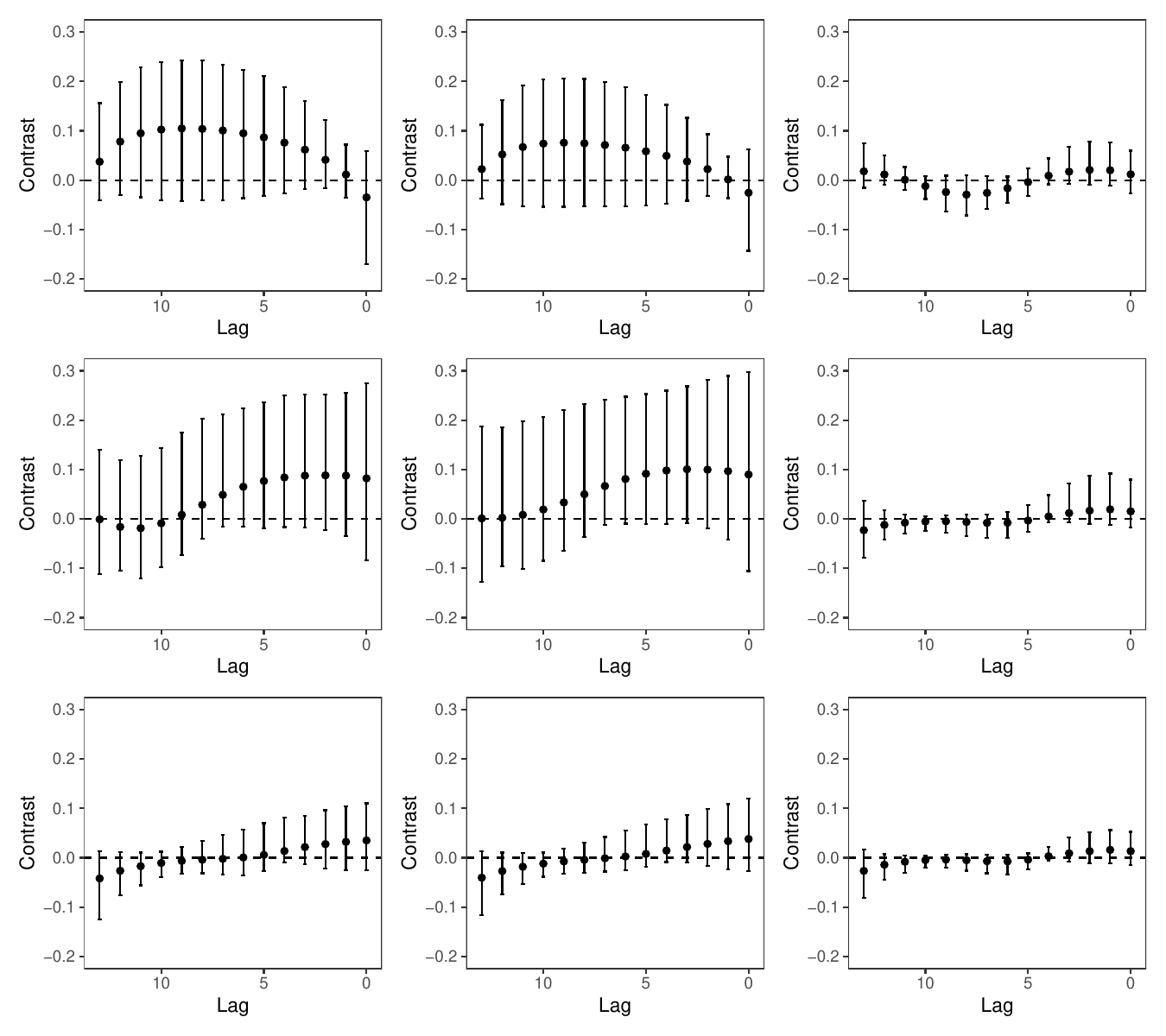}
    \caption{Mean difference in outcomes increasing PM$_{10}$ from 1SD below the mean to 1SD above at each lag (holding all else constant). Three rows correspond to (log) cardiovascular, respiratory and other (non-accidental) deaths. Left column is with separate model fits; middle is MV-DLNM without clustering; right is MV-DLNM with cluster-inducing priors.}
    \label{fig:contrast1}
\end{figure}

\begin{figure}[htbp!]
    \centering
    \includegraphics[width=0.9\linewidth]{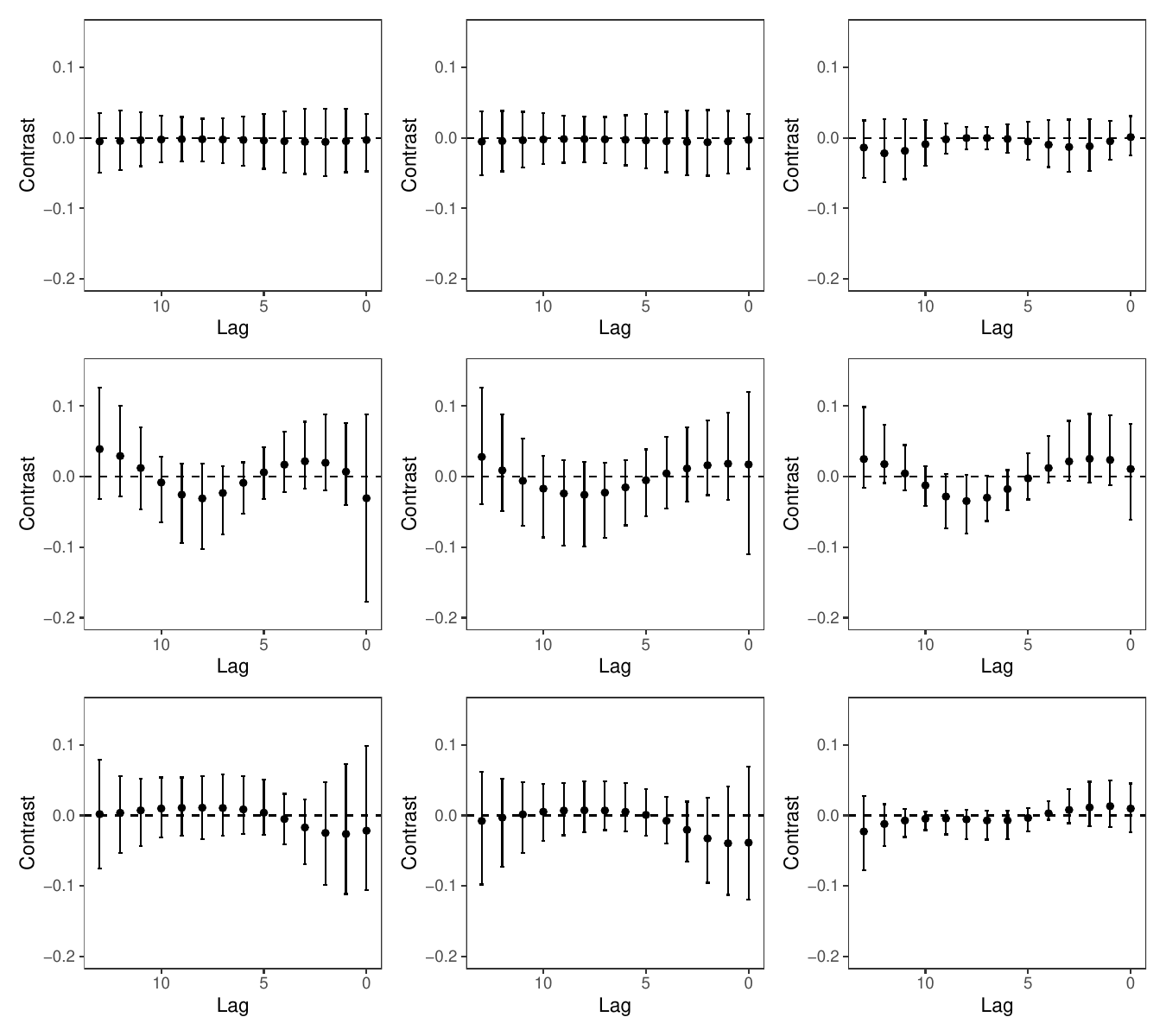}
    \caption{Mean difference in outcomes increasing O$_3$ from 1SD below the mean to 1SD above at each lag (holding all else constant). Three rows correspond to (log) cardiovascular, respiratory and other (non-accidental) deaths. Left column is with separate model fits; middle is MV-DLNM without clustering; right is MV-DLNM with cluster-inducing priors.}
    \label{fig:contrast2}
\end{figure}

In Figures \ref{fig:contrast1} and \ref{fig:contrast2} we report mean outcome differences corresponding to an increase in each exposure from 1SD below their mean to 1SD above at each lag $l$, holding exposures at all other lags constant.  Results for separate models are reported on the left hand column; MV-DLNM without clustering in the middle; results with clustering on the right. 

In Table \ref{tab:CIwidth} we quantify relative widths of credible intervals, comparing both the separate model fits and the MV-DLNM without clustering to the MV-DLNM with cluster-inducing priors.
\begin{table}[htbp!]
\centering
\caption{Relative interval width for lagged contrasts vs. proposed approach. ``Sep.'' represents separate model fits for each outcome; ``MV-DLNM (no clustering)'' represents the multivariate model but without cluster-inducing priors. Values greater than 1 indicate wider intervals than the proposed approach}
\label{tab:CIwidth}
\begin{tabular}{rrll}
  \hline
Exposure & Outcome & Sep. & MV-DLNM (no clustering) \\ 
  \hline
PM10 & Cardio. & 3.60 & 3.27 \\ 
       & Resp. & 4.68 & 5.29 \\ 
       & Other & 2.13 & 2.02 \\ 
\midrule
O3   & Cardio. & 1.39 & 1.41 \\ 
     & Resp.   & 1.42 & 1.54 \\ 
     & Other   & 2.33 & 2.12 \\ 
\midrule
Average & & 2.59 & 2.61 \\
\hline
\end{tabular}
\end{table}

\clearpage
\section{Alternative Model Specification: Non-Separable DLNM}\label{app:nonseparableDLNM}

We consider here the DLNM specification of \cite{gasparrini2017penalized}:
\begin{align}
y_{ik}&=g_{k1} ({\mathbf{x}}_{i1})+\cdots+g_{kP}({\mathbf{x}}_{iP})+\boldsymbol{z}_i^T\boldsymbol{\beta}_{Zk}+\xi \sigma_k u_i+\epsilon_{ik}, ~~k=1,\dots,K \nonumber \\ 
&\approx \sum_{l=1}^L h_{k1}({{x}}_{i1l},l)+\cdots+\sum_{l=1}^L h_{kP}({{x}}_{iPl},l)+\boldsymbol{z}_i^T\boldsymbol{\beta}_{Zk}+\xi \sigma_k u_i+\epsilon_{ik} \nonumber
\end{align}
where $h_{kj}(x,l)$ is a bi-dimensional dose-lag-response function \citep{gasparrini2017penalized}. 

We then adopt a tensor product basis representation, which requires specifying marginal bases for the $x$- and time-dimensions. We use the same bases as described in the main text. Applying these (known) basis functions to $\boldsymbol{x}_{ij}$ and $[1,\dots,L]^T$ yields corresponding basis matrices $\mathbf{R}_{ij}$ (of dimension $L\times d$) and $\boldsymbol{\Psi}$ (of dimension $L\times m$). This yields the representation:
\begin{align}
y_{ik}&\approx \beta_{k0} + \mathbf{b}_{i1}^T\boldsymbol{\beta}_{k1}+\cdots+\mathbf{b}_{iP}^T \boldsymbol{\beta}_{kP}+\boldsymbol{z}_i^T\boldsymbol{\beta}_{Zk}+\xi \sigma_k u_i+\epsilon_{ik}  \nonumber
\end{align}
where
\begin{align}
\mathbf{b}_{ij}^T&= \mathbf{1}^T_L \left[\left(\mathbf{R}_{ij} \otimes \mathbf{1}_{m}^{\top}\right) \odot\left(\mathbf{1}_{d}^{\top} \otimes \boldsymbol{\Psi}\right) \right], \nonumber
\end{align} is a known vector, and $\boldsymbol{\beta}_{kj}$ is an unknown coefficient vector of length $d\cdot m$. Note as in the main text we have included an intercept and imposed centering constraints to the basis representations \citep{wood2017generalized}.

To encourage smoothness in the bidimensional function we can adopt Gaussian priors for $\boldsymbol{\beta}_{kj}$ with mean $\mathbf{0}$ and precision matrix $\lambda^\beta \widetilde{\mathbf{P}}$ where $\widetilde{\mathbf{P}}= (\boldsymbol{\Sigma}_0^{-1}\otimes \mathbf{I}_{m}) + ( \mathbf{I}_{d}\otimes \boldsymbol{\Sigma}_{\theta}^{-1}) $, where $\boldsymbol{\Sigma}_0^{-1}$ and $\boldsymbol{\Sigma}_{\theta}^{-1}$ are penalty matrices as in the main text, and $\lambda^\beta$ and is a smooth penalty parameters defined in the main text. (Here we use a single penalty parameter for simplicity, but this can be extended to more complex penalties for different dimensions; sampling follows as in \citealp{bach2025anisotropic}.)

\subsection{Proposed Cluster-Inducing Priors}
\label{app:nonseparableDLNM_priors}
As in the main model specification, we aim to borrow strength across exposures and outcomes. A key departure for this model specification, however, is that there is no $\boldsymbol{\theta}_{kj}$; all parameters related to the dose-lag-response function for the $k^{th}$ outcome and the $j^{th}$ exposure is captured by the single (albeit much larger) vector $\boldsymbol{\beta}_{kj}$. An advantage of this non-separable specification is that we don't need to propose a dependent co-clustering strategy; we can simply place cluster-inducing priors on the larger coefficient vectors $\boldsymbol{\beta}_{kj}$. A disadvantage, however, is this doesn't allow separate levels of clustering lag profiles and exposure response relationships.  Specifically we propose DP mixture priors $\boldsymbol{\beta}_{kj}\sim G_{\beta}$ where $G_{\beta} \sim DP(\alpha_{\beta}, G_{0, \beta})$, and we use the same parameterization as in the main manuscript. As centering distribution $G_{0, \beta}$, we adopt the smoothness-inducing Gaussian described above. We use the same default priors on all other parameters.

\subsection{Modified MCMC Sampler}
\label{app:nonseparableDLNM_MCMC}

\begin{enumerate}

		\item For $k=1,\dots,K$, $j=1,\dots,P$, sample cluster IDs:
 
	\begin{align*}
P(Z^{\beta}_{kj}=a|...)&=\frac{\pi_{a}^{\beta} \prod_{i=1}^n f(y_{ik};\boldsymbol{\beta}^*_{Z^{\beta}_{k1}},\dots,\boldsymbol{\beta}^*_{Z^{\beta}_{k(j-1)}},\boldsymbol{\beta}^*_{a},\boldsymbol{\beta}^*_{Z^{\beta}_{k(j+1)}},\dots,\boldsymbol{\beta}^*_{Z^{\beta}_{kP}},u_i)}{\sum_{c=1}^C \pi_{c}^{\beta} \prod_{i=1}^n f(y_{ik};\boldsymbol{\beta}^*_{Z^{\beta}_{k1}},\dots,\boldsymbol{\beta}^*_{Z^{\beta}_{k(j-1)}},\boldsymbol{\beta}^*_{c},\boldsymbol{\beta}^*_{Z^{\beta}_{k(j+1)}},\dots,\boldsymbol{\beta}^*_{Z^{\beta}_{kP}},u_i)}, ~~a \in \{1,\dots,C\}.
	\end{align*}

\item  For $c=1,\dots,C$, if $\sum_{kj} Z^{\beta}_{kj}=0$, then draw $\boldsymbol{\beta}^*_c$ from the prior. Otherwise sample $\boldsymbol{\beta}^*_c$ from full conditional: 
\begin{align*}
\boldsymbol{\beta}^*_c|... &\sim N(\boldsymbol{\mu}^*,\boldsymbol{\Sigma}^*) \\
\boldsymbol{\mu}^*&= 
\boldsymbol{\Sigma}^* \times 
\left( \lambda^{\beta}\boldsymbol{\mu}_0^T \widetilde{\mathbf{P}} +\sum_{k}^K \sigma^{-2}_k\left[ \sum_{j:Z^\beta_{kj}=c}\mathbf{B}_{j}^T\right]\left[\mathbf{y}_{ k}-\beta_{0k}\mathbf{1}-\xi \sigma_k \mathbf{u}-\sum_{j:Z^\beta_{kj}\neq c}\mathbf{B}_{j} \boldsymbol{\beta}_{kj} \right]\right) \\
\boldsymbol{\Sigma}^*&=
\left(\lambda^{\beta}\widetilde{\mathbf{P}}+\sum_{k}^K \sigma^{-2}_k\left[ \sum_{j:Z^\beta_{kj}=c}\mathbf{B}_{j}^T\right]\left[\sum_{j:Z^\beta_{kj}=c}\mathbf{B}_{j}\right]\right)^{-1}
\end{align*}

\item Sample penalty term $\lambda^{\beta}$ from full conditional:
\begin{align*}
( (\lambda^{\beta})|\dots) &\propto Gamma(a_{\lambda}^{\beta}+\frac{C}{2}\text{rank}(\widetilde{\mathbf{P}}),~~b_{\lambda}^{\beta}+\frac{1}{2}\sum_{c=1}^C{\boldsymbol{\beta}_c^{*}}^T\widetilde{\mathbf{P}} \boldsymbol{\beta}_c^{*})
\end{align*}

\end{enumerate}

All other updates---for $V^{\beta}_c$, $\alpha_{\beta}$, $u_i$, $\xi$, $\sigma^2_k$, $\beta_{0k}$ and $\boldsymbol{\beta}_{Zk}$---are the same as in Appendix \ref{app:sampler},  replacing $\sum_{j=1}^p\mathbf{b}_{i,\boldsymbol{\theta}_{kj}}^T \boldsymbol{\beta}_{kj}$ with $\sum_{j=1}^p\mathbf{b}_{ij}^T \boldsymbol{\beta}_{kj}$.

\subsection{Non-Separable DLNM: Simulations}

In this section we evaluate the proposed approach within the nonseparable distributed lag framework described above. We explore three distinct approaches that vary in their degree of clustering:
\begin{enumerate}
    \item The proposed nonseparable formulation with clustering across exposures and outcomes (DLNM)
    \item The proposed nonseparable formulation without any clustering (DLNM no clustering)
    \item The proposed nonseparable formulation applied separately to each outcome, which allows for clustering across exposures within an outcome, but not across outcomes (DLNM separate)
\end{enumerate}

\begin{figure}[hbtp!]
    \centering
    \includegraphics[width=0.7\linewidth]{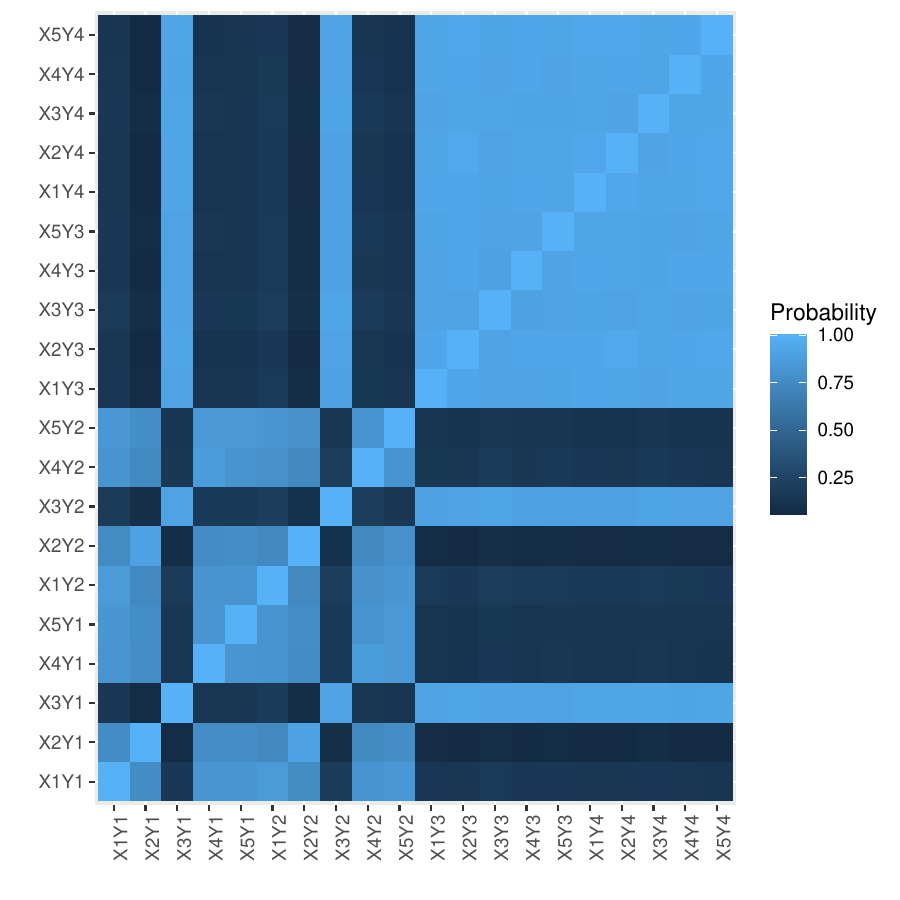}
    \caption{Posterior probabilities of clustering for model parameters, averaged across simulated datasets in the non-separable setting.}
    \label{fig:Nonsep_Clustering}
\end{figure}

\begin{figure}[hbtp!]
    \centering
    \includegraphics[width=0.9\linewidth]{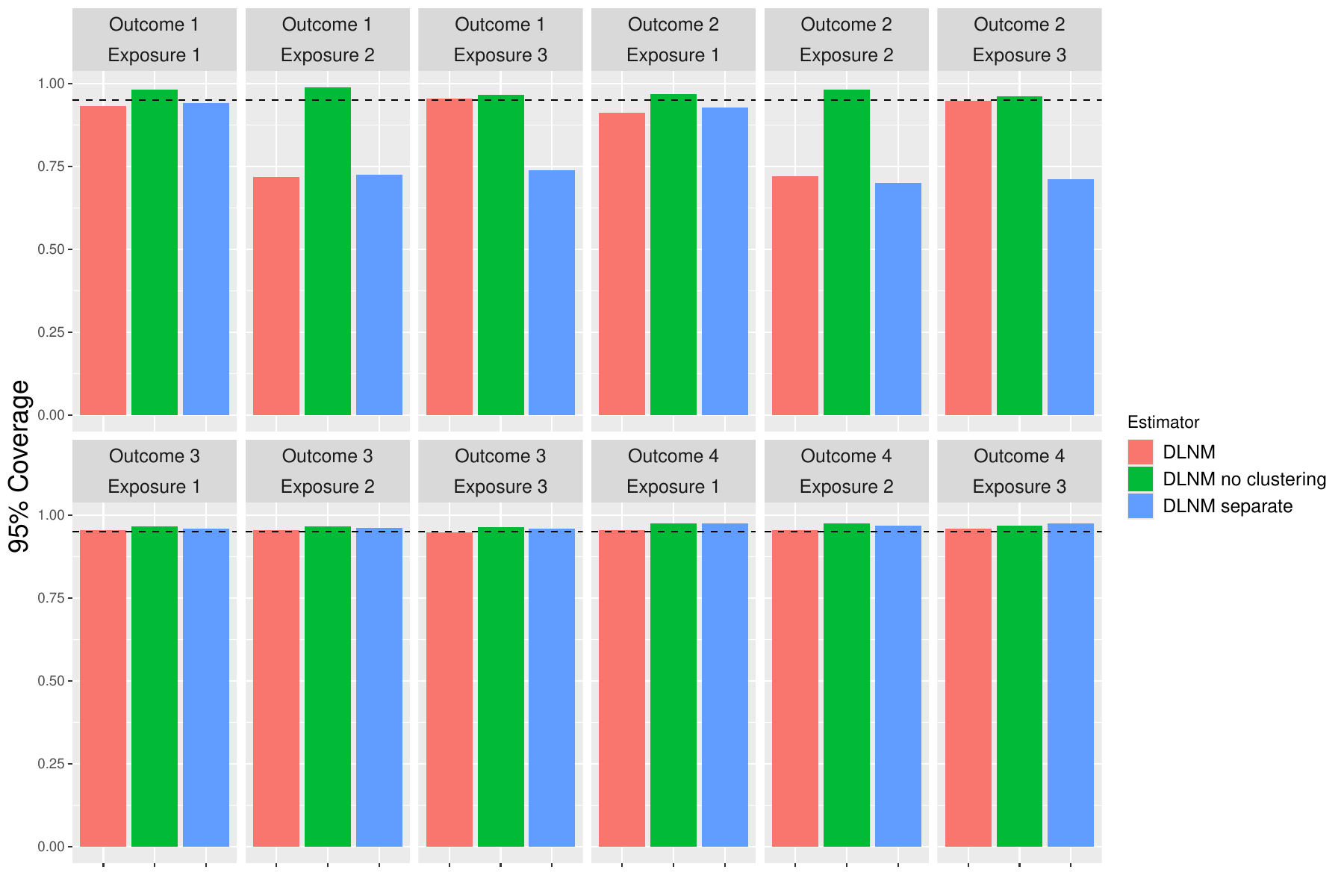}
    \caption{95\% credible interval coverage for estimating $f_{kj}(\cdot)$ across all estimators in the non-separable setting.}
    \label{fig:Nonsep_Cov}
\end{figure}

\begin{figure}[htbp!]
    \centering
    \includegraphics[width=0.9\linewidth]{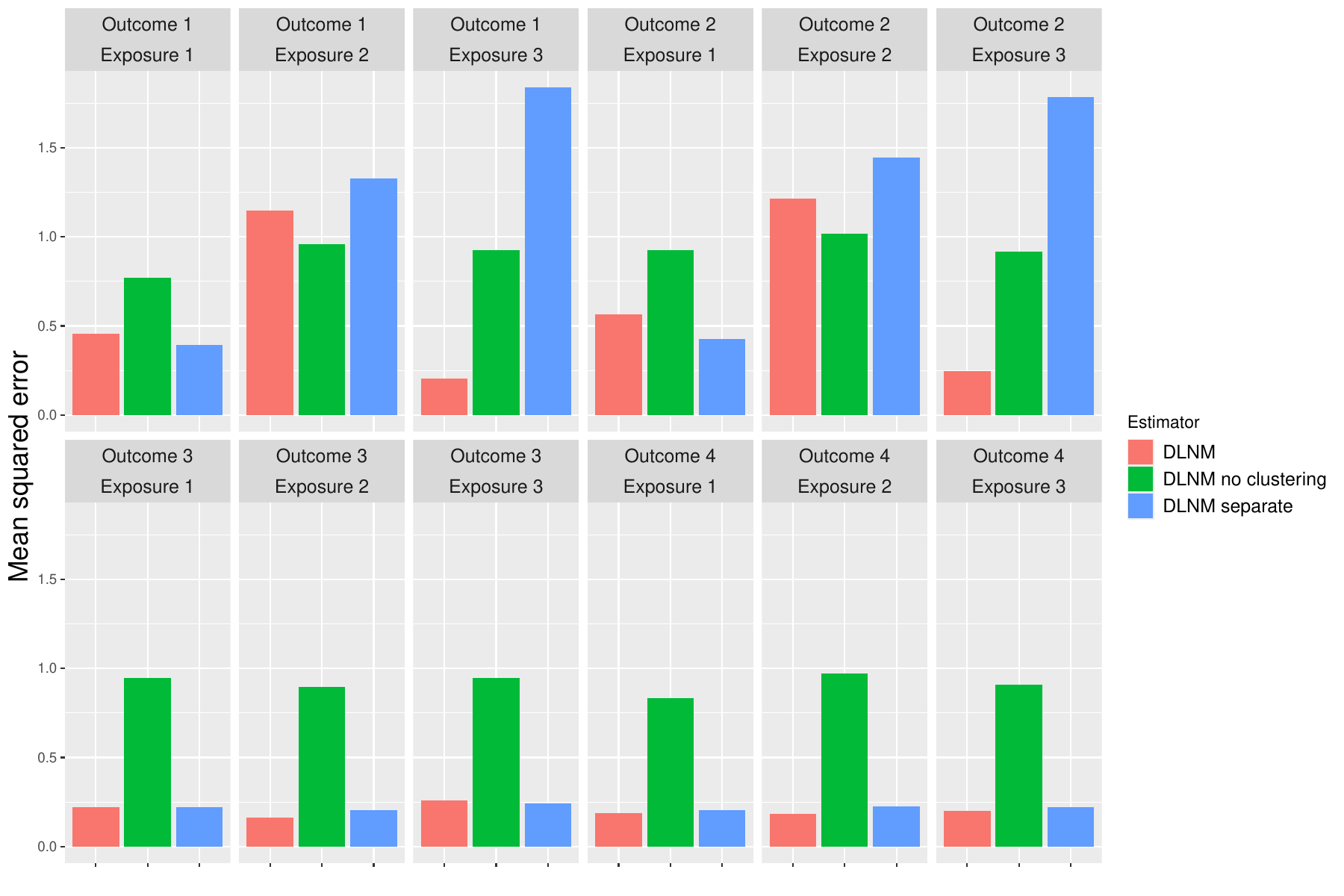}
    \caption{Mean squared error for estimating $f_{kj}(\cdot)$ across all estimators in the non-separable setting. }
    \label{fig:Nonsep_MSE}
\end{figure}

As in the manuscript simulations, we set the sample size to be $n=500$, the number of time points to be $L=52$, the number of exposures to be $P=5$, and the number of outcomes to be $K=4$. The exposures have an autoregressive correlation structure that ensures exposures at successive time points have correlation 0.85, while the exposures are uncorrelated with each other at each point in time. Responses were generated from a multivariate normal distribution with mean:
\begin{align*}
E[\boldsymbol{Y} \mid \boldsymbol{X}]&=\left[\begin{array}{l}
g_1(\mathbf{x}_{i1})+g_2(\mathbf{x}_{i2})+g_3(\mathbf{x}_{i3})+g_1(\mathbf{x}_{i4})+g_1(\mathbf{x}_{i5})\\
g_1(\mathbf{x}_{i1})+g_2(\mathbf{x}_{i2})+g_3(\mathbf{x}_{i3})+g_1(\mathbf{x}_{i4})+g_1(\mathbf{x}_{i5})\\
g_3(\mathbf{x}_{i1})+g_3(\mathbf{x}_{i2})+g_3(\mathbf{x}_{i3})+g_3(\mathbf{x}_{i4})+g_3(\mathbf{x}_{i5})\\
g_3(\mathbf{x}_{i1})+g_3(\mathbf{x}_{i2})+g_3(\mathbf{x}_{i3})+g_3(\mathbf{x}_{i4})+g_3(\mathbf{x}_{i5})\\
\end{array}\right]
\end{align*}
where $$g_m(\mathbf{x}_{ij})=\sum_{\ell=1}^L f_m \cdot w_m\left(x_{ij\ell}, \ell\right)  $$,
and
$$\begin{aligned} & f_1 \cdot w_1\left(x_{ij\ell}, \ell\right)=0.1 \cdot f_{1}\left(x_{ij\ell}\right) \cdot w_{1}(\ell), \\ & f_2 \cdot w_2\left(x_{ij\ell}, \ell\right)={l}f_{2}\left(x_{ij\ell}\right) \cdot w_{2}(\ell) \\ & f_3 \cdot w_3\left(x_{ij\ell}, \ell\right)=\left\{\begin{array}{l}f_{3}\left(x_{ij\ell}\right) \cdot w_{1}(\ell) \quad \text { if } x \geq 0 \\ f_{3}\left(x_{ij\ell}\right) \cdot w_{3}(\ell) \quad \text { if } x<0\end{array},\right.\end{aligned}$$
where $w_1(\ell)$ is flat, $w_2(\ell)$ is decreasing,  
$w_3(\ell)$  is increasing, and:
\begin{align*}
 f_{1}(x)&=0.04 \cdot(x-2) \\
 f_{2}(x)&=\sum_{p=0}^4 0.3 \delta_p (x^p-5^p) \\
 f_{3}(x)&=0.4 \left(\phi(x;\mu=1.5,\sigma=2)+\phi(x;\mu=7.5,\sigma=1)\right)\\
&~~-0.4\left(\phi(5;\mu=1.5,\sigma=2)+\phi(5;\mu=7.5,\sigma=1)\right)
\end{align*}
where $\boldsymbol{\delta}=[0.2118881,0.1406585,-0.0982663,0.0153671,-0.0006265]^{T}$. Note that these functional forms here were motivated by those of \cite{gasparrini2017penalized}. The results can be found in Figures \ref{fig:Nonsep_Clustering}, \ref{fig:Nonsep_Cov}, and \ref{fig:Nonsep_MSE}. Overall, we see that clustering dramatically improves performance within the non-separable model formulation as well. The model with clustering significantly outperforms the other two models in terms of mean squared error. We see that as in the manuscript, this can at times lower the 95\% interval coverage rate due to bias incurred from clustering, but this comes with large reductions in MSE. Lastly, we can see from the posterior probabilities of clustering that exposure-outcome pairs with similar effects on the outcome are indeed clustered together with high probability, allowing for borrowing of information across exposures and outcomes.

\clearpage
\section{Alternative Model Specification: Non-Additivity} \label{app:nonadditivity}

Here we discuss extensions of our approach to allow for non-additivity, with a particular focus on the distributed lag application of our general modeling approach. We do not discuss non-additivity in the adaptive multiple index model setting described in Section 4, because the effects of the exposures in that setting are already allowed to be non-additive. We focus on the distributed lag nonlinear model described in Section 3, as our current approach does not allow for the effects of one exposure to depend on the values of the others. In this setting, interactions between exposures can occur in complicated ways as exposure $j$ at time point $t$ can interact with exposure $j'$ at time point $t'$. To allow for arbitrary interactions such as this, we can fit the proposed extension of our model given by:
\begin{align*}
y_{ik}&=\sum_{j=1}^P f_{kj}({\mathbf{x}}_{ij}^T{\boldsymbol{\theta}}_{kj})+\sum_{j=1}^P\sum_{j'=j+1}^P \tilde{f}_{kjj'}({\tilde{\mathbf{x}}}_{ijj'}^T{\tilde{\boldsymbol{\theta}}}_{kjj'})+\boldsymbol{z}_i^T\boldsymbol{\beta}_{Zk}+\xi \sigma_k u_i+\epsilon_{ik},
\end{align*}
where we now have that ${\tilde{\mathbf{x}}}_{ijj'}$ contains all of the $L^2$ values for exposures $j$ and $j'$ at all possible pairs of time points, and ${\tilde{\boldsymbol{\theta}}}_{kjj'}$ is the corresponding parameter vector. The parameters ${\tilde{\boldsymbol{\theta}}}_{kjj'}$ can be viewed as a (vectorized) two-dimensional matrix of dimension $L \times L$ that corresponds to all possible pairs of interactions between exposures $j$ and $j'$. The same co-clustering ideas developed in the manuscript for $f_{kj}(\cdot)$ and $\boldsymbol{\theta}_{kj}$ could be applied in a similar manner for $\tilde{f}_{kjj'}(\cdot)$ and ${\tilde{\boldsymbol{\theta}}}_{kjj'}$. The main difficulty for implementation of this approach is the high-dimensional nature of the parameter vector as we would have $L^2$ parameters for each pair of exposures, and there are $\binom{P}{2}$ such pairs. One option is to do as we did for the $\boldsymbol{\theta}_{kj}$ parameters and assume smoothness over time, which can be incorporated through appropriate prior distribution variances for these parameters. A similar approach could be taken for ${\tilde{\boldsymbol{\theta}}}_{kjj'}$ although it is not trivial to develop a penalty matrix that incorporates smoothness in two dimensions, though this could be a topic for future research. Alternative approaches would be to impose smoothness by approximating ${\tilde{\boldsymbol{\theta}}}_{kjj'}$ as a linear combination of two-dimensional basis functions as done in \cite{antonelli2024multiple}, though this requires specifying basis functions a priori and may not work well if the true coefficient matrix is not well captured by the chosen basis functions. This can be viewed as a special case of the dimension reduction that we applied in the manuscript, and the same co-clustering ideas could apply immediately. In practice, however, one should be careful implementing such a model because interaction effects in distributed lag settings for environmental mixtures are likely very small and difficult to detect with even moderate sample sizes. Additionally, increasing the complexity of the model unnecessarily may lead to less stable and highly variable results unless appropriate regularization or variable selection is implemented. However, such regularization or variable selection will likely remove any interaction effects, even if they exist, due to their small signal.

\clearpage
\section{Identifiability \& Orthogonalization}
\label{app:identifiability}

As in a typical mixed effects model, the fixed effects are of primary interest, and the random effects serve to model residual correlation that remains after adjusting for exposure/covariate fixed effects. The clustering strategy aims to borrow strength in those exposure effects across different outcomes, whereas inclusion of a random effect $u_i$ serves to absorb residual correlation not captured by between-subject variation in exposure levels. In our case, however, the fact that the exposure effects are encouraged by the cluster-inducing priors to be similar means there could be an identifiability problem. This is analogous  to a phenomenon known as spatial confounding (not to be conflated with the causal notion of confounding), wherein near non-identifiability can arise when spatially varying covariates are included in a model in addition to spatially varying random effects. In the spatial statistics literature, several methods have been proposed to solve this issue, including restricted spatial regression (RSR; \citealp{reich2006effects}). The RSR approach reparameterizes the model such that the random effects are orthogonal to the column space of the fixed effects. For example replacing the vector of random effects $\boldsymbol{u}$ in the model
\begin{align*}
\mathbf{y}_{\cdot k}&\approx \beta_{k0}\mathbf{1} + \mathbf{B}_{\cdot k}\boldsymbol{\beta}_{k\cdot}+\boldsymbol{Z} \boldsymbol{\beta}_{Zk}+\xi \sigma_k \boldsymbol{u}+\boldsymbol{\epsilon}_{\cdot k}  \\ 
\boldsymbol{u}&{\sim} N(\mathbf{0},\mathbf{I}) 
\end{align*}
with $\boldsymbol{u}^*=[\mathbf{I}-\mathbf{P}]\boldsymbol{u}$, where $\mathbf{P}$ is the usual projection matrix onto the column space of $\widetilde{\mathbf{B}}:=[\mathbf{1}|\mathbf{B}|\mathbf{Z}]$. A computationally efficient approach for implementing the RSR approach is conditional kriging \citep{hanks2015restricted,rue2005gaussian}, wherein the MCMC sampler draws random effects from the usual un-constrained posterior $\boldsymbol{u}\sim N(\boldsymbol{\mu}_u,\boldsymbol{\Sigma}_u)$ and then orthogonalizes them:
$\boldsymbol{u}^*=\boldsymbol{u}-\boldsymbol{\Sigma}_u \widetilde{\mathbf{B}}[\widetilde{\mathbf{B}}^T\boldsymbol{\Sigma}_u \widetilde{\mathbf{B}}]^{-1}\widetilde{\mathbf{B}} \boldsymbol{b}$. 

A challenge in our setting is that the equivalent column space here depends on unknown parameters that are being sampled, $\boldsymbol{\theta}_{kj}$, so MCMC becomes very challenging. 
As an approximation we simply change the $u$ update to be orthogonal  to $\mathbf{B}_{\cdot, \boldsymbol{\theta}_{1,\theta}}$ (but under the non-separable parameterization of \ref{app:nonseparableDLNM} the exact update is straightforward). 

\subsection{Identifiability Simulations}
Here we assess the potential impact of the identifiability issue described above, and the extent to which the conditioning by kriging approximation is able to address this issue. Throughout these simulations we focus on the version of our approach that implements clustering across both exposures and outcomes, and does incorporate dimension reduction when estimating $\boldsymbol{\theta}_{kj}$. We explore our approach both with and without conditioning by kriging, and we evaluate its performance in two simulation scenarios. The first simulation scenario is identical to the distributed lag scenario explored in the manuscript, where there should be no identifiability issues due to the fact that the exposure effects vary across the outcomes. The second simulation is identical to the first simulation with respect to the distribution of the exposures, though it varies with respect to the relationship between the exposures and the outcome. It represents the most extreme scenario where the effect of each exposure on each of the outcomes is identical, both with respect to the distributed lag parameters $\boldsymbol{\theta}_{kj}$ and the exposure response function that is captured by the $\boldsymbol{\beta}_{kj}$ parameters. Specifically, we generate the outcome from a multivariate normal distribution with mean:
\begin{align*}
E[\boldsymbol{Y} \mid \boldsymbol{X}]&=\left[\begin{array}{l}
f_2\left(\boldsymbol{x}_1^{T} \boldsymbol{\omega}_2\right)+f_2\left(\boldsymbol{x}_2^{T} \boldsymbol{\omega}_2\right)+f_2\left(\boldsymbol{x}_3^{T} \boldsymbol{\omega}_2\right)+f_2\left(\boldsymbol{x}_4^{T} \boldsymbol{\omega}_2\right)+f_2\left(\boldsymbol{x}_5^{T} \boldsymbol{\omega}_2\right) \\
f_2\left(\boldsymbol{x}_1^{T} \boldsymbol{\omega}_2\right)+f_2\left(\boldsymbol{x}_2^{T} \boldsymbol{\omega}_2\right)+f_2\left(\boldsymbol{x}_3^{T} \boldsymbol{\omega}_2\right)+f_2\left(\boldsymbol{x}_4^{T} \boldsymbol{\omega}_2\right)+f_2\left(\boldsymbol{x}_5^{T} \boldsymbol{\omega}_2\right) \\
f_2\left(\boldsymbol{x}_1^{T} \boldsymbol{\omega}_2\right)+f_2\left(\boldsymbol{x}_2^{T} \boldsymbol{\omega}_2\right)+f_2\left(\boldsymbol{x}_3^{T} \boldsymbol{\omega}_2\right)+f_2\left(\boldsymbol{x}_4^{T} \boldsymbol{\omega}_2\right)+f_2\left(\boldsymbol{x}_5^{T} \boldsymbol{\omega}_2\right) \\
f_2\left(\boldsymbol{x}_1^{T} \boldsymbol{\omega}_2\right)+f_2\left(\boldsymbol{x}_2^{T} \boldsymbol{\omega}_2\right)+f_2\left(\boldsymbol{x}_3^{T} \boldsymbol{\omega}_2\right)+f_2\left(\boldsymbol{x}_4^{T} \boldsymbol{\omega}_2\right)+f_2\left(\boldsymbol{x}_5^{T} \boldsymbol{\omega}_2\right)
\end{array}\right], 
\end{align*}
where, as in the manuscript, we have that $f_2(a)=0.24(0.3a)^2$ and $\boldsymbol{\omega}_2$ is the same increasing function used in the manuscript. 

\begin{figure}[htbp!]
    \centering
    \includegraphics[width=0.9\linewidth]{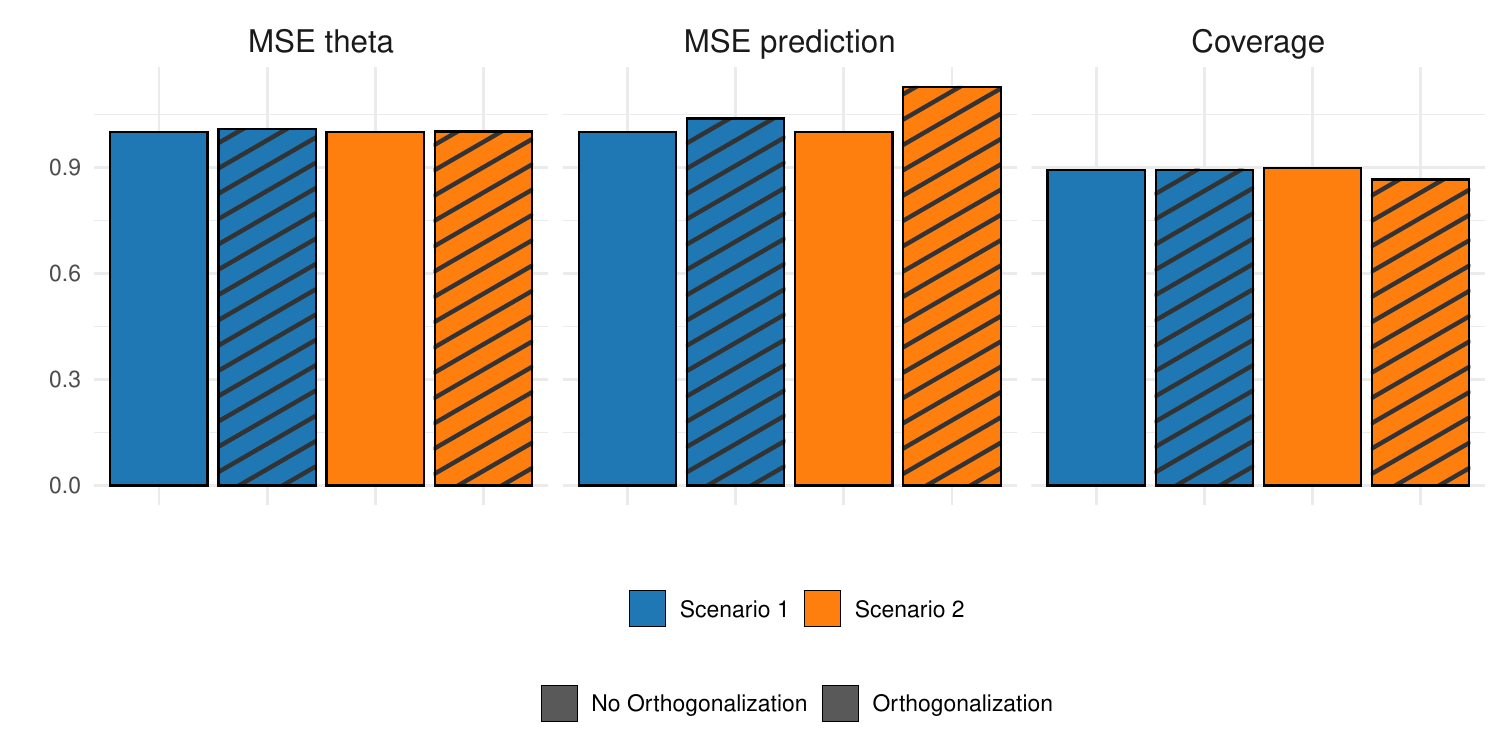}
    \caption{Comparison of results with and without orthogonalizing the random effects. }
    \label{fig:KrigingResults}
\end{figure}

The results can be seen in Figure \ref{fig:KrigingResults}, where we see that the decision about whether to orthogonalize the random effects does not substantially change the results. In both scenarios, the mean squared error for estimating the weight parameters or the entire mean function are largely similar, though slightly worse when implementing conditioning by kriging, and the coverage rates are quite similar between the two approaches in both scenarios. This is expected in scenario 1 where we should not have any identifiability issues, but is somewhat surprising for scenario 2 where it is expected that we would have an identifiability issue between the magnitude of the random effect variance and the magnitude of the exposure effects. While this identifiability issue could manifest in principle, it is not one that we encounter even in this extreme scenario. The reason for this is likely that it takes a reasonable number of MCMC samples for the clustering indicators to center on their correct values, which in scenario 2 would have all exposure and exposure-response parameters to be clustered together. Identifiability only becomes an issue when the exposure effects are identical across all outcomes, and it appears that the estimate of the random effect variance centers around the correct value in the early MCMC samples while the exposure effects are still not identical, which leads to the performance seen here. Overall this shows that identifiability is likely not a huge concern for our approach, particularly because it only is an issue in extreme scenarios.

\end{document}